\documentclass[graybox, envcountchap]{svmult}

\usepackage{mathptmx}        
\usepackage{amsmath}
\usepackage{amssymb}
\usepackage{color}
\usepackage{helvet}          
\usepackage{courier}         
\usepackage{dirtree}

\usepackage{makeidx}        
\usepackage{graphicx}        
\usepackage{subfig}

\usepackage{multicol}        
\usepackage[bottom]{footmisc}

\usepackage{xurl}
\usepackage[breaklinks]{hyperref}        
\hypersetup{colorlinks=true,urlcolor=blue}

\usepackage[misc]{ifsym}
\usepackage{wrapfig}
\usepackage{enumitem}
\usepackage{calrsfs}
\usepackage[font=small,labelfont=bf]{caption}

\makeindex             
                       
\usepackage[numbers]{natbib}                       
\usepackage[T1]{fontenc}
\usepackage[utf8]{inputenc}
\usepackage{txfonts}

\DeclareMathAlphabet{\pazocal}{OMS}{zplm}{m}{n}

\renewcommand\a{\alpha}
\renewcommand\b{\beta}
\newcommand\del{\delta}
\newcommand\ka{\kappa}
\newcommand\vk{\varkappa}
\newcommand\la{\lambda}
\renewcommand\r{\rho}

\renewcommand\t{\tau}
\newcommand\y{\upsilon}
\renewcommand\j{\psi}
\renewcommand\th{\theta}

\newcommand\e{\epsilon}
\newcommand\g{\gamma}

\newcommand\m{\mu}
\newcommand\n{\nu}

\newcommand\p{\pi}
\newcommand\h{\eta}
\newcommand\s{\sigma}
\newcommand\f{\phi}
\newcommand\w{\omega}
\newcommand{\wH}{\omega_{{\mkern-1mu}_\textit{H}}}

\newcommand\ve{\varepsilon}
\renewcommand\vk{\varkappa}

\newcommand\vf{\varphi}

\newcommand\La{\Lambda}

\renewcommand\S{\Sigma}

\renewcommand\D{\Delta}
\newcommand\G{\Gamma}

\newcommand{\lag}{\langle}
\newcommand{\rag}{\rangle}

\newcommand{\cA}{{\pazocal A}}
\newcommand{\cC}{{\pazocal C}}

\newcommand{\tF} {{\widetilde F}}

\newcommand{\cJ}{{\pazocal J}}

\newcommand{\cL}{{\pazocal L}}
\newcommand{\cM}{{\pazocal M}}
\newcommand{\cO}{{\pazocal O}}
\newcommand{\cS}{{\pazocal S}}

\newcommand{\pa}{\partial}

\newcommand{\nn}{\nonumber \\}
\newcommand{\na}{\nabla}
\newcommand{\sdfrac}[2]{\mbox{\small$\displaystyle\frac{#1}{#2}$}}

\newcommand{\GN}{G}
\newcommand{\rH}{r_{\!_H}}
\newcommand{\rM}{r_{\!_M}}
\newcommand{\rS}{r_{\!_S}}
\newcommand{\bn}{{\bf n}}
\newcommand{\bK}{{\bf K}}

\newcommand{\bea}{\begin{eqnarray}}
\newcommand{\eea}{\end{eqnarray}}
\newcommand{\be}{\begin{equation}}
\newcommand{\ee}{\end{equation}}
\newcommand{\bes}{\begin{subequations}}
\newcommand{\ees}{\end{subequations}}

\def\nbox#1#2{\vcenter{\hrule \hbox{\vrule height#2in
\kern#1in \vrule} \hrule}}
\def\sq{\,\raise0.8pt\hbox{$\nbox{.10}{.10}$}\,}
\def\sqb{\,\raise.5pt\hbox{$\overline{\nbox{.09}{.09}}$}\,}

\newcommand{\sumi}{\raisebox{-1.5ex}{$\stackrel{\textstyle\sum}{\scriptstyle i}$}}
\newcommand{\st}{{\star}}

\allowdisplaybreaks

\begin{document}

\title*{Gravitational Vacuum Condensate Stars}
\author{Emil Mottola}
\institute{Emil Mottola (\Letter) \at  Dept. of Physics \& Astronomy, Univ. of New Mexico \ \email{emottola@unm.edu}}
\maketitle

\abstract{Gravitational vacuum condensate stars, proposed as the endpoint of gravitational collapse
consistent with quantum theory, are reviewed. Gravastars are cold, low entropy, maximally compact objects 
characterized by a surface boundary layer and physical surface tension, instead of an event horizon.
Within this thin boundary layer the effective vacuum energy $\La_{\rm eff}$ changes rapidly, such that the interior 
of a non-rotating gravastar is a non-singular static patch of de Sitter space with eq.~of state $p=-\r$. 
Remarkably, essentially this same result is obtained by extrapolating Schwarzschild's 1916 constant density interior 
solution to its compact limit, showing how the black hole singularity theorems and the Buchdahl compactness 
bound are evaded. The surface stress tensor on the horizon is determined by a modification of the Lanczos-Israel 
junction conditions for null hypersurfaces, which is applied to rotating gravastar solutions as well.
The fundamental basis for the quantum phase transition at the horizon is the stress tensor 
of the conformal anomaly, depending upon a new light scalar field in the low energy effective action 
for gravity. This scalar conformalon field allows the effective value of the vacuum energy, described as a 
condensate of an exact 4-form abelian gauge field strength $F = dA$, to change at the horizon.
The resulting effective theory thus replaces the fixed constant $\La$ of classical general relativity, 
and its apparently unnaturally large sensitivity to UV physics, with a dynamical condensate whose ground state 
value in empty flat space is $\La_{\rm eff} =0$ identically. This provides both a solution to the cosmological
constant problem and an effective Lagrangian dynamical framework for the boundary layer and interior of
gravitational condensate stars. The status of present observational constraints and prospects for detection of 
gravastars through their gravitational wave and echo signatures are discussed.
}


\section{The Issue of the Final State of Gravitational Collapse}
\label{sec:BHs}

The possibility that a gravitationally bound system could be so compact that not even light can escape its surface was first imagined
in the eighteenth century by Rev. J. Michell~\cite{Michell:1784}, and P.-S. Laplace~\cite{Laplace:1796,HawkEllis:1973}. 
Each reasoned (apparently independently) that since the escape velocity $\rm v$ from a spherical mass $M$ of radius $r$ 
is given by ${\rm v}^2 \!= \! 2GM/r$ in Newtonian gravitation, light traveling at speed $c$ would not be able 
to escape the surface if emitted from a radius less than $\rM \!=\! 2GM/c^2$. Since this radius is much smaller than the
actual size of astronomical bodies known at the time, such as the sun or the planets, and with no astronomical evidence 
of the existence of the highly compact objects for which $\rM$ would be relevant, the possibility of such `dark stars' from 
which light itself could not escape was not taken very seriously, nor developed further for well over a century. 

In 1915 A. Einstein introduced the theory of general relativity (GR).  Soon thereafter, K. Schwarzschild found an exact 
solution of the sourcefree Einstein's eqs. in the case of spherical symmetry. The general static, spherically symmetric 
metric line element in GR can be expressed in the form
\be
ds^2 = -f(r)\,c^2 dt^2 +  \frac{dr^2\!\!}{\,h(r)}+  r^2\left( d\th^2 + \sin^2 \th\,d\f^2\right)
\label{sphsta}
\vspace{-2mm}
\ee
in terms of two functions of radius, $f(r)$ and $h(r)$. For the sourcefree asymptotically flat Schwarzschild solution, 
$f(r)$ and $h(r)$ are equal and given by
\be
f(r) = h(r) = 1 - \frac{2G  M}{c^2\,r}  = 1 -\frac{\rM\!}{\!r}
\label{Sch}
\ee
determining the static geometry of empty sourcefree space exterior to a spherically symmetric body of mass $M$.
In these $(t,r)$ coordinates the metric has two apparent singularities, one at the origin $r=0$ of spherical coordinates,
and the other of quite a different character on a sphere of finite radius $r=\rM$, the same radius discussed by Michell 
and Laplace in Newtonian gravity.

The singularity at $r=0$ is similar to that of a point charge in Coulomb's law where the electric field diverges. Likewise for
(\ref{sphsta})-(\ref{Sch}) the Riemannian curvature and Kretschmann scalar $R_{\a\b\g\la}R^{\a\b\g\la} = 12 \rM^2/r^6$ diverge
as $r\to 0$. In contrast the curvature remains finite at $r=\rM$, and of order $1/\rM^2$, which is relatively small for large $M$. 
However a light wave emitted from any $r \ge \rM$ with local frequency $\w_{\,\rm loc}(r)$ is gravitationally redshifted according to the 
relation
\be
\w_{\infty} = \w_{\,\rm loc}(r) \,f^{\frac{1}{2}}= \w_{\,\rm loc} (r)\sqrt {1 - \frac{\rM\!}{\! r}} = const.
\label{redshift}
\ee
Thus light emitted at $r =\rM$ with any finite local frequency $\w_{\,\rm loc}(\rM)$ becomes redshifted to zero frequency and cannot 
propagate outwardly at all. The radius $r=\rM$ therefore defines the critical surface or {\it event horizon} from beyond which no 
event can ever be observed by an outside observer. Thus the wave description of light in GR recovers exactly the `dark star' 
surface that Michell and Laplace had imagined within a Newtonian framework. Conversely and equivalently, (\ref{redshift}) 
also implies that an inwardly directed light wave with the finite frequency $\w_\infty$ far from the horizon is blueshifted 
to an {\it infinite} local frequency upon reaching $r=\rM$.

One would expect to be able to remove at least the singularity at $r=0$, and the infinite curvatures it produces by
allowing for a stress tensor source to Einstein's eqs.~of finite extent, rather than one concentrated all at the origin. Indeed Schwarzschild 
also found such a regular {\it interior} solution~\cite{Schwarzs:1916}, shortly after finding the well-known exterior solution 
(\ref{sphsta})-(\ref{Sch}) most often associated with his name. 

To simplify matters, Schwarzschild assumed a model of the interior matter composed of an incompressible fluid with isotropic pressure 
and constant mass density 
\be
\r = \bar \r \equiv \frac{3M}{4 \p \rS^3} \equiv \frac{\,3H^2\!\!}{8 \p G}\qquad {\rm so \ \ that} \qquad H^2 = \frac{\rM}{\rS^3}
\label{constden}
\ee
where $\rS$ is the radius of the star's surface.\footnote{Here and often in the following, units are employed where the speed of light 
$c=1$,  unless otherwise needed for emphasis.} At this surface the pressure $p(\rS)=0$, and the interior solution for $r \le \rS$ 
is matched to the exterior one (\ref{sphsta})-(\ref{Sch}) for $r \ge \rS$ by requiring $f(\rS) =h(\rS) = 1 - \rM/\rS >0$ to be continuous 
there (although their derivatives need not be). With these boundary conditions the metric for Schwarzschild's interior solution can 
be expressed in the form (\ref{sphsta}), but with
\bes
\begin{align}
&h(r) = 1 - H^2r^2\\
&f(r) = \frac{D^2\!}{\!\!4}\, \ge 0
\end{align}
where
\be
D \equiv 3 \sqrt{\smash[b]{1 - H^2 \rS^2}} - \sqrt{1 - H^2 r^2}
\ee
is assumed to be non-vanishing, so that the pressure 
\be
p(r) = \frac{\raisebox{1pt}{$\displaystyle \bar\r$}}{\raisebox{-1pt}{$\displaystyle D$}} \,
\left[\sqrt{1 - H^2 r^2} - \sqrt{\smash[b]{1 - H^2 \rS^2}}\, \right] 
\label{pressint}
\ee
\label{intsoln}\ees
is also finite (and positive) on the interval $r \in [0,\rS]$. This assumption holds for
\be
\rS > \frac{9}{8}\, \rM
\label{rsbound}
\ee
in which case the solution (\ref{intsoln}) is everywhere non-singular and finite, and the first 
derivative $f'(\rS)=\rM/\rS^2$ also turns out to be continuous at the surface $r=\rS$.

Conversely, if the inequality (\ref{rsbound}) is {\it not} satisfied, then $D$ vanishes at 
\be
r_0 =3\rS \,\sqrt{1 - \frac{8}{9}\frac{\rS}{\rM}}\quad \in\  [0,\rS]\quad {\rm for} \quad \rM \le \rS \le \frac{9}{8}\, \rM
\label{r0def}
\ee
where $f(r_0)=0$, and the pressure (\ref{pressint}) $p(r_0) \to \infty$ diverges. In this case the solution 
(\ref{intsoln}) becomes singular and may no longer be deemed acceptable.  Since $M = 4 \p \bar\r \rS^3/3$, the lower 
bound (\ref{rsbound}) on $\rS$ translates to an {\it upper} bound on the mass 
\be
\overline {\!M}_{\rm crit} = \frac{4 c^3\!\!}{9G} \left(\!\frac{1}{3\p G \bar \r}\!\right)^{\!\!\frac{1}{2}}\,,
\label{Mcrit}
\ee
that is, $M < \overline {\!M}_{\rm crit}$ is the condition for a non-singular Schwarzschild star of mass $M$ to exist with 
a given fixed constant density $\r =\bar \r$ interior. 

An incompressible fluid would seem to be matter that is the most resistant to further gravitational contraction, and it is already 
extreme at that for implying an infinite speed of sound (v$_s> c$), violating relativistic causality. Thus \,$\overline{\!M}_{\rm crit}$ 
would seem to be an upper bound for the mass of any self-gravitating body to contain a non-singular interior. If $\bar \r$ is taken 
to be $5$ times the density $2.3 \times 10^{14} {\rm gm/cm}^3$ of normal nuclear matter, about the value expected in the cores 
of neutron stars, then $\overline{\!M}_{\rm crit} \simeq 3.4\, M_{\odot}$. Any higher values of $\bar \r$ lead only to smaller values 
of $\overline{\!M}_{\rm crit}$ from (\ref{Mcrit}). Thus any self-gravitating system composed of ultra dense nuclear matter with a mass larger than 
a few solar masses would appear to be unstable and liable to complete gravitational collapse to the singular state of the original 
solution (\ref{sphsta})-(\ref{Sch}), in which all the matter is crushed to infinite density at the central singularity. 

J.~A.~Wheeler and his collaborators carried out calculations utilizing models of the eqs.~of state of nuclear matter, 
and their stability, obtaining upper bounds of order of a solar mass $M_{\odot}$~\cite{HarThorWakWheel:1965}. Current 
theoretical values for this Tolman-Oppenheimer-Volkoff (TOV) limit have been estimated to lie between 
$2.2\,M_{\odot}$ and $2.9\, M_{\odot}$~\cite{KaloBaym:1996}, the range due to the still considerable uncertainty in the high 
density nuclear eq.~of state. The LIGO/LSC observation of GW170817, the first gravitational wave (GW) event due to merging neutron 
stars, placed the limit in the range of $2.01\,M_{\odot}$ to $2.17\,M_{\odot}$~\cite{RezzMostWeih:2018,ChoNSmass:2018}
suggesting that higher central nuclear densities and the lower end of the theoretically allowed mass range is preferred by the GW data.

\begin{svgraybox}
The key point Wheeler made clear, and emphasized, is the reason that such an upper TOV bound exists at all in GR is due to the fact 
that unlike Newtonian gravity, positive internal pressure $p > 0$ only increases gravitational attraction in Einstein's theory~\cite{Wheeler:1964}. 
Formally taking the limit $c \to \infty$ in (\ref{Mcrit}) shows that the upper bound on $M$ is removed to infinity in the non-relativistic 
Newtonian limit where this pressure effect disappears. Thus in full GR even very `hard' eqs.~of state of the internal matter such as 
Schwarzschild's incompressible fluid only serve to promote rather than resist gravitational collapse, making singularities of 
the kind exhibited in (\ref{sphsta})-(\ref{Sch}) for the final state seem to be inevitable.  Wheeler hoped that consistent 
inclusion of quantum effects might provide an escape from this difficulty, since as he said: {\it ``proper physical variables 
do not and cannot go to infinity."} ~\cite{Wheeler:1964}
\end{svgraybox}

Wheeler thus brought to many physicists' attention the seriousness of the issue of the final state of complete 
gravitational collapse in classical GR. Despite strong misgivings, but unable to see any apparent way out of the difficulty,
Wheeler came to believe in the inevitability of singularities in gravitational collapse, and then popularized the term {\it black hole}
to both physicists and non-physicists alike.

\begin{wrapfigure}{hr}{.6\textwidth}
\vspace{-5mm}
\includegraphics[width=7cm, viewport=0 0 854 568,clip]{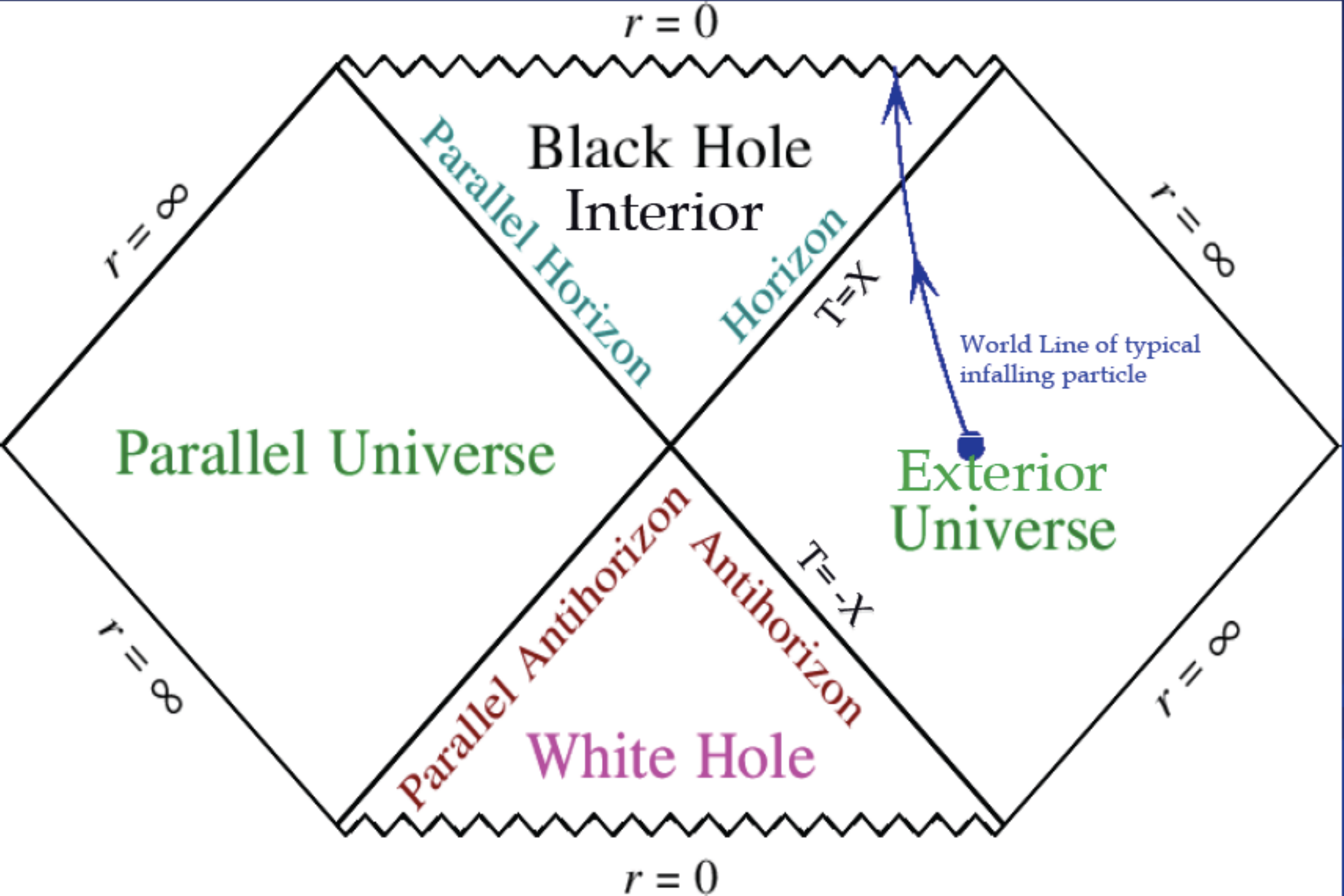}
\vspace{-7mm}
\caption{The Carter-Penrose conformal diagram of the maximal Kruskal analytic extension
of the Schwarzschild geometry. Radial light rays are represented in this diagram
as $45^{\circ}$ lines. The angular coordinates $\th,\f$ are suppressed.}
\label{Fig:SchwKruskal}
\vspace{-7mm}
\end{wrapfigure}
\noindent
The nature of the Schwarzschild singularity at $r= \rM$ as a causal boundary only became clear from the work of 
D. Finkelstein~\cite{Finkel:1958}, who showed that the coordinate singularity in the line element (\ref{sphsta})-(\ref{Sch}) at $r=\rM$ 
could be removed by a change of coordinates associated with the trajectories of freely falling light rays (a set of coordinates 
actually found earlier by A.~S.~Eddington~\cite{Edding:1924}). Since the worldlines of timelike trajectories starting at finite $
r > \rM$ also arrive at the horizon at a finite proper time, it seems natural to extend the coordinates through the horizon to the interior,
$r < \rM$, so that the proper time of such a timelike observer could continue uninterrupted. This leads to the maximal analytic extension
of the Schwarzschild geometry in Kruskal-Szekeres coordinates~\cite{MTW}, illustrated in the Carter-Penrose conformal diagram of 
Fig.~\ref{Fig:SchwKruskal}. The causal boundary at which light hovers indefinitely, unable to escape to infinity, which in the 
Schwarzschild solution (\ref{sphsta})-(\ref{Sch}) is the spherical surface of radius $r=\rM$, at $T=X$ in Fig.~\ref{Fig:SchwKruskal}
is the (future) event horizon, following terminology introduced by W.~Rindler~\cite{Rindler:1956}. 

\begin{svgraybox}
In order to remove the coordinate singularity at $r=\rM$, the transformation from (\ref{sphsta}) to Eddington-Finkelstein 
or Kruskal-Szekeres $(T,X)$ coordinates must itself be singular at the horizon. Although not often stated explicitly, this means 
that the geometries before and after the singular coordinate transformation are not necessarily physically equivalent at $r=\rM$, 
or if extended beyond it. This is because the mathematical procedure of analytic continuation through a null hypersurface 
involves a {\it physical assumption}, namely that the stress-energy tensor $T_\m^{\ \n}$ is exactly vanishing there. Even in classical GR, 
the hyperbolic character of Einstein's eqs.~allows generically for sources and discontinuities on the horizon which would violate 
this assumption, invalidating analytic continuation and potentially altering also the geometry of the interior from what the analytic 
continuation of Fig.~\ref{Fig:SchwKruskal} would predict.

It is relevant here to recall that Einstein's original conception of the Equivalence Principle between gravitational 
and inertial mass, and its subsequent mathematical formulation in terms of Riemannian geometry, requires
physics to be independent of {\it real} and {\it regular} local coordinate transformations~\cite{Einstein:1912}. 
Analytic continuation through BH horizons appends to this the much stronger mathematical hypothesis of 
continuation in the space of {\it complex} metrics, admitting {\it singular} coordinate transformations that lead to 
globally extended spacetimes as in Fig.~\ref{Fig:SchwKruskal}, which may or may not be realized in Nature. 
Occasional statements that free fall through the event horizon implied by continuation of worldlines to $r<\rM$
is required by the Equivalence Principle are incorrect, just because of the possibility of sources and 
discontinuities on the horizon, which would interrupt that free fall. The Equivalence Principle does not prevent 
Einstein's elevator from coming to an abrupt end of its free fall at the surface of the earth, or a surface at $r=\rM$,
if a surface exists there.
\end{svgraybox}

When R.~Kerr found an exact solution of the sourcefree Einstein eqs.~for a rotating BH~\cite{Kerr:1963}, the interior was found 
to contain not only singularities where the curvature diverges, located now on a ring of finite radius, but also 
{\it closed timelike curves}~\cite{HawkEllis:1973}. It is significant that this paradoxical and acausal feature, widely 
believed to be unphysical, occurs at {\it macroscopic} distance scales of order of the horizon scale 
$\rM\! \simeq \! 3\, (M/M_\odot)$\,km, far larger than the microscopic Planck scale 
$L_{\rm Pl}\!= \!\!\sqrt{\hslash G/c^3}\! \simeq\!1.6 \times\! 10^{-33}$ cm,~at which quantum gravity effects might 
be expected to become important or come to the rescue. The maximal analytic extension of the Kerr 
solution also results in global features such as an infinite number of asymptotically flat regions~\cite{HawkEllis:1973}, 
beyond the two in Fig.~\ref{Fig:SchwKruskal}. Thus endowing collapsing matter with angular momentum not only does 
not avoid the appearance of spacetime singularities, as Einstein apparently had hoped~\cite{Einstein:1939}, 
but rather leads to additional unphysical features of analytically extended BH solutions. 

An open question at the time was whether singularities might be a special property of highly symmetric exact solutions
of Einstein's eqs. To address this question, in 1965 R.~Penrose introduced the notion of a {\it closed trapped surface}~\cite{Penrose:1965}, 
within which outgoing light waves are actually bent backwards by gravity and forced to become ingoing, such as occurs in 
(\ref{sphsta})-(\ref{Sch}) for $r < \rM$.  He then showed that {\it if} such a trapped surface exists, and if matter within it satisfies 
the weak energy condition $\r + p_i \ge 0$, then Einstein's eqs.~imply that a singularity would necessarily result, independently of any 
symmetries. Here $p_i, i=1,2,3$ are any of the three principal pressure components of the matter stress tensor. The important 
assumption made in \cite{Penrose:1965} is that some continuation of the geometry beyond the horizon into the interior is 
necessary to have a trapped surface at all. Related singularity theorems were subsequently proven by S.~W.~Hawking, and 
Hawking and Penrose, with slightly different assumptions, but also relying either on the existence of a closed trapped surface 
or a stronger condition $\r + \sum_{i =1}^3 p_i \ge 0$  on the matter stress tensor (the strong energy condition)~\cite{HawkEllis:1973}.
Since these singularity theorems are independent of any symmetries and the existence of trapped surfaces in gravitational collapse
was widely assumed, they reinforced the conviction that BH singularities will arise inevitably in the final state of gravitational collapse 
in classical GR.

Also in 1965, a quite different possibility for the final state of gravitational collapse was proposed by E. Gliner, based on his 
speculation that superdense matter could arrive at an eq.~of state with all three principal pressures equal and {\it negative, i.e.} 
$p_i = - \r < 0$~\cite{Gliner:1966}. This corresponds to a cosmological $\La$ term in Einstein's eqs., but localized within the high 
density matter, rather than everywhere constant. With this eq.~of state and negative pressure, the strong energy condition is {\it violated,  i.e.} 
$\r + 3p <0$. This has the consequence that timelike geodesics become {\it defocused} rather than focused down to a singularity, or in 
more colloquial terms, that gravitational attraction in GR becomes effectively {\it repulsion} instead. The key point about
such an eq.~of state is that it reverses the sign of the pressure effect that led previously to the critical mass bound (\ref{Mcrit}).
The $p=-\r <0$ eq.~of state is that which A. D. Sakharov had speculated also could arise to avoid a singularity at the initial
conditions of the universe~\cite{Sak:1966}, implying instead a non-singular expanding de Sitter phase (well before the term 
`inflation' was coined). Today this eq.~of state is believed to be that of the cosmological dark energy driving the Hubble expansion 
to accelerate, rather than decelerate, again intuitively a kind of repulsion. 

\begin{svgraybox}
With an interior $p=-\r >0$ eq.~of state, BH singularities and the conclusions of the Penrose singularity theorem can be avoided, 
but if Einstein's eqs.~hold, only if no true closed trapped surface is ever formed. This shows that the crucial assumption of the 
Penrose theorem involves the physics of the near-horizon region, since once any trapped surface at any $r<\rM$ is assumed, 
it is `game over,' and a singularity must occur in GR. The logical loopholes in all the singularity theorems and resolution of 
Wheeler's difficulty for the final state of gravitational collapse can be realized in Einstein's theory, or its semi-classical extension, 
only if the stress tensor of quantum matter can produce a large enough effective positive $\La_{\rm eff}$ term with negative pressure, 
{\it and} prevent the formation of a true trapped surface in the collapse. 
\end{svgraybox}
 
\section{What's the (Quantum) Matter with Black Holes?}
\label{Sec:QM}

Additional difficulties beyond the singularities and acausal closed timelike curves, arising in strictly classical GR, appear 
when quantum effects in BH spacetimes are considered. The genesis of these is Hawking's argument that a non-rotating, 
uncharged BH would emit thermal radiation, at a temperature \cite{Hawking:1974,Hawk:1975}
\be
T_{_H} = \frac{\hslash\ka}{2\p c k_{_B}} = \frac{\hslash c^3}{8 \p G k_{_B} M}
\label{TH}
\ee
where $\ka = c^4/4 GM$ is the surface gravity.  With the temperature (\ref{TH}), it was argued that thermodynamics
could be applied to BHs, and a `First Law'~\cite{BarCarHawk:1973,HartHawk:1976}
\be
dE = dM c^2 = \frac{\ka\, c^2}{8\p G} \, d A_{_H} = T\!_{_H} \, dS\!_{_{BH}} 
\label{dETdS}
\ee
was deduced, with the BH assigned the Bekenstein-Hawking BH entropy,
\be
S\!_{_{BH}} = k_{_B}  \frac{A_{_H}}{4 \,L_{\rm Pl}^2\!}\,,
\label{SBHg}
\ee
equal to $1/4$ of its area $A_{_H} = 4 \p \rM^2$ in Planck units. This identification of entropy with area supported
J. Bekenstein's speculation that BHs be assigned an entropy in order to avoid violations of the second law 
of thermodynamics~\cite{Beken:1973}, since ordinary matter with non-zero entropy apparently can fall into BHs and so 
disappear without a trace. The hypothesis that BH entropy is proportional to the area of its event horizon was guided by 
D.~Christodoulou's classical result that horizon area can never decrease in any process of interaction with matter, 
and will positively increase if the process irreversibly sends matter with positive mass-energy into the BH~\cite{Christo:1970}.

\begin{svgraybox}
The curious feature of (\ref{dETdS}) is that $\hslash$ cancels out between $T\!_{_H}$ and $dS\!_{_{BH}}$. 
This is because the first form of (\ref{dETdS}) is nothing but the {\it classical} Smarr formula
\cite{Smarr:1973}
\be
dM = \frac{\ka}{8 \p G}\, dA_{_H}
\label{Smarr}
\ee
relating the change of a BH mass $M$ to the change in area $A_{_H}$ of its horizon, a simple geometric relation
in which $\hslash$ does not appear at all. As in Christodoulou's area law~\cite{Christo:1970}, quantum mechanics 
plays no role in the classical Smarr relation (\ref{Smarr}). Multiplying and dividing by $\hslash$ should not be expected 
to turn a stictly classical relation into a valid quantum one. 
\end{svgraybox}

Related to this, if the identification of the classical area rescaled by $k_{_B}/4L_{\rm Pl}^{2}$ with entropy 
and the thermodynamic interpretation of (\ref{dETdS}) is to be valid in the quantum theory, then the 
classical limit $\hslash \to 0$ (with $M$ fixed) which yields an arbitrarily low Hawking temperature (\ref{TH}), 
assigns to the BH an arbitrarily large entropy, completely unlike the zero temperature limit of any other cold 
quantum system. 

Another paradoxical consequence of temperature in (\ref{TH}) inversely proportional to $M = E/c^2$ is the
implication that the heat capacity of a Schwarzschild BH
\be
\frac{\!dE}{\,dT\!_{_H}} = - \frac{8 \p G k_{_B} M^2}{\hslash c} = - \frac{Mc^2}{T\!_{_H}} < 0
\label{dEdT}
\ee
is {\it negative}, as was pointed out by Hawking himself \cite{HawkSpecHeat:1976}, and incidentally also very large in magnitude. 
In statistical equilibrium the heat capacity of a system is related to the energy fluctuations about its mean value $\lag E \rag$ by
\be
c_{_V} = \left(\frac{d\lag E\rag}{dT}\right)_{\!\!_V} = 
\frac{1}{\,k_{_B}T^2\!} \ \Big\lag \big(E- \lag E\rag\big)^2\Big\rag > 0
\label{cV}
\ee
at constant volume $V$. If pressure or some other thermodynamic variable is held fixed there is an analogous
relation. The positivity of the statistical average in (\ref{cV}) requires only the existence of a well defined ground 
state upon which the thermal equilibrium canonical ensemble is defined, but is otherwise quite independent of the details 
of the system or its interactions. Hence on general grounds of quantum statistical mechanics, the heat capacity 
of any system in stable equilibrium at a fixed temperature must be positive. Although a negative heat capacity
can be observed in self-gravitating systems in the microcanonical ensemble, that situation is temporary and not 
a true equilibrium state, but one subject to large fluctuations and indicative of a phase transition involving
a large latent heat~\cite{LyndBell:1977,Padman:1990}. 

The physical basis of the instability of a BH in equilibrium with its own Hawking radiation at $T= T_{_H}$ is easy to see.
If by a small thermal fluctuation the BH should absorb slightly more radiation in a short time interval than it emits, its mass 
would increase, $\D M > 0$, and hence from (\ref{TH}) its temperature would decrease, $\D T < 0$, so that it would 
now be cooler than its surroundings and be favored to absorb more energy from the heat bath than it emits in the 
next time step, decreasing its temperature further and driving it further from equilibrium. In this way a runaway 
process of the BH growing to absorb all of the surrounding radiation in the heat bath would ensue. Likewise, if 
the initial fluctuation has $\D M< 0$, the BH temperature would increase, $\D T > 0$, so that it would now be 
hotter than its surroundings and favored to emit more energy than it absorbs from the heat bath in the next time step, 
increasing its temperature further. Then a runaway process toward hotter and hotter evaporation of all its mass to its 
surroundings would ensue. In either case, the initial equilibrium is clearly unstable, and hence cannot be a candidate 
for the quantum equilibrium ground state for the system. The instability has also been verified from the negative eigenvalue 
of the fluctuation spectrum of a BH in a box of (large enough) finite volume~\cite{GrossPerryYaffe:1982,Allen:1984,York:1986}.

The time scale for this unstable runaway process to grow exponentially is the time scale for fluctuations 
away from the mean value of the Hawking flux, {\it not} the much longer time scale associated with the 
lifetime of the BH under continuous emission of that flux. The time scale for thermal fluctuations is easily estimated
to be the typical time between emissions of a single quantum with typical energy (at infinity) of $k_{_B}T$, 
of a source whose energy emission per unit area per unit time is of order $(k_{_B}T_{_H})^4/\hslash^3 c^2$. 
Multiplying by the area and dividing by the typical energy $k_{_B}T$,  the average number of quanta emitted 
per unit time is found. The inverse of this, namely
\be
\D t \sim \frac{1}{A_{_H}} \frac{ (\hslash c)^3}{(k_{_B} T_{_H})^3} \sim \frac{\rM}{c}
\sim 10^{-5}\left(\frac{M}{M_{\odot}}\right)\, {\rm sec}
\label{deltat}
\ee
is the typical time interval (as measured by a distant observer) between successive emissions of individual Hawking quanta. 
This time scale is quite short. Any tendency for the system to become unstable or undergo a phase transition to a stable phase 
would be expected to show up on this short a time scale, governing the fluctuations in the mean flux, which is of order of the 
collapse time itself and before a steady state flux could even be established. The interpretation of the classical relation of energy 
to area (\ref{Smarr}) as an equilibrium thermodynamic relation (\ref{dETdS}) of any kind is therefore placed in serious doubt.

It is also instructive to evaluate $S\!_{_{BH}}$ for typical astrophysical BHs. For a solar mass
$M_{\odot} \simeq 2 \times 10^{33}$ gm
\be
S\!_{_{BH}} \simeq 1.05 \times 10^{77}\, k_{_B}\, \left(\frac{\!\!M}{M_{\odot}\!}\right)^{\!\!2}
\label{SBH}
\ee
which is truly an enormous entropy. For comparison, the entropy of the sun as it is, a hydrogen burning 
main sequence star, is given to good accuracy by the entropy of a non-relativistic perfect 
fluid. This is of the order $N k_{_B}$ where $N$ is the number of nucleons in the sun 
$N \sim M_{\odot}/m_{_N} \sim 10^{57}$, times a logarithmic function of the density and temperature 
profile which may be estimated to be of the order of $20$ for the sun. Hence the entropy of the sun is 
approximately $S\!_{\odot} \sim 2 \times 10^{58} \, k_{_B}$ or nearly $19$ orders of magnitude smaller than (\ref{SBH}).  

A simple scaling argument shows that the entropy of a gravitationally bound relativistic fluid should scale like
$M^{3/2}$ and be of order of \cite{ZelNovbook, EMZak:2010}
\be
S \sim k_{_B} \left(\frac{\!\!M}{M_{\rm Pl}\!}\right)^{\!\!\frac{3}{2}} \sim 10^{57}\, k_{_B}  
\left(\frac{\!\!M}{M_{\odot}\!}\right)^{\!\!\frac{3}{2}}
\label{relstar}
\ee
which is less than $S_{\odot}$ because the relativistic radiation pressure in the sun is small compared 
to the non-relativistic fluid pressure. However, the entropy from the relativistic radiation pressure 
(\ref{relstar}) grows with the $3/2$ power of the mass, whereas the non-relativistic fluid entropy $S_{\odot}$
grows only linearly with $M$. For stars with masses greater than about $50\, M_{\odot}$ which are hot enough 
for their pressure to be dominated by the photon $T^4$ radiation pressure, (\ref{relstar}) gives 
the correct order of magnitude estimate of such a star's entropy at a few times $10^{59}\,k_{_B}$~\cite{ZelNovbook}. 

On the other hand, the BH entropy (\ref{SBH}) is proportional to $M^2$, so the discrepant factor with (\ref{relstar})
is $(M/M_{\rm Pl})^{1/2} \sim 10^{19}$ for $M= M_{\odot}$. Since (\ref{SBH}) makes no reference to how the BH
was formed, and a BH may always be theoretically idealized as forming from an adiabatic process, which 
keeps the entropy constant, (\ref{SBH}) states that this entropy must suddenly jump by a factor of order 
$10^{19}$ for a solar mass BH at the instant the horizon forms at $r=\rM$. When Boltzmann's 
formula 
\be
S = k_{_B} \ln W(E)
\ee
is recalled, relating the entropy to the total number of microstates in the system $W(E)$ at the fixed energy $E$, 
the number of such microstates of a BH satisfying (\ref{SBH}) must jump by $\exp(10^{19})$ at that instant 
that the event horizon is reached, a truly staggering proposition. How and from where this enormous number of microstates
can appear, when the horizon is supposed to be nothing but a mathematical surface only, with no independent degrees of
freedom or stress tensor of its own, has remained perplexing and the subject of numerous investigations spanning five
decades~\cite{Preskill:1992,Mathur:2009,AMPS:2013,Giddings:2013,MottVauPT,Marolf:2017,AHMST:2020,AHMald:2021},
often with the suggestion that new physics or modifying the laws of physics themselves might be required.

Hawking radiation emerging from the BH at temperature (\ref{TH}) in a mixed thermal state also seems to imply a 
breakdown of quantum unitary evolution, that is difficult if not impossible to recover at the late or final stages of the 
BH evaporation process, resulting in a severe `information paradox'~\cite{HawkUnit:1976,Preskill:1992,Page:1993,UnruhWald:2017}. 

As a consequence of the gravitational redshift (\ref{redshift}), it also follows that although the Hawking temperature $T\!_{_H}$ 
of the radiation far from the BH is very low, the local temperature of the radiation (\ref{Tol}) is arbitrarily high when extrapolated back
\begin{wrapfigure} {r}{.42\textwidth}
\vspace{-6mm}
\be
T\!_{\rm loc}(r) =\ \,\frac{T\!_{_H}}{\hspace{-3mm}\sqrt{\raisebox{-1pt}{$1\ $--}\ \sdfrac{\raisebox{-1.5ex}{$r_M$\!}}{r}}}\hspace{1cm}
\label{Tol}
\vspace{-8mm}
\ee
\end{wrapfigure}
to the vicinity of the horizon $r\! \to\! \rM$, even becoming transplanckian in this limit. Unlike classical test particles, when $\hslash \neq 0$ 
such extremely blue shifted photons are necessarily present in the vacuum as virtual quanta. Their effects upon the geometry 
depend upon the quantum state of the vacuum, defined by boundary conditions on the wave equation in a non-local way over 
all of space, and $\lag T_\m^{\ \n} \rag$ may be large at $r=\rM$, notwithstanding the smallness of the local classical curvature 
there \cite{Boul:1975,ChrFul:1977}. Since the limits $r\to \rM$ and $\hslash \to 0$ do not commute, non-analytic behavior near
the event horizon, quite different from that in the strictly classical ($\hslash \equiv 0$) situation is possible in the quantum 
theory. Since gravity couples to all energies, thermal fluctuations at large transplanckian temperatures and energies would 
be expected to have significant backreaction effects on the classical spacetime geometry near the horizon, giving rise
to just the sources and discontinuities that would violate the analytic continuation to the interior and global geometry
in Fig.~\ref{Fig:SchwKruskal} it leads to. Thus it is by no means clear why assuming a rigidly fixed classical BH background,
neglecting quantum fluctuations in the stress tensor down to $r\!=\!\rM$, as in Hawking's original semi-classical treatment, 
should be valid~\cite{Jacobson:1991}.

\begin{svgraybox}
These myriad difficulties and BH paradoxes suggest that some important element is missing, and needed to describe quantum 
effects or degrees of freedom on the horizon scale in BH spacetimes, which are not present in classical GR. Since the various 
problems and paradoxes arise at the macroscopic scale of the BH horizon, these additional degrees of freedom require an 
addition or modification of Einstein's theory at that scale, well before the microscopic Planck scale $L_{\rm Pl}$ or Planck 
scale curvatures are reached. The problem of the magnitude of cosmological $\La$ vacuum energy on the very largest macroscopic 
Hubble scale of the universe carries with it a similar implication. Together with the observation that the $p=-\rho$
eq.~of state can provide a resolution of both the BH and Big Bang singularities, this indicates that the two problems at the interface of 
classical GR and quantum theory, involving the nature of the quantum vacuum and vacuum energy, at macroscopic scales are related~\cite{EMEFT:2022}.
\end{svgraybox}

\section{The Proposed Solution: Gravitational Condensate Stars}
\label{Sec:gstar1}
\subsection{Background and Motivations for the First Gravastar Model}

One model that attempted to take backreaction of Hawking radiation on the BH into account, imagined it to be immersed 
in a Hawking radiation atmosphere, with an effective equation of state, $p =\kappaup \r$~\cite{tHooft:1998}. It was found that due 
to the blue shift effect (\ref{Tol}), the backreaction of such an atmosphere on the metric near $r=\rM$ is enormous, with the 
interior region quite different from the vacuum Schwarzschild solution. A large entropy of order of $S_{_{BH}}$ is obtained 
from the hot fluid alone in such a model, with $S = 4\,\frac{\kappaup + 1}{7 \kappaup + 1} S_{_{BH}}$, becoming equal to the BH
entropy (\ref{SBHg}) for $\kappaup = 1$. This suggested that the maximally stiff equation of state consistent with the causal limit 
$p=+\rho$ may play a role in the quantum theory of fully collapsed objects. 

Despite the interesting result of \cite{tHooft:1998}, indicating the importance of backreaction on the geometry, 
the model of \cite{tHooft:1998} cannot be viewed as a satisfactory solution to the final state of the collapse problem, 
since it involves huge Planckian energy densities near $r=\rM$, and a negative mass singularity at $r=0$,
indicating the breakdown of the semi-classical approximation in both regions. The negative mass singularity arises because 
a repulsive core is necessary to counteract the self-attractive gravity of the dense relativistic fluid with positive energy.
This suggests again that an effectively repulsive eq.~of state with a negative pressure $p <0$ (rather than a negative 
mass-energy singularity) should play a role in the BH interior.

A different proposal of a quantum phase transition at or near $r=\rM$ was made in \cite{ChaplineHLS:2001}, based on a
suggestive condensed matter analogy with the liquid-vapor critical point of a non-relativistic Bose fluid, where the wave eq.~for
acoustic excitations mimics the wave eq.~for light in the vicinity of a BH horizon. In this condensed matter analog
system the position dependent speed of sound ${\rm v}_s(r)$ takes the place of a position dependent speed of light $c_{\rm eff}(r)$, 
both of which formally vanish at the critical surface, where $c^2_{\rm eff}(r) = c^2 f(r)$ is the $g_{tt}$ metric 
coefficient of $dt^2$ in (\ref{sphsta}). The authors of \cite{ChaplineHLS:2001,Chap:2003} also argued
for a negative pressure on the interior side of the critical surface, and an effective $p=-\r$ eq.~of state, as Gliner had
35 years earlier, but did not tackle the delicate issue of joining a de Sitter interior to a Schwarschild exterior, in
a fully consistent treatment of the problem in GR.

Matching the Schwarzschild exterior solution to a non-singular de Sitter (dS) interior had a long
previous history. Continuous transitions between the two were studied {\it e.g.} in \cite{Dymn:1992}, while it was
recognized that joining the exact Schwarzschild and dS geometries directly at their mutual horizons $H^{-1}$ 
and $2GM$, requires some discontinuity or interposition of `non-inflationary material' \cite{PoisIsr:1988}. In addition 
to uncertainties of the physics involved, the earlier GR formalism~\cite{Lanczos:1924,OBrSyn:1952,Israel:1966a,Israel:1967}
for dealing with singular hypersurfaces when the normal to hypersurface becomes null, as it does at a BH horizon, was
recognized to be inadequate \cite{BarIsr:1991}. The necessity of some anisotropic matter at the joining of the
interior to exterior geometries was made explicit in \cite{CatVis:2005}. 

Motivated by these considerations, the first fully relativistic gravitational vacuum condensate star model was proposed
in \cite{gravastar:2001,Grav_Univ:2023}, with the critical surface of refs.~\cite{ChaplineHLS:2001} replaced 
by a thin shell of ultra-relativistic fluid with the maximally stiff eq. of state $p=\r$ and vacuum condensate interior with 
negative pressure, $p=-\rho$. Partly for the reason of avoiding the technical difficulties associated with singular null hypersurfaces, 
the proposal in the original paper \cite{gravastar:2001,Grav_Univ:2023} made use of two timelike hypersurfaces at 
$r_1$ and $r_2$ with an interposed fluid boundary layer of `non-inflationary material' obeying the eq.~of state $p=\r$. In addition
to the model of~\cite{tHooft:1998}, the choice of this eq.~of state at the causal limit where the speed of sound coincides with 
the speed of light, was motivated by physical considerations of a quantum phase transition produced by the infrared effects 
of dimensional reduction from $D=4$ to $D=2$ dimensions. Nevertheless the choice of $p=\r$ in \cite{Grav_Univ:2023} 
is certainly an ansatz, illustrating a proof of principle, but without a rigorous basis in fundamental physics. It therefore would be 
subject to modification as that fundamental physics came more clearly into view by subsequent 
developments~\cite{EMVau:2006,EMZak:2010,EMEFT:2022}, and it also became possible to treat joining at null horizon 
hypersurfaces directly~\cite{MazEM:2015,BelGonEM1:2022}.

\subsection{The First Gravastar Model}

The general form of the stress-energy tensor in the static, spherically symmetric geometry of (\ref{sphsta}) for
a non-rotating body is
\be
T^\m_{ \ \,\n} = \left(\begin{array}{cccl}\!\! -\rho\ & 0\ & 0\ & 0\\
0 & p & 0& 0\\ 0 & 0 & p_\perp& 0\\ 0 & 0 & 0 & p_\perp \end{array}\right) 
\label{Tgen}
\ee
so that the Einstein equations in the static spherical coordinates of (\ref{sphsta}) are
\begin{subequations}
\begin{align}
-G_{\ t}^t &= \sdfrac{1}{r^2} \sdfrac{d}{dr}\,\big[r\left(1 - h\right)\big] = -8\p \GN T_{\ t}^t = 8\p \GN \,\rho\,,\\[1ex]
G_{\ r}^r &= \sdfrac{h}{r f}\sdfrac{d f}{dr}  + \sdfrac{1}{r^2} \,\big(h -1\big) = 8\p \GN T_{\ r}^r = 8\p \GN\, p\
\end{align}
\label{Eins}\end{subequations} 
together with the conservation equation
\be
\na\!_\la\, T^\la_{\ r} = \frac{d p}{dr} + \frac{\rho + p}{2f} \,\frac{d f}{dr} + \frac{2}{r} \,(p-p_\perp) = 0
\label{cons}
\ee
which ensures that the other components of Einstein's eqs.~are satisfied.  In (\ref{cons}) the transverse pressure 
$p_\perp \!\equiv\! T^{\theta}_{\ \theta} \!=\! T^{\f}_{\ \f}$ is allowed to be different from the radial pressure 
$p \equiv T^r_{\ r}$. For a perfect fluid $p_\perp \!=\! p$ and the last term of (\ref{cons}) vanishes. In that case 
(\ref{Eins})-(\ref{cons}) are three first order equations for the four functions, $f, h, \rho$, and $p$, 
which become closed when an eq.~of state for the fluid relating $p$ and $\rho$ is specified.

Because of the considerations of the previous sections, the first gravastar model was assumed to contain three different regions
with the three different eqs.~of state
\be
\begin{array}{clcl}
{\rm I.}\ & {\rm de\ Sitter\ Interior:}\ & 0 \le r < r_1\,,\ &\r = - p \,,\\
{\rm II.}\ & {\rm Thin\ Shell:}\ & r_1 < r < r_2\,,\ &\r = + p\,,\\
{\rm III.}\ & {\rm Schwarzschild\ Exterior:}\ & r_2 < r\,,\ &\r = p = 0\,.
\end{array}
\ee
At the interfaces $r\!=\!r_1$ and $r\!=\!r_2$, the metric functions $f$ and $h$ are required to be 
continuous, although the first derivatives of $f$, $h$ and $p$ are generally discontinuous from the first 
order eqs.~(\ref{Eins}) and (\ref{cons}).

In the interior region $\r = - p$ is a constant from (\ref{cons}). Labelling this constant $\r_{_V} = 3H^2/8\p G$,
and requiring that at the origin the solution is free of any mass singularity determines the interior to be a region of dS
spacetime in static coordinates, {\it i.e.}
\be
{\rm I.}\qquad f(r) = C\,h(r) = C\,\big(1 - H^2\,r^2\big)\,,\quad 0 \le r \le r_1
\label{fdS}
\ee
where $C$ and $H$ are constants, which at this point are arbitrary. Note that the static dS horizon where both $f$ and $h$
vanish is at $r=\rH=H^{-1}$.

The unique solution in the exterior vacuum region where $T_\m^{ \ \,\n}=0$ that approaches flat Minkowski space
as $r \to \infty$ is the Schwarzschild solution (\ref{Sch})
\be
{\rm III.} \qquad f(r) = h(r) = 1 - \frac{2 G M}{r} = 1-\frac{\rM}{r}\,,\qquad  r_2 \le r
\ee
where the mass $M$ can take on any (positive) value.

The only non-vacuum region is the thin shell interface region II. In this region it proves useful to define the dimensionless 
variable $w$ by 
\be
w \equiv 8\p G r^2 p
\label{wdef}
\ee 
so that eqs.~(\ref{Eins})-(\ref{cons}) with $\r = p$ may be recast in the form
\begin{subequations}
\begin{align}
\frac{dr}{r} &= \frac{dh}{1-w-h}\,, \label{ueqa}\\[0.9ex]
\frac{dh}{h} &= -\frac{1-w-h}{1 + w - 3h}\, \sdfrac{dw}{w}\,.
\label{ueqb}
\end{align}
\label{ueq}\end{subequations}
together with $p f \propto wf/r^2$ a constant. The first eq.~(\ref{ueqa}) is equivalent to the definition of the (rescaled) 
Misner-Sharp mass function $\m(r)=2Gm(r)$, with $h = 1 - \m/r$ and $d\m(r) = 8\p G\, \r r^2\, dr = w\, dr$ within the shell. 
The second eq.~(\ref{ueqb}) can be solved only numerically in general. However, it is possible to obtain an analytic solution 
in the thin shell limit $0 < h \ll 1$, since in this limit $h$ may be set to zero on the right side of (\ref{ueqb}) to leading order.
Assuming $w$ remains finite, (\ref{ueqb}) can then be integrated immediately to obtain
\be
h \equiv 1- \frac{\m}{r} \simeq  \ve\ \frac{(1 + w)^2}{w \ }\ll 1
\label{hshell}
\ee
in region II, where $\ve$ is an integration constant. Because of the condition $h \ll 1$, we require 
$\ve \ll 1$, if $w$ is of order unity. Making use of eqs. (\ref{ueq}) and (\ref{hshell}) then gives
\be
\frac{dr}{r} \simeq - \ve\, dw\, \frac{(1 + w)}{w^2}
\label{req}
\ee
so that because $\ve \ll 1$ the radius $r$ hardly changes within region II, and $dr$ is of order $\ve \,dw$. 
The final unknown function $f$ is given by (\ref{cons}) to be $f = (r/r_1)^2 (w_1/w) f(r_1)$ so that
\be
f \simeq \frac{w_1}{\!w}\, f(r_1)
\label{fsoln}
\ee
to leading order in $\ve$ for $\ve \ll 1$.

Continuity ($\cC^0$) of the metric functions $f$ and $h$ at $r_1$ and $r_2$ gives the conditions
\begin{subequations}
\begin{align}
f(r_1) &= C\, h(r_1) = C\, (1-H^2 r_1^2) \simeq C\, \ve \,\frac{(1+ w_1)^2\!\!}{w_1}\label{match1}\\ 
f(r_2) & = h(r_2) = 1 - \frac{2GM}{r_2} \simeq  \ve \,\frac{(1+ w_2)^2\!\!}{w_2}
\label{match2}
\end{align}\label{match}
\end{subequations}
which together with (\ref{fsoln}) evaluated at $r=r_2,w=w_2$ provides
\be
C\,(1+w_1)^2 = (1 + w_2)^2
\label{Cw1w2}
\ee
and hence three independent relations among the eight integration constants $(r_1, r_2, \\w_1, w_2, H_0, M, C, \ve)$. 
Assuming that $(r_1, r_2, w_1, w_2, H_0, M, C)$ all remain finite as $\ve\to 0$, {\it i.e.} they are all of order $\ve^0$, 
then $r_1\! \to\!  \rH =H^{-1}$ and $r_2\!  \to \! \rM$ with $r_2-r_1=\D r$ of order $\ve$, so that $\rH \simeq \rM$ 
to leading order in $\ve$. Thus the boundary layer {\rm II} straddles the location of the classical Schwarzschild and dS
horizons, and $r_1 \to r_2$ coincide at $\rH = \rM$, becoming no longer independent in the limit $\ve \to 0$. Since the 
mass $M$ is a free parameter there remain three undetermined integration constants $C, w_1 ,w_2$ which satisfy 
the one relation (\ref{Cw1w2}) in addition to $\ve \ll 1$ itself. 

\begin{svgraybox}
The important feature of this solution is that for any $\ve>0$ both $f$ and $h$ are of order $\ve$ but nowhere 
vanishing.  Hence there is no event horizon or trapped surface, and $t$ is a global Killing time. A photon experiences 
a very large, ${\cal O}(\ve^{-\frac{1}{2}})$ but finite blue shift in falling into the shell from infinity.
\end{svgraybox}

The physical proper thickness of the shell in the metric (\ref{sphsta}) is
\be
\ell = \int_{r_1}^{r_2}\frac{dr}{\!\!\!\sqrt{h}} = \rM\! \sqrt{\ve}\int_{w_2}^{w_1}\! dw\, w^{-\frac{3}{2}}
= 2 \rM\!\sqrt{\ve}\ \Big(w_2^{-\frac{1}{2}} - w_1^{-\frac{1}{2}}\Big)
\label{thickness}
\ee
to leading order in $\ve$, and hence $\ell$ is ${\cal O}(\ve^{\frac{1}{2}})$ and small compared to $\rM$. 
The magnitude of $\ve$ and hence of $\ell$ can be fixed only by consideration of the quantum effects that give rise 
to the phase transition boundary layer, which will be discussed in Sec.\ref{Sec:Anom}.

The entropy of the thin shell is obtained from its eq.~of state, which is that of a relativistic fluid in $1+1$ dimensions,
and can be written in the form $p = \r = (a^2/8\p G) (k_B T/\hslash)^2$, where $G$ is introduced for dimensional 
reasons, so that $a^2$ is a dimensionless constant.  By the standard Gibbs relation, $T s = p + \r$ 
for a relativistic fluid with zero chemical potential, the local specific entropy density is 
\be 
s(r) = \frac{a^2k_B^2 T(r)}{4\p \hslash^2 G} =\frac{ak_B}{\hslash} \left(\frac{p}{2\p G}\right)^{\!\!\frac{1}{2}}
= \frac{ak_B}{4\p \hslash Gr} \,w^{\frac{1}{2}}
\label{localent}
\ee
for local temperature $T(r)$. The entropy of the fluid within the shell is
\be
\hspace{-3mm}S =4\p\! \int_{r_1}^{r_2} \frac{s\, r^2\,dr}{\!\!\!\sqrt{h}\ }
= \frac{ak_{_B}\rM^2}{\hslash G} \sqrt{\ve} \,\ln \Big(\frac{w_1}{w_2}\Big)
\sim  a\, k_B \frac{M\ell}{\hslash}
\label{entsh}
\ee
and of order $k_B M \ell/\hslash$ to leading order in $\ve$, assuming $a,w_1,w_2$ are $\cO(1)$. Since the interior region I 
has $\rho_{_V} = -p_{_V}$, $(T s)_{_V} = p_{_V} + \r_{_V}=0$ there. This is in accord with a gravitational Bose-Einstein condensate
(GBEC) being a single macroscopic quantum state with zero entropy. Thus the entropy of the entire compact quasi-black hole (QBH)
is given by the entropy of the shell alone. By (\ref{entsh}) this is of order $k_B (\rM/L_{\rm Pl})^{\frac{3}{2}}$ for 
$\ell \!\sim \!\sqrt{L_{\rm Pl}\rM}$, or $S\!\sim\! \sqrt{\ve} S_{BH} \ll S_{BH}$, far smaller than the Bekenstein-Hawking entropy
(\ref{SBHg}). The $M^{3/2}$ scaling of (\ref{entsh}) furthermore makes it comparable to the entropy of typical stellar progenitors 
of mass $M$, in the range of $10^{57} k_B$ to $10^{59} k_B$ for a solar mass and $M_\odot/m_N \sim 10^{57}$ nucleons. Thus 
there is no information paradox arising from an enormous entropy unaccountably associated with a BH horizon, if the horizon 
is replaced by a thin boundary layer of this kind. Since $w$ is of order unity in the shell, the {\it local} temperature of the fluid 
within the shell is of order $T_{_H}  \sim \hslash/k_B \rM$ and quite cold, so that the typical quanta are {\it soft} with wavelengths 
of order $\rM$, and there is no transplanckian problem. 

\begin{svgraybox}
Because of the global timelike Killing field $K_{(t)} =\pa_t$ and absence of either an event horizon or an interior singularity, 
a gravitational condensate star shows no loss of unitarity, information paradox, or any conflict with either quantum theory
or general principles of statistical mechanics.  As a static solution, neither the interior nor the thin shell boundary layer 
emit Hawking radiation. A gravitational condensate star is both cold and dark, and hence in its appearance to distant 
observers and in its external geometry, in most respects indistinguishable from a BH.
\end{svgraybox}

\subsection{The Lanczos-Israel conditions at the $r_1$ and $r_2$ boundaries}

In the original gravastar model of \cite{gravastar:2001,Grav_Univ:2023,MazEMPNAS:2004} the Lanczos-Israel junction conditions
\cite{Lanczos:1922,OBrSyn:1952,Israel:1966a,Israel:1967}
\be
[K_a^{\ b}] \equiv K_{a\, +}^{\ b} - K_{a\, -}^{\ b} =- 4 \pi G\, \Big( 2  S_a^{\ b} - \del_a^{\ b} S_c^{\ c}\Big)
\label{Kdisc}
\ee
were used to relate the discontinuity in the extrinsic curvature
\be
K_{ab} =-n_\m  \y_{(b)}^{\ \n} \na_{\!\n} \y_{(a)}^{\ \m}=
- n_\m  \y_{(b)}^{\ \n}\big(\pa_\n \y_{(a)}^{\ \m} + \G^\m_{\ \ \,\n\la}\y_{(a)}^{\ \la}\, \big)
\label{Kdef}
\ee
to the surface stress tensor $S_a^{\ b}$, where $n^\m$ is the spacelike normal to the surface at fixed $r$, normalized to 
\be
n^\m n_\m =1
\label{nnorm}
\ee
in the full four-dimensional spacetime. The indices $a,b$ in (\ref{Kdisc}) are intrinsic to the surface and thus range over $t,\th,\f$ only,
while $\y_{(a)}$ are a set of three mutually orthogonal basis vectors, orthogonal also to $n_\m$, {\it i.e.} satisfying 
$\y_{(a)}^{\ \m} n_\m = 0$, and normalized so as to project the four-dimensional metric onto the hypersurface of fixed $r$. Thus
\be
\g_{ab} = \y_{(a)}^{\ \m} \y_{(b)}^{\ \n} g_{\m\n}
\ee
is the induced metric on the three-dimensional hypersurface, the inverse of which must be used to raise the indices $a,b,c$.
Since the normal to the surface at fixed $r$ and basis vectors so defined have components
\be
n^\m = \sqrt{h(r)} \,\del^\m_{\ \, r}\,,\qquad \y^{(a)}_\m = \left\{ \begin{array}{cl}\del^{\ a}_\m\,, \quad& \m = t, \th,\f\\
0\,,\quad& \m= r\end{array}\right.
\label{nvcomponents}
\ee
in the coordinates (\ref{sphsta}), the non-vanishing components of $K_a^{\ b} = K_{ac}\g^{cb}$ in these 
coordinates are
\bes
\bea
K_t^{\ t} &= \displaystyle {\frac{\!\!\sqrt{h}}{f}}\, \G_{\, rtt} = \frac{\!\!\sqrt{h}}{2f}\,\frac{d\! f\!}{dr}\\
K_\th^{\ \th}= K_\f^{\ \f} &= - g^{\th\th} \sqrt{h}\,  \G_{r\th\th} = \displaystyle{\frac{\!\!\sqrt{h}}{r}}
\eea
\label{Knon0}\ees
with the result that the non-vanishing components of the surface stress tensor from (\ref{Kdisc}) are
\bes
\bea
&S\!_t^{\ t} = \displaystyle{\frac{1}{4\p G}}\left[\displaystyle{\frac{\!\!\sqrt{h}}{r}}\right]\\
&S\!_\th^{\ \th}= S\!_\f^{\ \f} =\displaystyle{\frac{1}{8 \p G}} \left( \left[ \displaystyle{\frac{\!\!\sqrt{h}}{2f}\,\frac{d\! f\!}{dr}}\right] + \left[\displaystyle{\frac{\!\!\sqrt{h}}{r}}\right]\right)
\eea
\label{SKdisc}\ees
on the timelike surface interfaces at $r_1$ and $r_2$. Since the metric function $h$ is continuously matched at the interfaces,
$[K_\th^{\ \th}]$ and $S_t^{\ t}$ vanish, while $S_\th^{\ \th}$ is of order $\ve^{-\frac{1}{2}}$. Making use of the 
eqs.~(\ref{hshell})-(\ref{match})
\be
\frac{\sqrt{h}}{2f}\frac{d\!f\!}{dr} \simeq \frac{1}{2\rM} \sqrt{\frac{w}{\ve}}\qquad \quad\rm in\ region\ II \hspace{-1.5cm}
\label{regionII}
\ee
to leading order in $\ve$, which enables evaluation of the discontinuities in (\ref{SKdisc}). The non-zero angular components are 
then~\cite{Grav_Univ:2023}~\footnote{The sign conventions in \cite{gravastar:2001,MazEMPNAS:2004} are such 
that $\s_{1,2}$ there are the {\it negative} of the surface stress tensors $S^{\ \th}_\th = S^{\ \f}_\f$ properly defined here. 
Eqs.~(C5) and (C7) of \cite{MazEM:2015} also have an overall sign change from the Lanczos-Israel eq.~(\ref{Kdisc}) for 
$S^{\ b}_a$, such that $\h,\s$ of (C7) have the same values as $\h,\s$ in \cite{gravastar:2001,MazEMPNAS:2004}.}
\vspace{-2mm}
\begin{subequations}
\begin{align}
&S^{\ \th}_{\!\th}\big\vert_{r=r_1} = S^{\ \f}_{\!\f}\big\vert_{r=r_1}\equiv -\s_1= \frac{1}{32\p G^2M}\frac{(3 + w_1)}{(1 + w_1)}\left(\frac{w_1}{\ve}\right)^{\!\!\frac{1}{2}}\\[0.8ex]
&S^{\ \th}_{\!\th}\big\vert_{r=r_2}  = S^{\ \f}_{\!\f}\big\vert_{r=r_2}\equiv -\s_2= -\frac{1}{32\p G^2M} \frac{w_2}{(1 + w_2)}\left(\frac{w_2}{\ve}\right)^{\!\!\frac{1}{2}}
\end{align}
\label{surf}\end{subequations}
respectively, to leading order in $\ve$, at $r_1$ and $r_2$. 

The signs of these surface stresses correspond to the inner surface at $r_1$ exerting an outward force and the outer surface 
at $r_2$ exerting an inward force, {\it i.e.} both surfaces exert a confining force on the thin shell layer in region II. Clearly 
these large transverse surface stresses violate the perfect fluid ansatz at the interfacial boundaries. Nevertheless, as it will 
turn out from semi-classical estimates in Sec.~\ref{Sec:Anom},  $\ve^{-\frac{1}{2}} \sim (M/M_{\rm Pl})^\frac{1}{2}$, so 
that the surface tensions (\ref{surf}) are of order $M^{-\frac{1}{2}}$ and far from Planckian. Thus the matching of the metric 
at the phase interfaces $r_1$ and $r_2$, analogous to that across stationary shocks in hydrodynamics, should be 
reliable in the mean field semi-classical approximation. The time component of the surface stress tensor at $r_1$ and $r_2$ 
vanishes and makes no contribution to the Misner-Sharp mass function $\m(r)=2Gm(r)$ at either of the two interfaces. 

The Misner-Sharp mass-energy within the shell 
\be
E_{\rm II} = 4\p\! \int_{r_1}^{r_2}\!\rho\,r^2 dr = \ve M\! \int_{w_2}^{w_1}\frac{dw}{w}\,(1 + w)
= \ve M\, \Big[\!\ln \Big(\sdfrac{w_1}{w_2}\Big) + w_1 - w_2\Big]
\ee
to leading order in $\ve$, is of order $M_{\rm Pl}$ and also extremely small. In this accounting essentially all of the mass 
of the object comes from the energy density of the vacuum condensate in the interior, even though the shell is responsible 
for all of its entropy.

\subsection{Thermodynamic Stability}

In \cite{Grav_Univ:2023,MazEMPNAS:2004} a thermodynamic argument for the stability of the first gravastar model was provided.
This was based on analyis of the entropy functional which can be expressed in the form 
\be
S=\frac {a k_{_B}}{\hslash G}\int_{r_1}^{r_2}r \,dr\, \left(\frac{d\m}{dr}\right)^{\!\!\frac{1}{2}}
\left(1 - \frac{\m(r)}{r}\right)^{\!\!-\frac{1}{2}}
\label{entropyfn}
\ee
in the thin shell region II where the eq.~of state $p=\rho$, (\ref{localent}), and the relation $d\m(r) = 8 \p G \rho r^2 dr = w dr$
has been used. The first variation of (\ref{entropyfn}) with respect to $\m(r)$ with  the endpoints $r_1$ and $r_2$ fixed vanishes, 
{\it i.e.} $\del S =0$ by the Einstein eqs.~(\ref{Eins}) for a static, spherically symmetric star. Thus any solution of 
eqs.~(\ref{Eins})-({\ref{cons}) is guaranteed to be an extremum of $S$ \cite{Cocke:1965}. This is consistent with 
regarding Einstein's eqs. as an effective field theory (EFT) of low energy hydrodynamics, strictly valid only at long wavelengths.

The second variation of (\ref{entropyfn}) is 
\be
\del^2 S=\frac{a k_B}{4\hslash G}\int_{r_1}^{r_2}\!r\,dr\, \left(\frac{d\m}{dr}\right)^{\!\!-\frac{3}{2}}
h^{\!-\frac{1}{2}}\left\{-\left[\frac{d(\del\m)}{dr}\right]^2 + \frac{(\del\m)^2}{r^2h^2}
\frac{d\m}{dr}\left(1+ \frac{d\m}{dr}\right) \right\}
\label{varent}
\ee
when evaluated on the solution. Associated with this quadratic form in $\del\m$ is a second order linear differential 
operator $\cL$ of the Sturm-Liouville type, {\it viz.}
\be
\cL \,\chi = \frac{d}{dr} \left\{r\left(\frac{d\m}{dr}\right)^{\!\!-\frac{3}{2}}\!
h^{\!-\frac{1}{2}} \frac{d\chi}{dr}\right\}+ \frac{h^{-\frac{5}{2}}}{r} 
\left(\frac{d\m}{dr}\right)^{\!\!\frac{1}{2}}\left(1+ \frac{d\m}{dr}\right)\, \chi \,.
\label{stli}
\ee
This operator possesses two solutions satisfying ${\cL}\chi_{_0} = 0$, obtained by variation of the classical solution, 
$\m (r; r_1, r_2)$ with respect to the parameters $(r_1, r_2)$. Indeed by changing variables from $r$ to $w$ and using 
the explicit solution (\ref{hshell})-(\ref{req}) it is readily verified that one solution to $\cL\chi_{_0}=0$ is $\chi_{_0}=1-w$, 
from which the second linearly independent solution $(1-w)\ln\,w + 4$ may be obtained. Since these correspond to 
varying the positions of the $r_1,r_2$ interfaces, neither $\chi_{_0}$ vanishes at $(r_1, r_2)$ and neither is a true zero 
mode.  However, by setting $\del \m = \chi_{_0}\, \psi$, where $\psi$ does vanish at the endpoints and inserting this into the 
second variation (\ref{varent}) one obtians 
\be
\del^2 S= -\frac{a k_B}{4\hslash G}\int_{r_1}^{r_2}\!r\, dr\, \left(\frac{d\m}{dr}\right)^{\!\!-\frac{3}{2}}
h^{-\frac{1}{2}}\chi_{_0}^2\,\left(\frac{d\psi}{dr}\right)^{\!2} < 0
\label{secvar}
\ee
after integration by parts and using $\del \m =0$ at the endpoints and $\cL \chi_{_0} = 0$.

Since (\ref{secvar}) is negative definite, the entropy of the solution is maximized with respect to radial variations that 
vanish at the endpoints, with fixed total energy. Since deformations with non-zero angular momentum decrease the 
entropy even further, stability under radial variations is sufficient to demonstrate that the solution is stable to all small 
perturbations with the given boundary conditions. On the other hand, the boundary conditions that the radial variations 
$\del\m$ vanish at the interface boundaries at $r_1$ and $r_2$ may themselves be questioned. This apparent shortcoming
of the thermodynamic stability proof provided in \cite{Grav_Univ:2023,MazEMPNAS:2004} can only be addressed within
a a fully dynamical theory where the boundary conditions follow from the dynamical variational principle itself,
rather than being imposed, as they were in the original gravastar model of \cite{Grav_Univ:2023,MazEMPNAS:2004}.
The EFT introduced in \cite{EMEFT:2022} and discussed in Sec.~\ref{Sec:EFT} is expected to settle the question of
stability with a full dynamical treatment based on an effective Lagrangian whose Euler-Lagrange eqs.~possess
a gravitational condensate star solution, removing any additional assumptions of boundary conditions beyond
those of a non-singular interior and an asymptotically flat exterior.

Even from the preliminary analysis of \cite{Grav_Univ:2023,MazEMPNAS:2004}, it should be clear that the main qualititative 
results do not depend crucially on the specific eq.~of state used in the thin shell transition region II. The important features 
are the interior dS and exterior Schwarzschild regions and the location of the thin shell at or straddling the mutual 
Schwarzschild and dS horizons, where they are joined. The assumption of the specific eq.~of state $p=\r$ of the
thin shell in the first gravastar model of \cite{gravastar:2001,MazEMPNAS:2004} should therefore be regarded as 
an ansatz, illustrating a proof of principle of joining the interior dS geometry of a GBEC condensate with $\La_{\rm eff} >0$
to the exterior Schwarzschild solution, with the possibility left open to modification of the thin shell transition region when 
a more complete or fundamental theoretical framework became available~\cite{Grav_Univ:2023}. A proposal 
for that more fundamental EFT framework will be described in Secs.~\ref{Sec:Anom} and \ref{Sec:EFT}.

\section{The Schwarzschild Constant $\bar \r$ Interior Solution Revisited: Evading the Buchdahl Bound and Determination of $C$}
\label{Sec:SchwInt}

A major step and refinement of the original gravastar proposal took place in 2015, with the re-analysis of the
Schwarzschild constant density interior solution (\ref{intsoln}) in the limit of $\rS \to \rM$, {\it i.e.} for $\ve =0$,
leading also to an improved understanding and formulation of the Lanczos-Israel conditions for null hypersurfaces.

In 1959 H. A. Buchdahl proved that in order for a static, spherically symmetric solution of Einstein's eqs.~of mass $M$
and radius $\rS$ to remain everywhere finite, its compactness $GM/c^2 \rS$ must not be greater than $4/9$, 
or equivalently 
\be
\rS \ge \frac{9}{8} \rM 
\label{Buch}
\ee
the same as from the limiting case (\ref{rsbound}) of the constant density interior Schwarzschild
solution (\ref{intsoln}). The Buchdahl bound (\ref{Buch}) depends upon three necessary conditions:
\begin{enumerate}[label= (\roman*)]
\item The pressure is everywhere isotropic: $p_{\perp}(r) = p(r)$,
\item The density profile is monotonically non-increasing outward: $\r'= \frac{d\r}{dr} \le 0$,
\item The metric functions $f(r),h(r)$ and the derivative $\frac{df}{dr}$ are continuous at $r=\rS$,
\end{enumerate}
in addition to Einstein's eqs.~holding everywhere. The first condition (i) is satisfied by all ordinary fluid eqs.~of state. The second condition (ii) is physically reasonable 
since an outer layer of higher density material would generally result in an instability to the denser material falling towards 
the center. The third condition (iii) precludes a discontinuity at the surface of the star. 

When the inequality (ii) is saturated, {\it i.e.} $\r = \bar \r$ is a constant, one recovers the Schwarzschild interior solution 
(\ref{intsoln}), which actually disallows the equality in (\ref{Buch}), since in that marginal case the pressure (\ref{pressint})
diverges at $r_0 = 0$. This is consistent with the bound on the central pressure \cite{MartinVisser:2003}
\be
p(0) \ge \bar \r \,\left[ \frac{ 1 - \sqrt{1 - \rM/\rS}} {3 \sqrt{1 -\rM/\rS} - 1} \right]
\label{p0}
\ee
when condition (ii) holds, since this lower bound diverges when $\rS = \frac{9}{8} \rM$.

Because the bound for the general spherically symmetric static solution obeying conditions 
(i)-(iii) is saturated by the limiting case of $\r'=0$, the behavior of the constant density solution (\ref{intsoln}) itself is 
of fundamental interest as the limit (\ref{Buch}) is reached and $\rS$ is then reduced further.

When one does consider the constant density solution (\ref{intsoln}) for $\rS< \frac{9}{8} \rM$, some of its rather 
remarkable features quickly become apparent. First, the pressure divergence which first appears at the origin moves 
out to a spherical surface of finite radius $r_0$ given by (\ref{r0def}), and a {\it new regular solution} for $0\le r < r_0$ 
opens up behind it, with $D <0$, {\it negative} pressure, and $\bar \r + 3p <0$, violating the strong energy condition. 
The metric functions $f,h$ remain non-negative in the interior, with $f(r_0) =0$ at $r=r_0$ only, {\it cf.} 
Figs.~\ref{Fig:Pressuresing}-\ref{Fig:hvarious}.

\begin{figure}[ht] 
\begin{center}
\includegraphics[height=5cm, trim=6cm 0cm 2cm 0cm, clip=false]{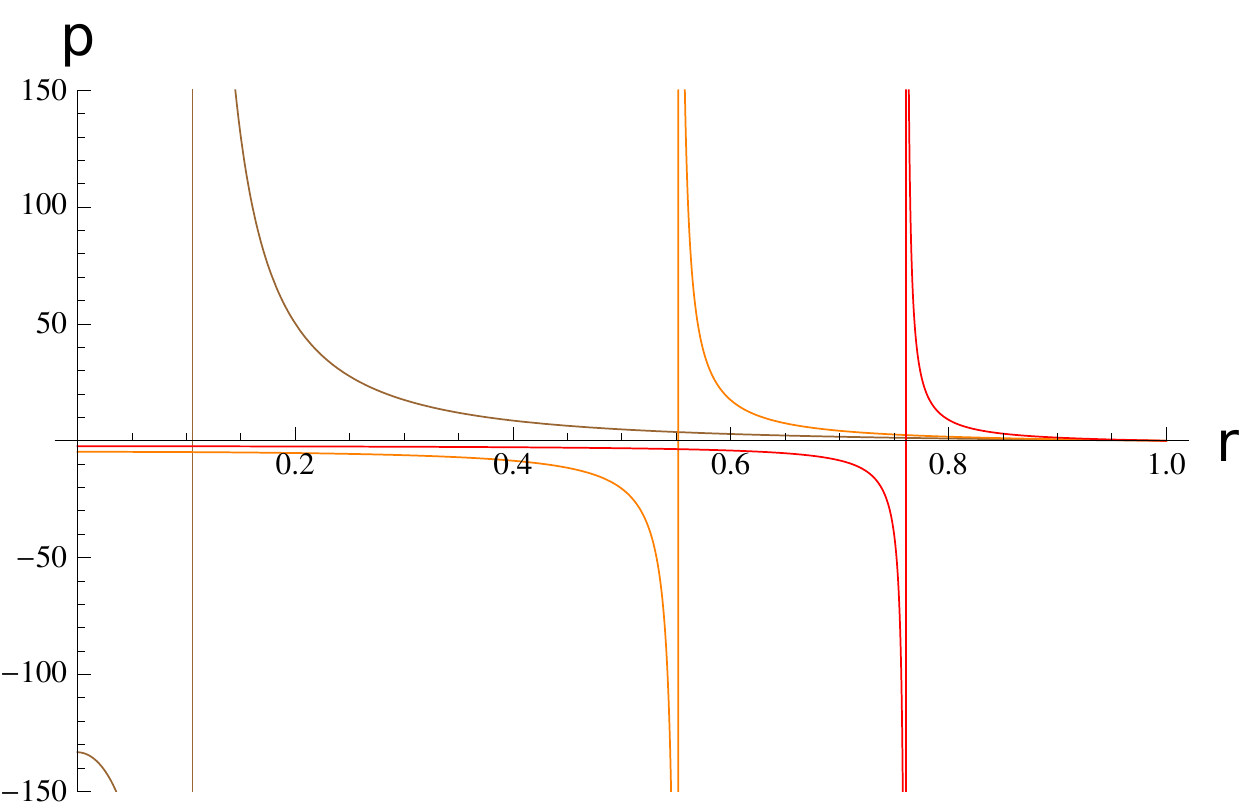}\\[2ex]
\includegraphics[height=3.7cm, trim=2cm 0cm 6cm 1cm, clip=false]{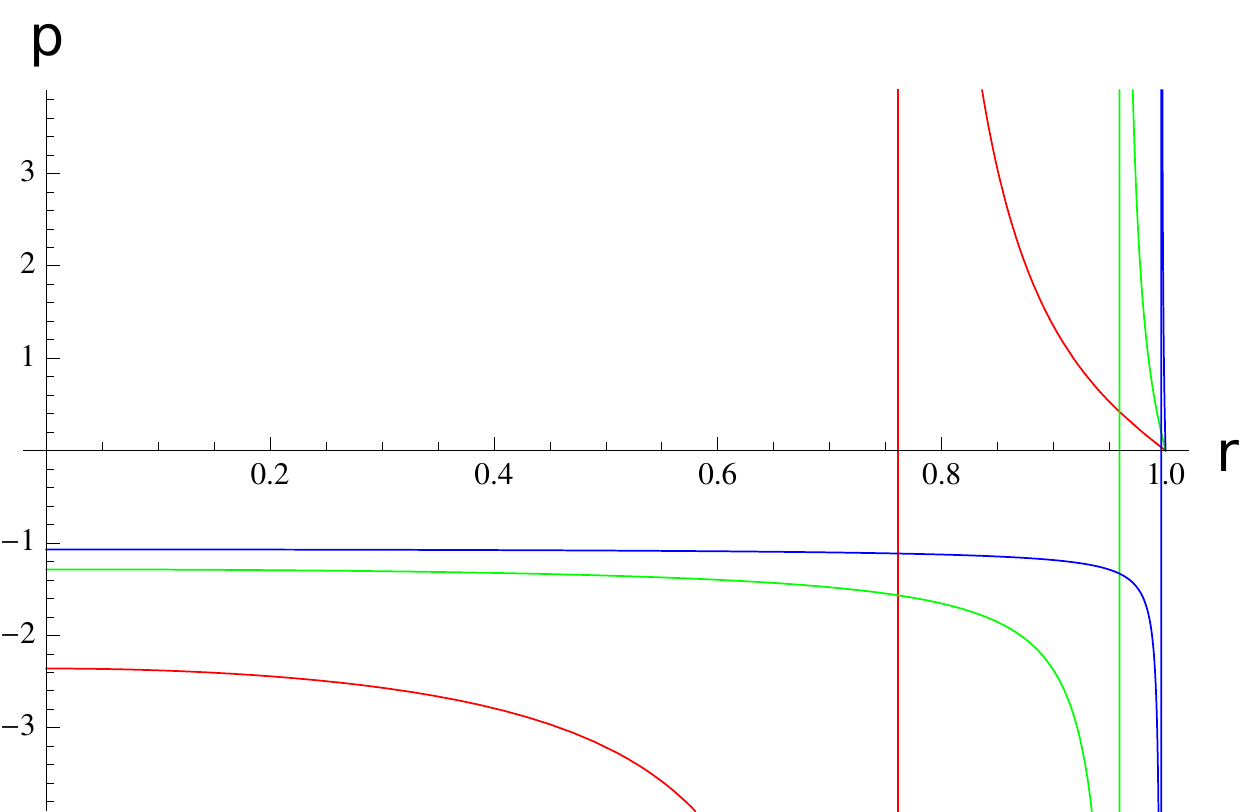}
\end{center}
\vspace{-3mm}
\caption{Pressure (in units of $\bar\rho$) as a function of $r$ (in units of $\rS$) of the interior Schwarzschild
solution (\ref{intsoln}) for various values of the parameter $\rS/\rM < 9/8 = 1.125$. The upper plot shows $p(r)$
for the values $\rS/\rM = 1.124, 1.087, 1.053$ (brown, orange, red curves), where the divergence in the pressure 
occurs at $r_0/\rS = 0.106, 0.552, 0.761$ respectively. The lower plot shows $p(r)$ for the values $\rS/\rM= 1.053, 1.010, 1.001$ 
(red, green, blue curves), where the divergence in the pressure occurs at $r_0/\rS = 0.761, 0.959, 0.996$ respectively. 
For $r <r_0$ the pressure and $\bar \r + 3p$ are negative. Note the change of vertical scale in the plots (the red curves 
are the same in each) and the pointwise approach of the negative interior pressure $p \rightarrow -\bar\rho$
as $\rS$ approaches the Schwarzschild radius $\rM$ from above and $r_0$ approaches $\rM$ from below.}
\vspace{-5mm}
\label{Fig:Pressuresing} 
\end{figure}

\begin{figure}[ht] 
\begin{center}
\includegraphics[height=5cm, trim=2cm 7mm 2cm 0cm, clip=false]{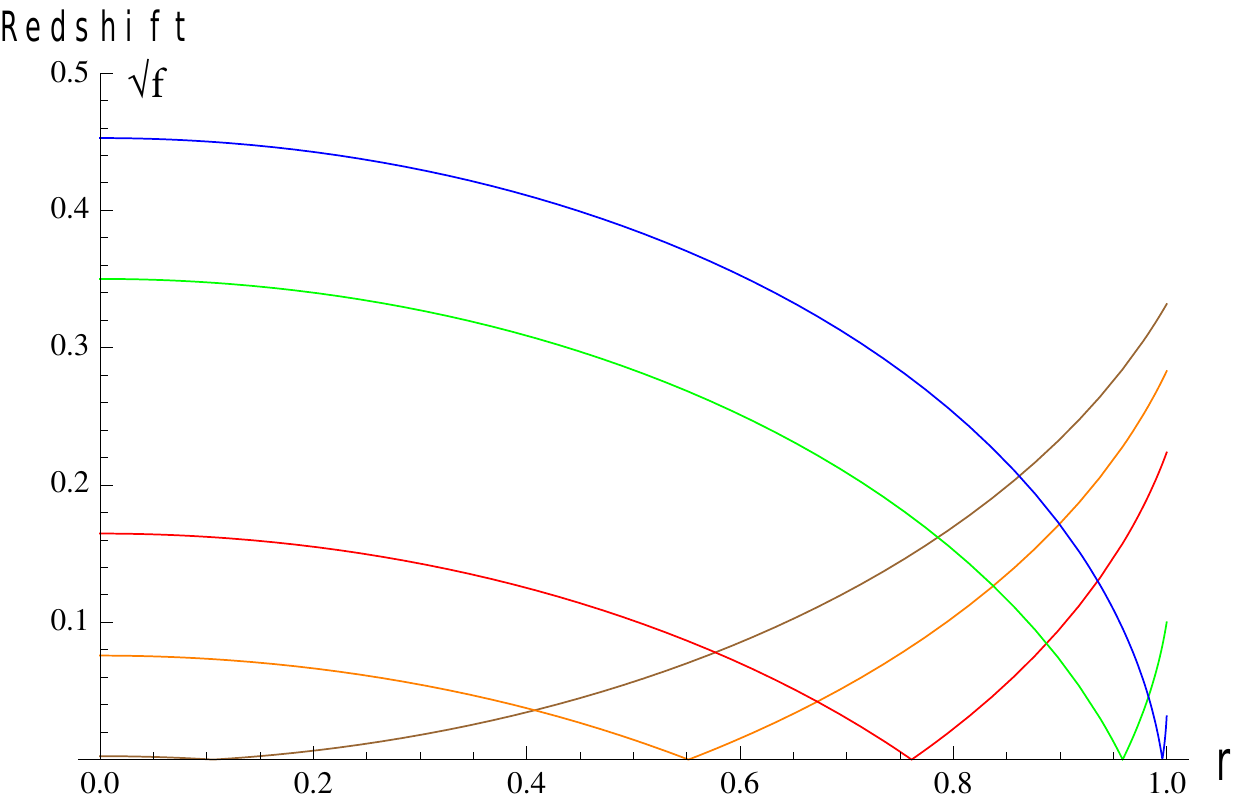}
\end{center}
\caption{The redshift factor $\sqrt{f}$ as a function of $r$ (in units of $\rS$) of the interior Schwarzschild
solution (\ref{intsoln}) for the same values of the parameter $\rS/\rM< 9/8 = 1.125$ as in Figs. \ref{Fig:Pressuresing}.
The brown, orange, red, green and blue curves are for the values $\rS/\rM = 1.124, 1.087, 1.053,1.010, 1.001$
respectively. Note the approach of the zero of $\sqrt{f}$ at $r_0$ towards $\rS$ from below as $\rS$ 
approaches the Schwarzschild radius $\rM$ from above. In this limit $\sqrt{f(0)} \to 1/2$.}
\vspace{-1mm}
\label{Fig:Redshiftsing} 
\end{figure}

\begin{figure}[ht] 
\begin{center}
\includegraphics[height=5cm, trim=2cm 7mm 2cm 0cm, clip=false]{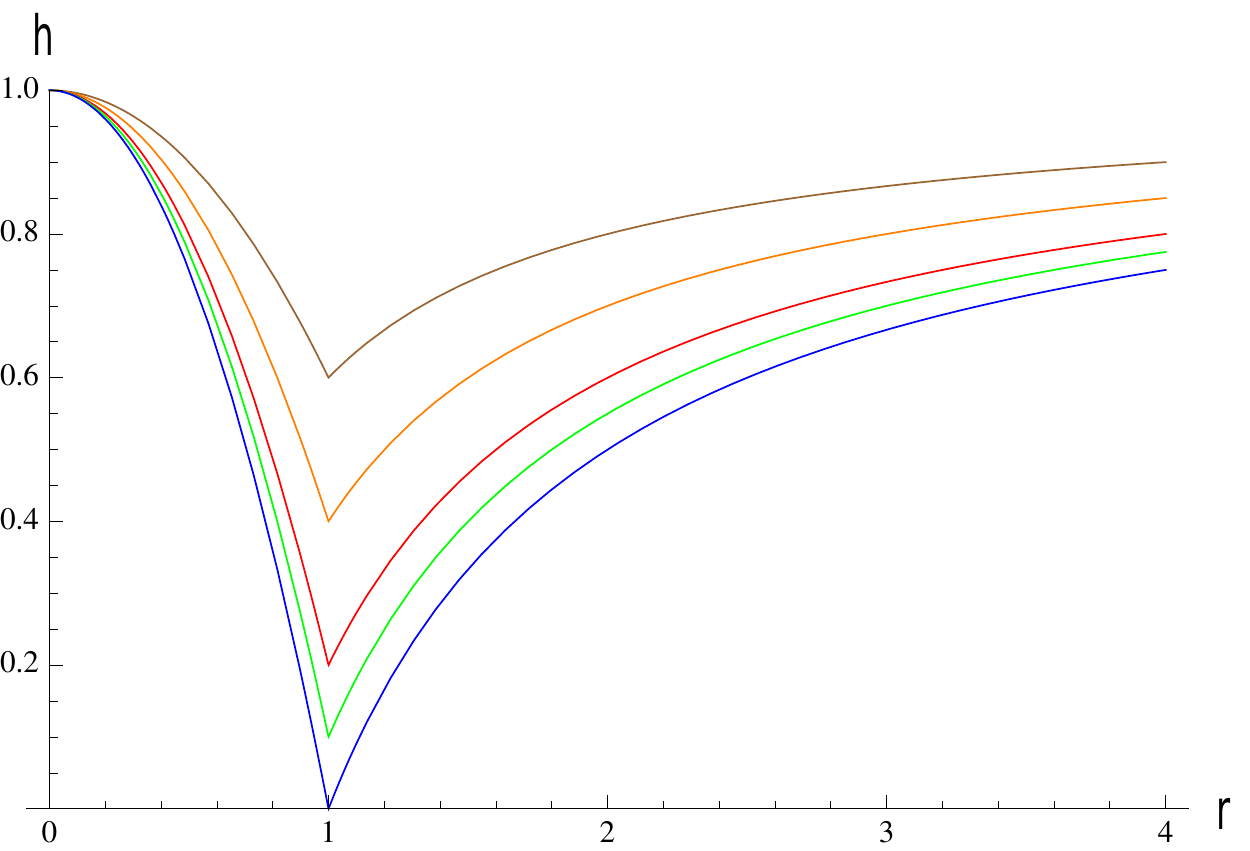}
\end{center}
\caption{The metric function $h$ as a function of $r$ (in units of $\rS$) of the interior and exterior Schwarzschild 
solution for the values of the parameter $\rS/\rM  = 2.500, 1.667, 1.250,1.111, 1.000$  (brown, orange, red, green 
and blue curves) respectively. The minimum of $h$ approaches zero at $r=\rS$, as $\rS$ approaches the 
Schwarzschild radius $\rM$ from above. Note the cusp-like discontinuity of the derivatives $dh/dr$ at $\rS$
and $df/dr$ at $r_0$ in this and the previous figure, which coincide at $r_0 =\rS$, when $\rS=\rM$.}
\label{Fig:hvarious} 
\vspace{-5mm}
\end{figure}

As the star is compressed beyond the Buchdahl bound and its radius approaches the Schwarzschild radius 
$\rS \rightarrow \rM^+$ from outside, (\ref{r0def}) shows that the radius of the sphere where the pressure 
diverges and $f(r_0) = 0$ moves from the origin to the outer edge of the star, {\it i.e.} $r_0 \to \rM^-$, 
and in that limit the interior solution with negative pressure comes to encompass the entire interior 
region $0 \le r < \rS$, excluding only the outer boundary at $\rS= \rM$. Finally, and most remarkably of all, 
since in this limit $H^2 \rS^2 = \rM/\rS \rightarrow 1$, inspection of (\ref{pressint}) shows that
the entire interior solution then has {\it constant} negative pressure
\be
p = - \bar \rho\,,\qquad {\rm for} \qquad r < \rS=r_0 = \rM = 2GM
\ee
with the metric functions
\be
f(r) = \sdfrac{1}{4}\, (1 - H^2 r^2) = \sdfrac{1}{4}\, h(r) = \sdfrac{1}{4}\, \Big(1 - \sdfrac{r^2}{\rM^2}\Big), \qquad H = \frac{1}{\rM}
\label{fhint} 
\ee
corresponding to a regular static patch of pure dS space, although one in which $g_{tt}$ is $\frac{1}{4}$ its usual value, 
so that the passage of time in the interior is modified from what would be expected in the usual static coordinates of dS space. 
In other words the constant density interior solution which Schwarzschild found in 1916 becomes essentially the gravitational condensate 
star solution of 2001 and Sec.~\ref{Sec:gstar1}, in which the thin shell boundary layer is of infinitesimal thickness ($\ve =0$), residing
exactly at the null hypersurface $r=\rM$, fixing the value of $C=\frac{1}{4}$ of (\ref{fdS}) in the interior dS patch unambiguously.

Typical profiles of the pressure $p(r)$ and $f^{\frac{1}{2}}$ for values of the radius $\rS$ in the range $\rM < \rS < \frac{9}{8}\rS$ 
and the approach of $\rS\to \rM$ are shown in Figs. \ref{Fig:Pressuresing} and \ref{Fig:Redshiftsing}.
Insight into the pressure divergence at $r=r_0$ and vanishing of $f(r_0)$ is obtained by a careful examination of
the covariant Komar mass-energy for a time independent geometry possessing the static Killing field $K_{(t)}=\pa_t$, with 
components $K^\m_{(t)} = \del_t^{\ \m}$ in coordinates (\ref{sphsta}). This Komar mass is defined by
\be
M = \int_V\! \big(\!-2\, T_{\m}^{\ \n} + T_{\la}^{\ \la}\, \del_{\mu}^{\ \n}\big)\, K^{\m}_{(t)}\,d^3\S_\n
\ +\ \frac{1}{4 \p G} \int_{\pa V_-}\!\!\! \ka\, dA 
\label{KomarM}
\ee
expressing the total mass-energy of the system $M$ in terms of a three-volume integral of the matter stress-energy, 
plus a possible surface flux contribution from the inner two-surface, where
\be
\ka = -\na^{\m}\!K^{\n}_{(t)} \,e^0_{\ [\mu}e^1_{\ \nu]} = \displaystyle \frac{1}{2} \sqrt{\frac{h}{f}}\ \frac{d\! f\!}{dr}
\label{kapdef}
\ee
is the surface gravity, defined in terms of the vierbeins 
$e^0_{\ t} = \sqrt{-g_{tt}}= f^{\frac{1}{2}},\, e^1_{\ r} = \sqrt{g_{rr}}=h^{-\frac{1}{2}}$ with the areal integration measure
\be
dA=  r^2\sin\th\, d\th\,d\f
\label{areameas}
\ee
on the spherical two-surface of constant $t$ and $r$. 

Since the volume integration measure in (\ref{KomarM}) is
\be
d^3\S_{\n} =\del^t_{\ \n}\, \displaystyle{\sqrt{\frac{f}{h}}}\ r^2\sin\th\, dr\,d\th\,d\f 
\label{vol}
\ee
it contains a factor of $\sqrt{f}$ which vanishes at exactly the same $r=r_0$ as where $p(r_0)$ diverges, so that the
pressure singularity at $r=r_0$ and in fact the total volume integrand in (\ref{KomarM}) is \cite{MazEM:2015}
\be
\frac{1}{G}\frac{d}{dr} \big[r^2 \ka(r) \big] = 4\p \sqrt{\frac{f}{h}} \ r^2\, \big(\r + p + 2 p_{\perp}\big) =
4 \p r^2 \bar\r \ {\rm sgn} (D) + \frac{8\p}{3} \,r_0^{\ 3}\, \bar \r\ \del (r-r_0)
\label{Mintegrand}
\ee
with a well-defined $\del$-function contribution having support at $r=r_0$, which when integrated over $dr$ gives a finite result.
This $\del$-function contribution to the volume integral in (\ref{KomarM}) is a result of the discontinuity in the surface gravities
\be
\ka_{\pm} \equiv \lim_{r \to r_0^{\pm}}  \ka(r) =  \pm \frac{4 \p G}{3}\, \bar\r\,r_0 = \pm \frac{r_M\,r_0}{2\rS^3}
\label{kappm}
\ee
on either side of the $r=r_0$ surface, which results in a transverse pressure and contribution to the integrand (\ref{Mintegrand}) of
\be
8 \p \sqrt{\frac{f}{h}} \ r^2\,  (p_{\perp} - p) = \frac{8 \p }{3} \, \bar \r r_0^{\ 3}\,\del (r-r_0) = 2M\, \bigg(\!\frac {r_0}{\rS}\!\bigg)^{\!\!3} 
\,\del (r-r_0)
\label{pdiff}
\ee
localized on the spherical surface at $r=r_0$, and hence a breakdown of the isotropic pressure assumption 
on that surface.  This gives the finite contribution 
\be
E_{_S} = 2M\, \bigg(\!\frac {r_0}{\rS}\!\bigg)^{\!\!3} \to 2M
\label{Esurf}
\ee
to the Komar energy (\ref{KomarM}) of the surface, which becomes $2M$ in the limit $r_0\to \rS$.  

\begin{svgraybox}
The Schwarzschild constant density interior solution (\ref{intsoln}) in the limit $\rS\to\rM$ therefore has the physical interpretation 
of a gravastar with a well defined surface tension energy (\ref{Esurf}) localized at $r=\rM=\rH$, precisely where the BH and dS horizons 
would be, and the two geometries are joined instead.
\end{svgraybox}

The dynamical stability of this sequence of constant density Schwarzschild stars beyond the Buchdahl limit 
has been studied in \cite{PosChir:2019}.

\subsection{Redshift Modified Boundary Conditions on a Null Hypersurface}
\label{Sec:modjunction}

It is clear that the presence of the $\sqrt{f} \to 0$ factor in the volume measure (\ref{vol}) is critical in rendering the Komar energy
on the null surface at $r=r_0$ finite. This points to the modification of the Lanczos-Israel junction conditions necessary to give
a finite, well-defined and physical surface stress tensor on a null surface where the normal $n^\m$ itself becomes null
and cannot be normalized as in (\ref{nnorm}). To handle this case one should define the redshifted extrinsic curvature tensor
\be
\bK_a^{\ b} =-\g^{bc} \bn_\m  \y_{(c)}^{\ \n} \na_{\!\n} \y_{(a)}^{\ \m}=
- \g^{bc} \bn_\m  \y_{(c)}^{\ \n} \G^\m_{\ \ \,\n\la}\y_{(a)}^{\ \la}
\label{bKdef}
\ee
where the normal $\bn$ to the surface at constant $r$ approaching $r_0$ satisfies 
\be
\bn \cdot \bn = (\bn^r)^2 g_{rr} = f(r)\,, \qquad {\rm so\ that} \qquad \bn^r = \sqrt{fh}
\label{bnnorm}
\ee
which goes to zero on the null surface, instead of (\ref{nnorm}) and the first member of (\ref{nvcomponents}).
This gives a redshift modified extrinsic curvature tensor (\ref{bKdef}) with the non-vanishing components
\bes
\bea
\bK_t^{\ t} &= \displaystyle {\sqrt{\frac{h}{f}}\, \G_{\, rtt} =\frac{1}{2} \sqrt{\frac{h}{f}}\,\frac{d\! f\!}{dr} = \ka}\\
\bK_\th^{\ \th}= \bK_\f^{\ \f} &= - g^{\th\th} \sqrt{fh}\,  \G_{r\th\th} = \displaystyle{\frac{\!\!\sqrt{fh}}{r}} \to 0
\eea
\label{bKnon0}\ees
instead of (\ref{Knon0}). This redshifted extrinsic curvature has the finite discontinuities
\bes
\begin{align}
\big[\bK_t^{\ t}\big] & = [\ka]\\
\big[\bK_\theta^{\ \theta} \big]&= \big[\bK_\f^{\ \f}\big] = 0
\end{align}
\label{discbK}\ees
on the null surface, so that applying the standard Lanczos-Isreal conditions, but to this redshift modified extrinsic curvature $\bK_a^{\ b}$, 
gives the surface stress tensor density 
\be
^{(\S)}\!T_a^{\ \,b}\, \sqrt{\frac{f}{h}} = \cS_a^{\ \,b} \, \del(r-r_0)\,,\qquad 
8 \pi G\, \cS_a^{\ b} =- \big[{\bf K}_a^{\ b}\big] + \del_a^{\ b} \, \big[\bK_c^{\ c}\big]
\label{modjunc}
\ee
instead of (\ref{SKdisc}), and a finite surface tension
\be
\cS_\theta^{\ \theta} = \cS_\f^{\ \f} = \t_{_S} = \frac{\D \ka}{8 \p G} = \frac{M\,r_0}{8\p\,\rS^3}
\label{surfphys}
\ee
of the null surface at $r=r_0$, instead of (\ref{surf}) which diverges as $\ve \to 0$.

The factor of $\sqrt{f/h}$ in (\ref{surfphys}) and modified extrinsic curvature (\ref{bKdef}) is clearly necessitated by the covariant
integration measure (\ref{vol}) in the Komar energy integral (\ref{KomarM}). Once $\bK_a^{\ b}$ is defined as in (\ref{bKdef}),
with the normal $\bn$ normalized as in (\ref{bnnorm}) so as to become a null vector in the horizon limit, the surface stress tensor 
(\ref{surfphys}) on a null hypersurface can be determined unambiguously by application of the junction conditions (\ref{modjunc}),
which are of the same form as the standard Lanczos-Israel conditions, but for $\bK_a^{\ b}$ rather than $K_a^{\ b}$ of (\ref{Kdef}),
and crucially for $\bK_a^{\ b}$ with one covariant and one contravariant index. Raising or lowering these indices would
introduce factors of $f$ or $1/f$ that would again render the formalism empty or ambiguous on a null hypersurface where $f(r_0)$
vanishes. There is no need for introducing a `transverse' normal and `transverse' curvature tensor as proposed in
\cite{BarIsr:1991}.

This procedure can also be justified and verified explicitly from the form of the Einstein tensor in the spacetime metric (\ref{sphsta}).
Since the metric functions $f,h$ must be equal on the null hypersurface approached from either side, but their derivatives may
not be, the first derivatives $f',h'$ can contain Heaviside step functions. Thus Dirac $\del$-functions signaling
integrable surface contributions can only appear from second derivative terms $f'', h''$. The angular components 
$G_\theta^{\ \theta} = G_\f^{\ \f}$ in the metric (\ref{sphsta}) indeed contain a second derivative of $f$ with respect to $r$,
but they contain singular $f'/f, h'/h$ and $h/f$ terms as well, which become infinite or ill-defined on the null hypersurface
where $f(r_0) = 0$. However, the tensor {\it densities}  
\bea
&\displaystyle {\sqrt{\frac{f}{h}}\,G^{\ \theta}_{\theta} = \sqrt{\frac{f}{h}}\,G^{\ \f}_\f = 
\frac{1}{2}\frac{d}{dr} \left(\sqrt{\frac{h}{f}} \frac{df}{dr}\right) +
 \frac{1}{2r}\left(\sqrt{\frac{f}{h}} \frac{dh}{dr} + \sqrt{\frac{h}{f}} \frac{df}{dr}\right)}
\label{Gththdens}\\
&= \displaystyle{\left[\frac{1}{2} \sqrt{\frac{h}{f}}\,\frac{df}{dr}\right] \del(r-r_0)  + \dots }
\label{discGthth}
\eea
combine these terms into a total $r$ derivative. It is clear that the remaining terms of (\ref{Gththdens}), although discontinuous, 
do not contain any $\del$-function contributions to the surface stress tensor at $r=r_0$, while the $\del$-function term coming from
the total derivative first term is precisely the well-defined discontinuity in the surface gravity $\ka$ of (\ref{kapdef}). The
$\del$-function comes automatically with the correct measure factor of the Komar mass integral, so that the surface stress tensor 
in the angular components on the null hypersurface at $r=r_0$ is defined by (\ref{modjunc}), directly from Einstein's eqs. 

\begin{svgraybox}
Thus once the technical issue of the correct modification of the Lanczos-Israel junction conditions on a null hypersurface is 
determined, with the finitely integrable distributional surface tensor at $r=r_0$ given by (\ref{modjunc}), the Schwarzschild constant 
density interior solution becomes not a pathological case arguing for the necessary collapse to a BH singularity for masses violating 
the Buchdahl bound (\ref{Buch}), but on the contrary a well-defined and physically sensible solution to Einstein's eqs. for 
$\rS= \rM$, which shows that evading the Buchdahl bound requires discarding the first condition (i) of isotropy assumed 
in its proof, and the appearance of a physical surface with surface tension (\ref{surfphys}), as well as allowing a negative 
pressure interior. Furthermore and remarkably, in the limit $\rS \to \rM^+$ from above, the spherical null surface becomes 
coincident with the Schwarzschild radius itself $r_0 \to \rM^-$ from below, and the negative pressure of the interior also 
becomes a constant $p = -\bar \r$, which is just the $\ve \to 0$ classical limit of the gravitational condensate star 
of \cite{Grav_Univ:2023,MazEMPNAS:2004}. Thus this viable non-singular alternative for the final state of 
gravitational collapse might have been found just a few months after Einstein introduced GR.
\end{svgraybox}

In the gravastar limit $\rS\to \rM$ the surface stress tensor and surface tension becomes
\be
\cS_\theta^{\ \theta} = \cS_\f^{\ \f} = \t_{_S} = c^2 \frac{\D \ka}{8 \p G} = \frac{Mc^2}{8\p\,\rM^2}=  \frac{c^6}{32\p\,G^2M}
\label{surfphysM}
\ee
after restoring the factors of $c$. The discontinuity of the derivatives $f',h'$ at $r =\rM$ and non-zero stress tensor there 
in this limit clearly does not satisfy the analytic continuation assumption, or `uneventful free fall' hypothesis used to obtain 
Fig.~\ref{Fig:SchwKruskal}. The conformal diagram of the idealized classical gravastar in which $\ve =0$ identically is shown in Fig.~\ref{Fig:Conformal} instead. 

It is important that in this limit $f(\rM) = h(\rM)=0$, vanishing at one radius only, but never becoming negative. Hence light rays 
can hover there indefinitely, but there is no trapped surface on which outgoing light rays are bent inward, so this critical assumption 
of the 1965 Penrose theorem is not satisfied. There is no singularity in the interior, only a regular static patch of dS space. Since 
$p= -\bar \r$ also violates the strong energy condition, the singularity theorems relying on this condition instead of a trapped 
surface are also evaded. 

\begin{wrapfigure}{hr}{.41\textwidth}
\vspace{-7mm}
    \centering
    \includegraphics[width=6.5cm,viewport=135 85 620 705,clip]{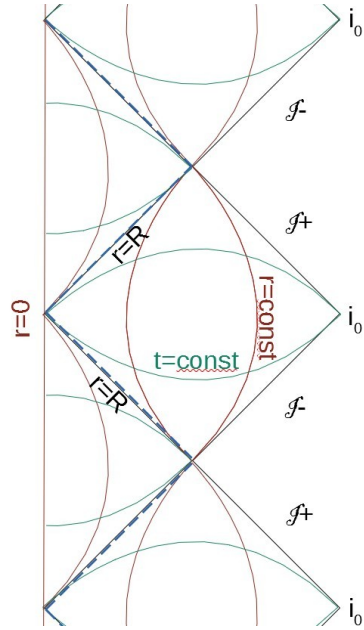}
\vspace{-6mm}
    \caption{Conformal diagram for the gravastar with angular variables $(\theta,\f)$ suppressed, in the idealized case 
$\ve\!\!=\!\!0$. Schwarzschild exterior (diamonds) and regular de Sitter static patch interior (triangles), are joined at their mutual 
null horizons at $r\!\!=R (=\!\rM\!\!=\!\rH)$.  A few typical constant $r$ timelike worldlines (red curves) and constant $t$ spacelike 
hypersurfaces (green curves) are also shown.}
    \label{Fig:Conformal}
\vspace{-1.2cm}
\end{wrapfigure}

More realistically one may expect that $f,h$ become small, $\cO(\ve)$, but never exactly vanishing when quantum effects are
included, as in the first gravastar model. Then light rays can also escape the interior in principle, and the multiple copies of the 
conformal diagram of Fig.~\ref{Fig:Conformal} would no longer appear. This shows again that global properties are very sensitive 
to small changes in the near-horizon region. The conformal diagram of a gravastar with any finite $\ve >0$ (however small) is one 
triangular wedge only of Fig.~\ref{Fig:Conformal}, resembling that of any regular body such as a star, albeit with a highly redshifted 
timelike tube, which is the thin transitional boundary layer between the Schwarzschild exterior and dS interior.

To relate (\ref{surfphysM}) to (\ref{surf}) of the earlier first gravastar model of Sec.~\ref{Sec:gstar1}, one observes that
multiplying (\ref{regionII}) by the redshift factor of $\sqrt{f(r)}$ needed in the definitions of $\bK$ and $\ka$ for
null hypersurfaces gives
\be
\ka_{\,\rm II}\!\simeq \frac{1}{2 \rM} \sqrt{\frac{w f}{\ve}} \simeq \frac{1}{2 \rM} \sqrt{\frac{w_1 f(r_1)}{\ve}}\simeq \frac{1+ w_1}{4\rM}
\nonumber\ee
in the entire region II, which is both finite and {\it constant} as $\ve \to 0$. Thus adding the contributions to the surface stress 
tensor from the discontinuities of $\bK_a^{\ b}$ at the two surfaces at $r_1$ and $r_2$, give equal and opposite contributions from 
the intermediate region II, which cancel. This leaves only the total discontinuity from the interior dS region I to exterior Schwarzschild region III, {\it i.e.}
\bea
&\sqrt{f(r_1)}\,S^{\ \th}_{\!\th}\big\vert_{r=r_1} + \sqrt{f(r_2)}\,S^{\ \th}_{\!\th}\big\vert_{r=r_2}= 
\displaystyle{\frac{3 + w_1}{64 \p G^2 M}-  \frac{w_2}{64 \p G^2 M} \frac{1+w_1}{1+w_2}} \nn
&= \displaystyle{\frac{1}{16 \p G^2 M} = \frac{\D \ka}{8 \p G} }
\label{totaltau}
\eea
where (\ref{Cw1w2}) with $C = 1/4$, implying $w_1 = 1 +2 w_2$ has been used. Hence the total surface tension 
is the very thin boundary layer on the null horizon is obtained, to leading order in $\ve \ll 1$ (even if non-zero),
 in which no reference to the intermediate region II, its eq.~of state or integration constants $w_1,w_2$ at all
appear in the final result. Thus the multiplication of the extrinsic curvature and surface stress tensor defined by 
(\ref{Kdisc})-(\ref{SKdisc}) by an additional factor of $\sqrt{f}$, as required by the volume measure in the
the Komar energy integral gives a finite result when the normal in the definition (\ref{Kdef}) declines into tangency with 
the hypersurface and $\bn$ itself becomes a null vector~\cite{BarIsr:1991,BelGonEM1:2022}. This result is finite
and universal, in the sense of being entirely independent of any assumptions of the eq.~of state or any other characteristics
of the thin shell boundary layer interposed between the exterior Schwarzschild and interior dS geometries,
joined at their mutual null horizons, in the classical limit $\ve \to 0$.

\subsection{The `First Law' for Spherically Symmetric Gravastars}

For the sourcefree exterior BH solution of (\ref{sphsta}), the volume term in (\ref{KomarM}) vanishes if integrated
from $r=\rM$ to $r =\infty$, but since
\bea
\ka_{ext} (r) = \frac{GM}{r^2}\to \ka_+ =\frac{1}{4GM}  = \frac{1}{2\rM}
\qquad as\qquad r\to\rM
\eea
the surface integral at the horizon boundary gives the entire mass $M$ of the BH. Already at the level of the
Smarr relation (\ref{Smarr}), the coefficient of $dA_{_H}$ appears to be some sort of surface tension,
but this interpretation is unclear since the BH horizon is assumed to be a causal boundary only, 
with vanishing surface stress tensor.

On the other hand, for the full interior + exterior Schwarzschild solution in the gravastar limit with $\rS=\rM$, there is a surface stress
tensor and physical surface tension (\ref{surfphysM}) at $r=\rM$. Since the solution is regular at the origin, one may integrate 
(\ref{KomarM}) from $r=0$ to $r=\infty$. There is no contribution from the last explicit surface area integral in (\ref{KomarM}), 
but the integrand (\ref{Mintegrand}) of the volume integral in (\ref{KomarM}) has a Dirac $\del$-function contribution at $r=\rM$ 
from the discontinuity (\ref{kappm}) in the surface gravities there, which becomes
\be
\D \ka \equiv \ka_+ - \kappa_- = \frac{1}{2GM} = \frac{1}{\rM} 
\label{DelkapM}
\ee
for $\rS=\rM$. This surface tension itself gives rise to a surface energy (\ref{Esurf}), which becomes just $2M$ in the gravastar limit.
The first term in (\ref{Mintegrand}) gives the volume contribution $-M$ to the Komar mass integral from the negative pressure in the interior. 
The sum of these and result from (\ref{KomarM}) is 
\be
E = E_v + E_{_S} = -M + 2M = M
\label{totE}
\ee
giving again just the total mass $+M$. The differential relation 
\be
dE = dE_v + \t_{\,S} \,dA_{_H} =-dM c^2 + \t_{\,S} \, dA_{_H}
\label{newfirst}
\ee
for the volume and surface terms in the Komar mass may be derived from this \cite{MazEM:2015}, showing that
identification of $\t_s$, the coefficient of $d A_{_H}$ with the surface tension of the physical surface
located at $r=\rM=\rH$ for the full (interior + exterior) Schwarzshild solution is fully justified. Since
$\t_{_S}$ depends on the discontinuity $\D \ka= \ka_ +-\ka_- = 2 \ka_+$ of equal and opposite surface
gravities at the surface it contributes twice the value from the Smarr BH formula (\ref{Smarr}), with the volume
term of the interior, absent in (\ref{Smarr}) making up the difference to the total $M$. Despite the
different local attributions of energy to the volume and surface in the Komar vis-a-vis the Misner-Sharp 
definition of energy, the total mass of the gravastar is $M$ in either definition, as it must be.

The constant density interior solution is the critical case that saturates and exceeds the Buchdahl bound. Its behavior 
including negative pressure and the dark energy vacuum eq.~of state $p_{_V}=-\r_{_V}$ 
are inherent in and can be described quite satisfactorily in strictly classical GR, providing an explicit counterexample
to the singularity theorems. This maximally compact classical solution has no trapped surface, assumed in the Penrose
singularity theorem. The discontinuity (\ref{DelkapM}) of the surface gravities at the would-be horizon
is instead the location of the thin shell of a gravitational condensate star, with a non-zero surface tension of
a physical surface. This possibility for a non-singular endpoint of gravitational collapse is and was always allowed
in classical GR, awaiting a more complete analysis of possible discontinuities at null surfaces, and physical 
realization of a quantum phase transition at the horizon capable of generating a change in the local vacuum
to $p_{_V}=-\r_{_V} <0$. 

A summary of the main results of the reanalysis of the Schwarzschild constant density interior solution presented in
this section is as follows.
\begin{svgraybox}
\begin{itemize}
\item
The Buchdahl bound (\ref{Buch}) on the compactness of a spherically symmetric self-gravitating mass $M$ of any kind
subject to Einstein's eqs.~is evaded and $\rS \to \rM < 9 \rM/8$, if the assumption of isotropy of the pressure everywhere
is removed, and instead $p_\perp - p \neq 0$ exhibiting a $\del$-function corresponding to a physical surface tension
at $r_0 = \rM$ when $\rS = \rM$ coincide. The constant negative pressure interior with $p = - \bar \r$ is a non-singular 
patch of static dS space, which is responsible for effective repulsion in general relativity rather than the fatal attraction 
that leads to the critical mass (\ref{Mcrit}), thereby resolving Wheeler's `difficulty' of the final state of complete gravitational 
collapse, which need not be a BH.
 \item 
The $\ve \to 0$ classical GR limit may be considered the {\it universal} gravastar limit, in the sense that it
is independent of any ansatz or additional assumptions of an eq.~of state of an interposed boundary layer of
`non-inflationary material' with a well-defined surface tension $\tau\!_{_S}$ of (\ref{surfphysM}). This is a clear specific example
of the generalization of the Lanczos-Israel matching conditions to null hypersurfaces, also dispensing with the need
to restrict to matching on timelike boundaries, as in the original gravastar proposal, and subsequent papers.
\item The determination of $C=\frac{1}{4}$ is just that required to make the surface gravities on the Schwarzschild
exterior and dS interior sides of the surface at $r=\rM$ equal and opposite, {\it i.e.} $\ka_-= -\ka_+$, which
is the physical condition of equality of the forces pressing inward and outward on the surface in static equilibrium.
This is seen also in the Euclidean formulation as the equality of the Euclidean time periodicities $\b_+ = 2 \p\ka_+= 8 \p GM$
and $\b_- = 2 \p \ka_- =4 \p/H$, so that the inferred Rindler-Hawking `temperatures' $1/\b_+ = 1/8\p GM$ and
$1/\b_- = H/4 \p$ (rather than $H/2\p$) are {\it equal} at $\rH=H^{-1} = \rM =2GM$, just as would be expected
in equilibrium.
\item
Because $\tau\!_{_S} >0$, the `First Law' of non-rotating gravastars (\ref{newfirst}) shows that increasing the area by $A_{_H}$ by
allowing for surface mode perturbations with non-zero angular momenta costs energy, so that the static equilibrium of the
surface is stable to such perturbations, independently of any eq.~of state assumptions. Similar to the Bohr atom,
these angular perturbations are expected to have a {\it discrete spectrum} which will distinguish gravastars with
a surface from BH horizons in gravitational wave (GW) signatures. 
\item The differential relation (\ref{newfirst}) between gravastar solutions parametrized by $M$ is purely a
{\it classical mechanical} relation, entirely within the domain of classical GR, rather 
than a quantum or thermodynamic one. The area $A_{_H}$ is simply the geometrical area of the condensate star 
surface with no implication of entropy. As (\ref{Gibbs}) shows there is no entropy at all associated with 
a macroscopic condensate at zero temperature. The Planck length $L_{\rm Pl} =\sqrt{\hslash G/c^3}$, Planck mass 
$M_{\rm Pl} = \sqrt{\hslash c/G}$,  or the Boltzmann constant $k_B$ do not enter these classical relations at all at $\ve = 0$.
\item Further, by the Gibbs relation
\be
p + \r = s\,T + \m\, n
\label{Gibbs}
\ee
and $\m=0$ here, since no chemical potential corresponding to a conserved quantum number has entered 
our classical considerations, the interior Schwarzschild-de Sitter solution with $p + \r = 0$ is a
{\it zero entropy} density $s=0$ and {\it zero temperature} single macroscopic state, justifying its designation 
as a gravitational Bose-Einstein condensate (GBEC). 
\item The matching of the metric interior to the exterior solution for $r=\rM$ has the cusp-like behavior shown 
in Figs.\,\ref{Fig:Redshiftsing}-\ref{Fig:hvarious}, with discontinuous first derivative with respect to the original 
Schwarzschild radial coordinate $r$. This is symptomatic of non-analytic behavior, invalidating the assumption 
of metric analyticity needed for deriving periodicity in complexified imaginary time $t\to -i\tau$ \cite{HartHawk:1976}. 
Unlike in the analytically extended vacuum Schwarzschild solution, where $f(r)$ becomes negative 
and the Killing vector $K=\pa_t$ becomes spacelike in the interior $r < \rM$ of a BH, in the negative pressure 
gravastar solution $f(r)$ is everywhere non-negative, and there is no requirement of any fixed periodicity 
in imaginary time of either the geometry or Green's functions of quantum fields in this geometry. The surface 
gravity $\ka$ therefore carries no implication of temperature or thermal radiation. The zero temperature 
vacuum state in the exterior Schwarzschild geometry is expected to be the Boulware vacuum, which is the
usual Minkowski vacuum in the asymptotically flat region $r\to \infty$, whose Green's functions also
have non-analytic cusp-like behavior at $r=\rM$  \cite{Boul:1975}, and which has no Hawking radiation \cite{ChrFul:1977}.
\item Non-analyticity at $r=\rM$ is exactly the property suggested by the analogy of BH horizons to phase boundaries 
and quantum critical surfaces in condensed matter physics \cite{ChaplineHLS:2001,Chap:2003},
although the equality of $|\ka_\pm|$ was not obtained in the condensed matter analogies, requiring as it does
a proper general relativistic treatment of matching on null hypersurfaces, which is given also for
horizons with non-zero angular momentum in \cite{BelGonEM1:2022}
\item 
Since $f(r) = 0$ corresponds to the `freezing' of local proper time at $r=\rS$, it also suggests critical slowing down
characteristic of a phase transition. The vanishing of the effective speed of light $c^2_{eff} =c^2f(r)$ is analogous 
to the behavior of the sound speed determined by the low energy excitations at a critical surface or phase boundary.
This and experience with Bose-Einstein condensates in other contexts suggest that gravitation and spacetime 
itself are `emergent' phenomena of a more fundamental microscopic many-body theory \cite{Laugh:2003, Mazur:2007,ChapMazur:2009}. 
\item The interpretation of $c^2_{eff}\ge 0$ as the effective speed of light squared, which must remain always non-negative, 
calls to mind Einstein's original papers on the local relativity principle for static gravitational fields, which led him to the 
general theory from the Minkowski metric $ds^2 = -c^2\, dt^2 + dx^2 + dy^2 + dz^2$, by allowing first the time 
component $-g_{tt} = c^2$ and eventually all other components of the metric to be functions of space (and 
then also time) \cite{Einstein:1912}. Thus it could be argued that the non-negativity of $c^2_{eff} =c^2f(r)$ 
in a static geometry is more faithful to Einstein's original conception of the Equivalence Principle, realized by real 
continuous coordinate transformations, rather than complex analytic extension around a square root branch point 
that would allow $c_{eff}^2 <0$. 
\item At the minimum, the matching of the $p=-\bar\r$ dS interior to Schwarzschild exterior provides a consistent 
alternative to analytic extension, entirely within the framework of classical GR, provided only that surface boundary 
layers on null boundary surfaces are admitted. A phase boundary and non-singular `BH' interior require no violation 
of the Equivalence Principle, at least in its weak form.
\item Finally, since $K_{(t)}$ remains timelike for a gravastar, $t$ is a global time and unlike in
the analytic continuation hypothesis, the spacetime is truly globally static. The $t=$ const. hypersurface 
is a Cauchy surface and is everywhere spacelike. This is exactly the property of a static spacetime necessary 
to apply standard quantum theory, for the quantum vacuum to be defined as the lowest energy
state of a Hamiltonian bounded from below, and for the Schr\"odinger equation to describe 
unitary time evolution, thus avoiding any conflict with unitarity or an `information paradox.'
\end{itemize}
\end{svgraybox}

\section{Slowly Rotating Gravastars: Junction Conditions \& Moment of Inertia}
\label{Sec:Rot}

A serendipitous consequence of the reanalysis of the 1916 Schwarzschild interior solution via the Komar mass integral
(\ref{KomarM}) is that it provides an explicit example  of `gluing' of two different geometries at their mutual null horizons, in which 
the surface stresses can be unambiguously determined, providing a clear interpretation of the surface tension of the null surface at 
$\rM=\rH$. This example serves as a general template to provide the general matching conditions for null surfaces with non-zero angular 
momentum, necessary to describing rotating gravastars as well \cite{BelGonEM1:2022}.

Consider the general axisymmetric and stationary metric line element
\cite{Lewis:1932,HartleSharp:1967,Bardeen:1970,FriedmanChandra:1972,ChandraBH} 
\be
ds^2 = -e^{2\n} dt^2 + e^{2\j} \big(d \f - \w\, dt\big)^2 + e^{2 \a} dr^2 + e^{2 \b} d\th^2
\label{axisymstat}
\ee
in coordinates adapted to the two Killing symmetries of time translation $K_{(t)} = \pa_t$ and axisymmetry $K_{(\f)} = \pa_\f$
around the axis of rotation. The five functions $\n, \j, \w, \a, \b$ are functions of the remaining $(r,\th)$ coordinates only.
The choice of those remaining two coordinates is still subject to some coordinate freedom, which if fully exploited reduces
the five functions appearing in (\ref{axisymstat}) to just four independent functions. However it is convenient to leave
the remaining coordinate freedom unfixed, in order to encompass various choices that may be made. 

Corresponding to the two Killing symmetries and Killing vectors $K_{(t)}$ and $K_{(\f)}$ are two conserved quantities
which may be constructed by Komar's general method, {\it viz.} the total mass-energy
\bea
&M =  \displaystyle\frac{1}{4 \p G}\int_{\pa V_+}\!  (\ka + \w \cJ)\, dA\nn
& \hspace{-1cm}=\displaystyle \int_V \sqrt{-g}\  \big(\! \!-T_t^{\ t} + T_r^{\ r}  + T_\th^{\ \th}  + T_\f^{\ \f}  \big)\, dr\,d\th\,d\f\, 
+\, \frac{1}{4 \pi G} \int_{\pa V_-} \! (\ka + \w \cJ)\, dA
\label{KomarMass}
\eea
and angular momentum
\bea
&\displaystyle J= \frac{1}{8 \p} \int_{\pa V_+}\!  \! \cJ \, dA = G \int_V  \sqrt{-g}\ T_t^{\,\f} \,dr\,d\th\,d\f 
+\frac{1}{8 \p} \int_{\pa V_-}\!  \!\cJ \, dA 
\label{KomarJ}
\eea
where \cite{BelGonEM1:2022}
\bes
\bea
&\ka =\displaystyle \sdfrac{1}{2} \, e^{-\a-\n}\ \frac{\pa  e^{2\n}\!}{\!\pa r} \, 
\label{kaprot}\\
&
\cJ =\displaystyle-\sdfrac{1}{2} \,e^{2\j} \, e^{-\a-\n}\ \frac{\pa \w}{\pa r} 
\eea
\ees
are the surface gravity and ($8 \p$ times) the angular momentum density per unit area respectively.
The volume measure factor and area element are given by
\bea
&\sqrt{-g}= \exp \,(\n + \j + \a + \b)\\
\label{detg}
&dA = \exp\,(\b + \j) \, d\th\,d\f\,.
\label{dArea}
\eea
in the coordinates of (\ref{axisymstat}). In the case of spherical symmetry $\w,T_t^{\ \f}$ and $\cJ$ vanish, 
(\ref{KomarMass})-(\ref{kaprot}) reduce to (\ref{KomarM})-(\ref{kapdef}), and (\ref{dArea}) becomes $\sqrt{f/h} \ r^2 \sin\th$.

A null horizon in the geometry (\ref{axisymstat}) occurs when $e^{2\n} \to 0$, which without loss of generality 
may be taken to be at a fixed $r=\rH$. The induced metric on the horizon
\be
ds^2\Big\vert_{r=\rH} = e^{2\j_{\!_H}} \big(d \f - \wH dt\big)^2 + e^{2 \b_{\!_H}} d\th^2
\label{surfmet}
\ee
involves the functions $\j,\w,\b$ which must be continuous there, whereas $\n$ and $\a$ need not be.
It is the inclusion of the$\sqrt{-g}$ factor of the volume integration measure in the Komar integrals that
leads to total second derivatives of $\n, \a$ with respect to $r$. These total derivative terms give rise both
to the areal surface terms in (\ref{KomarMass})-(\ref{KomarJ}), and the distributional Dirac $\del$-functions on
null hypersurfaces from the discontinuities of $\ka$ and $\cJ$ there. Thus, as in (\ref{Gththdens})-(\ref{discGthth})
it is necessary to consider the tensor density$\sqrt{-g} G_\m^{\ \n}$ and in particular its potentially
singular part $e^{\n + \a}G_\m^{\ \n}$, to find these $\del$-function and surface contributions~\cite{BelGonEM1:2022}. 

This is the fundamental physical basis for defining the redshifted extrinsic curvature for stationary, axisymmetric geometries by
\be
\bK^a_{\ b} = - \g^{ac} n^\m \G_{\m bc} =\sdfrac{1}{2} e^{-\a + \n} \g^{ac} 
\left(\frac{\pa g_{bc}}{\pa r} - \frac{\pa g_{rc}}{\pa x^b}  - \frac{\pa g_{rb}}{\pa x^c}\right)\,,\quad \{a,b,c\}= t,\th,\f
\label{upK}
\ee
with the normal vector having the components 
\be
\bn_\m = \del^r_{\ \m} \, e^{\a + \n} \,,\hspace{1.2cm} \bn^\m = \del^\m_{\ r} \, e^{-\a + \n}  
\label{nrot}
\ee
normalized to 
\be
\bn \cdot \bn = e^{2\n} \to 0
\label{bnnormnu}
\ee
on the null hypersurface, instead of (\ref{nnorm}) for a timelike hypersurface.
These relations and the inverse $\g^{ac}$ of the induced metric are to be evaluated for finite $e^{2\n}$ and the 
horizon limit $e^{2\n}\sim \ve \to 0$ taken at the end.

With $\bn$ defined by (\ref{nrot}),  $\bK^a_{\ b}$ has the components
\bes
\bea
&\displaystyle{\bK^t_{\ t} = \sdfrac{1}{2} \, e^{-\a-\n} \,\frac{\pa}{\pa r} e^{2\n}
-  \sdfrac{1}{2}\, \w\, e^{2\j}\, e^{-\a-\n} \,\frac{\pa \w}{\pa r} 
= \ka + \w\, \cJ}\\
&\displaystyle{\bK^t_{\ \f} = \sdfrac{1}{2}\, e^{2\j}\, e^{-\a-\n} \,\frac{\pa \w}{\pa r} = -\cJ}\\
&\displaystyle{\bK^\f_{\ t} = \sdfrac{1}{2} \,\w\, e^{-\a-\n} \,\frac{\pa}{\pa r} e^{2\n} 
- \sdfrac{1}{2}\, \w^2 e^{2\j}\, e^{-\a-\n} \,\frac{\pa \w}{\pa r} - \sdfrac{1}{2}\, e^{-2\j}\, e^{-\a+\n}} \,
\frac{\pa}{\pa r}\,\big(\w e^{2\j}\big) \nn 
&= \displaystyle{\w\,\ka + \w^2\,\cJ + \cO (\ve) }\\
&\displaystyle{\bK^\th_{\ \th} = \sdfrac{1}{2}\, e^{-\a+\n-2\b} \,\frac{\pa}{\pa r}e^{2\b}= \cO (\ve) }\\
&\displaystyle{\bK^\f_{\ \f} = \sdfrac{1}{2}\,\w\, e^{2\j}\, e^{-\a-\n} \,\frac{\pa \w}{\pa r} 
+ \sdfrac{1}{2}\,e^{-2\j}\, e^{-\a+\n} \,\frac{\pa}{\pa r}e^{2\j} = -\w\,\cJ + \cO (\ve)}
\eea
\label{extKrot}\ees
with the non-listed components vanishing. Thus the discontinuities on the null horizon surface are
\bes
\bea
&\left[\bK^t_{\ t}\right] =  [\ka] + \wH [\cJ]\\
&\left[\bK^t_{\ \f}\right]= - [\cJ]\\
&\left[\bK^\f_{\ t} \right]=  \wH[\ka] + \wH^2\,[\cJ] \\
&\left[\bK^\th_{\ \th}\right] = 0 \\
&\left[\bK^\f_{\ \f}\right]= -\,\wH[\cJ]
\eea
\label{extKrotdisc}\ees
where $\w \!=\!\wH$ on the horizon. The null horizon junction conditions are then of the same form as the Lanczos-Israel
conditions (with one contravariant and one covariant index), namely
\bes
\bea
&[\bK_a^{\ b}] =- 4 \pi G\, \Big( 2 \cS_a^{\ b} - \del_a^{\ b} \cS_c^{\ c}\Big)\\
&8\p G\, \cS_a^{\ b} = -[\bK_a^{\ b}] + \del_a^{\ b} [\bK_c^{\ c}]
\label{KSdisc}
\eea
\ees
but with this physical stress tensor on the null horizon hypersurface related to the original Lanczos-Israel (LI) one by
\be
\big(\cS_a^{\ b}\big)_{\rm here}  = \lim_{e^\n \to 0} \left\{e^\n \, \big(S_a^{\ b}\big)_{\rm LI}\right\}
\label{Isrmod}
\ee
which is finite in the horizon limit $e^\n \to 0$ since the discontinuities of $\bK^a_{\ b}$ in (\ref{extKrotdisc}) are finite 
in that limit. The surface stress tensor density is then found to be
\be
^{(\S )}T_a^{\ \,b}\, e^{\a + \n}  = \cS_a^{\ \,b} \, \del(r-\rH)
\label{SurfTS}
\ee
localized on the surface $r=\rH$ with the correct volume measure factor (to be multiplied by the continuous $e^{\j + \a}$ factors)
for the Komar mass and angular momentum integrals.

This improved understanding and generalization of the junction condition formalism to null hypersurfaces and in particular
rotating null horizons appropriate for the Kerr geometry opens the way to finding rotating gravastar solutions.
The formalism has been applied in the case of slow rotation following methods of~\cite{Hartle:1967,HartleThorne:1968,ChandraMiller:1974}
who express the slowly rotating line element in the (\ref{axisymstat}) form with
\bes
\bea
&e^{2\n} = f(r) \Big[ 1 + 2 h_0(r) + 2 h_2(r) \,P_2(\cos \theta)\Big] \label{eq:exp2nu} \\
&e^{2\j} =  r \sin\theta\,  \Big[1  + 2 k_2(r)\, P_2(\cos \theta)\Big]\\
&e^{2\a} =  \displaystyle{\frac{1}{h(r)} \left[1 +  2\ \frac{m_0(r)+m_2(r)\,P_2(\cos \theta)}{r- 2 m(r)}\right]}\\
&e^{2\b} =  r \Big[1  + 2 k_2(r) \, P_2(\cos \theta)\Big]\,.
\eea
\label{HarThorpert}\ees
The frame dragging angular velocity $\w = \w(r)$ is independent of $\th$ to lowest (first) order in the slow rotation expansion 
about the spherically symmetric solution, while $h_0, m_0$ are the monopole ($\ell =0$) and $h_2,m_2,k_2$ the quadupole ($\ell=2$) 
perturbations, which are second order in that expansion, and multiplied by the appropriate Legendre polynomials $P_0, P_2$. 
The Einstein eqs.~linearized  around the previously determined spherically symmetric solution are then solved assuming that the 
eq.~of state for the fluid composing a slowly rotating star is unchanged from its form in the non-rotating solution. In the case of 
a slowly rotating gravastar the solutions are separated again into an exterior sourcefree region, and interior dS region with 
the same $p=-\r$ eq.~of state, and the solutions are matched at their mutual horizons using the discontinuities and surface stress 
tensor (\ref{KSdisc}).

Requiring that the solution be asymptotically flat as $r \to \infty$, the induced metric on the horizon (\ref{surfmet}) to
be continuous ($\cC^0$), and the interior solution not to contain $\del$-function source terms at the origin $r=0$,
severely restricts the solutions for the functions $h_0,m_0,h_2,m_2,k_2$. With these conditions one finds~\cite{BelGonEM2:2022}
\footnote{Here $J$ is the angular momentum in geometric units $GJ/c^3$, with dimensions of (length)$^2$.}
\bes
\begin{align}
\hspace{-2cm}r> \rH:\qquad \qquad\w &= \frac{2J}{r^3}\\
m_0 &= J^2 \left( \frac{1}{\rH^3\!} - \frac{1}{r^3}\right) \\
h_0 &=  - \frac{m_0}{r- \rH} = - \frac{J^2\!}{\rH^3r^3} \left(r^2 + r\rH + \rH^2\right)\\
m_2 &=  \frac{J^2}{r^3} \left(r- \rH\right)\left( \frac{5}{r} -\frac{2}{\rH\!} \right)\\
h_2 &= \frac{J^2}{r^3} \bigg( \frac{1}{r}  + \frac{2}{\rH\!\!}\, \bigg) \\
k_2 &= -\frac{2J^2}{r^3} \bigg(\frac{1}{r} + \frac{1}{\rH\!\!}\, \bigg) 
\end{align}
\label{extfinal}\ees
for the exterior solution $r>\rH$, and
\bes
\begin{align}
\hspace{-4cm} r< \rH:\qquad\qquad\w&=\frac{2J}{\rH^3}= \w_H \label{omfin}\\
m_0&= 0 \label{m0fin}\\
h_0&=C_{_I}\label{h0fin}\\
m_2&=-\frac{2J^2(r^2-\rH^2)^2}{r^2\rH^5}\label{m2fin}\\
h_2&= \frac{2J^2}{r\rH} \left(\frac{1}{r^2} - \frac{1}{\rH^2}\right)\label{h2fin}\\
k_2&=-\frac{2J^2}{r\rH}\left(\frac{1}{r^2} + \frac{1}{\rH^2}\right)\label{k2fin}
\end{align}
\label{intfinal}\ees
for the interior solution, with $\rH = \rM$ unchanged from the non-rotating solution to this order.

With these requirements the exterior solution for a slowly rotating gravastar (\ref{extfinal}) is in fact the same as that
of a slowly rotating Kerr BH, up to a coordinate transformation from the $(r,\theta)$ Hartle-Thorne coordinates 
used in slow rotation expansion of the metric in (\ref{HarThorpert}) to the Boyer-Lindquist $(r_{_{\rm BL}},\th_{_{\rm BL}})$ 
coordinates in which the Kerr solution is more commonly expressed, and then expanded to second order in $J$.  This
coordinate transformation is given explicitly by eqs.~(4.12) of \cite{BelGonEM2:2022}. In particular the
ergosphere boundary is given by $r_{\rm ergo}(\theta) = \rH [ 1 +4 J^2 \sin^2\theta/\rM^4] \ge \rH$ in the
Hartle-Thorne coordinates of (\ref{HarThorpert}). The reason for the two exterior solutions of a Kerr BH and 
slowly rotating gravastar to be identical is that although the metric functions $\a,\n,\w$ have discontinuous derivatives 
on the gravastar null horizon surface, the induced surface metric (\ref{surfmet}) is continuous, and all the metric functions 
have been required to be finite there, which are the same conditions required for the horizon of a Kerr BH, that
lead to the `no hair' theorems~\cite{Carter:1971,Carter:1973,Carter:2009b,Robinson:1975}.

The solution (\ref{extfinal})-(\ref{intfinal}) is such that the discontinuities of two quantities which appear in (\ref{extKrotdisc}) and
the surface stress-energy tensor (\ref{SurfTS}) are
\bes
\bea
& \displaystyle{[\ka] = \frac{1}{\rM} \left( 1 - \frac{3J^2}{\rM^4} + \frac{C_{_I}}{2} \right)}\label{kapdisc}\\
& [\cJ] =\displaystyle{ \frac{3J}{\rM^2}\, \sin^2\th}\label{Jdisc}
\eea
\label{kapJdisc}\ees
with the result that surface energy of the horizon boundary at $r=\rH$ in the Komar mass-energy is
\bea
&\displaystyle{E_{_S} = \frac{1}{4 \p G} \int_H \Big( [\ka] + \w [\cJ] \Big) dA 
= \frac{[\ka]}{4\p G} \,A_{_H} + \frac{3\w_{_H}\, J}{2\,G} \int_0^\p d\theta \, \sin^3\theta}\nn
&\displaystyle{= \frac{[\ka] \rM^2}{G} + \frac{2\w_{_H} J}{G} = 2M\left( 1 + \frac{C_{_I}}{2} +  \frac{J^2}{\rM^4} \right)}
\label{EsurfJ}
\eea
where (\ref{omfin}) has been used. The Komar volume energy of the interior solution is
\be
E_{_V} = -M\, \left( 1 + C_{_I} \right)
\label{Evol}
\ee
so that the total energy of the slowly rotating gravastar specified by  (\ref{extfinal})-(\ref{intfinal})  is
\be
E \equiv \cM = E_{_V} + E_{_S} = M\left(1 +  \frac{2J^2}{\rM^4}\right)
\label{Etot}
\ee
where $M$ is the mass of the non-rotating solution. Since this is the irreducible mass for a non-rotating BH, (\ref{Etot})
will be recognized as equivalent to the Christodoulu formula 
\be
\cM^2 = M^2 + \frac{J^2}{4G^4M^2} = M^2\left( 1 + \frac{4J^2}{\rM^4}\right)
\label{Mirr}
\ee
when $E=\cM$ is expanded to first order in $J^2/\rM^4$. Thus $C_{_I}$ drops out of the total mass-energy $\cM$ of the slowly rotating
solution, when the volume and surface contributions are added, which then obeys exactly the same relation as a Kerr BH 
of the same mass and angular momentum to this order in $J^2$.

For the Komar angular momentum of the gravastar, the integrand of the volume term is proportional to $T^{\ t}_\f$, which
vanishes in the interior under the assumption of the $p=-\r$ eq.~of state being unchanged from the non-rotating case
in order to apply the formalism of~\cite{Hartle:1967,HartleThorne:1968,ChandraMiller:1974}. Using (\ref{Jdisc}),
the entire angular momentum is carried by the surface contribution 
\be
J_{\rm tot} = J_{_S} =\frac{1}{8 \p} \int_H [\cJ] \,dA =  \frac{3J}{4}\, \int_0^\p d\th \, \sin^3\th = J
\label{Jtot}
\ee
which is due to the surface stress tensor $\cS^{\ t}_\f$. From (\ref{Kdisc}) and (\ref{kapJdisc}) the
components of the surface stress tensor are
\bes
\begin{align}
8 \pi  \cS^t_{\ t} &=-\frac{6 J^2}{\rH^5}\,  \sin^2\!\theta \\
8 \pi  \cS^t_{\ \f} &=\frac{3J}{\rH^2}\,  \sin^2\!\theta \, \\
8 \pi  \cS^\f_{\ t} &=-\frac{2 J}{\, \rH^4} \\
8 \pi  \cS^\f_{\ \f} &=\frac{1}{\rH} +  \frac{C_{_I}}{2 \rH} - \frac{3J^2}{\rH^5} +\frac{6 J^2}{\rH^5}\,  \sin^2\!\theta  \\
8 \pi  \cS^\th_{\ \th} &= \frac{1}{\rH} + \frac{C_{_I}}{2 \rH}  - \frac{3 J^2}{\rH^5}\,.
\end{align}
\label{surfstressfin}\ees
by a straightforward application of the modified Lanczos-Israel junction conditions on the null horizon hypersurface
at $\rH = \rM$~\cite{BelGonEM1:2022,BelGonEM2:2022}.

The moment of inertia $I$ is defined as the ratio of its angular momentum $J/G$ to angular velocity $\wH$, and given by
\be
I = \frac{J/G}{\wH\!\!} = M \rH^2 = 4G^2 M^3
\label{MomIner}
\ee
identical to its value for a Kerr BH~\cite{Posada:2017,BeltrPosad:2023}. This might have been expected from the fact that the external 
geometry is identical to the Kerr BH geometry, and the interior dS condensate carries no angular momentum, all of $J$ being concentrated 
on the rotating null horizon surface from (\ref{Jtot}).  The result (\ref{MomIner}) may be compared to the moment of inertia of a thin shell 
of uniform surface density with total mass $M$ and radius $\rH$ in Newtonian mechanics, which is $2/3$ of (\ref{MomIner}). The factor 
of $3/2$ larger for the moment of inertia of a rotating gravastar is evidently a relativistic effect due to the surface density contributing 
to the mass-energy localized on the surface being $2M$ rather than $M$ and the non-uniform distribution of the surface stress tensor 
components $-\cS^t_{\ \,t} + \cS^\th_{\ \,\th} + \cS^\f_{\ \,\f}$ (\ref{surfstressfin}) contributing to the Komar surface energy (\ref{EsurfJ}).
in the non-rotating inertial frame. Nonetheless, with the moment of inertia (\ref{MomIner}) the contribution of the rotational kinetic
energy to the total mass $\cM$ in (\ref{Etot}) is
\be
E_{\rm rot} = 2M\, \frac{J^2}{\rM^4} = \frac{1}{2}\, I \wH^2
\label{Erot}
\ee
exactly what would be expected from the Newtonian mechanics of a rotating body. 
\begin{svgraybox}
Unlike a BH where it is supposed that there is no mass-energy at all at the horizon to rotate or to give rise to the large moment 
of inertia (\ref{MomIner}), the rotating gravastar has a well-defined Komar stress-energy (\ref{surfstressfin}) and angular
momentum on the rotating surface at $r=\rH$ (\ref{Jtot}), where all the angular momentum resides, and this gives rise 
to (\ref{MomIner}) and (\ref{Erot}) by straightforward evaluation of the relevant surface integrals (\ref{EsurfJ}), (\ref{Jtot}).
\end{svgraybox}

The results for the external solution (\ref{extfinal}), (\ref{MomIner}) and (\ref{Erot}) being identical to that of a slowly rotating Kerr BH 
is consistent with the BH `no hair' theorems being extended to rotating gravastars as 
well~\cite{Carter:1971,Carter:2009b,Maztalk:2020,BarCarRubLib:2019}, at least in the strictly classical limit of an infinitesimally 
thin shell ($\ve \to 0$). This again is the result of the boundary conditions on the metric perturbations in (\ref{HarThorpert}) at 
the null horizon being finite and identical in the two cases. Thus in this limit slowly rotating gravastars are expected to be 
indistinguishable from Kerr BH's in their external geometry, and have the same tidal deformability and Love number as a Kerr BH, 
namely none at all~\cite{ChirPosGued:2020}. Any gravastar `hair'  would be limited to the very thin quantum phase transition boundary 
layer of thickness $\ell$ (\ref{thickness}), and be very short.

As welcome as these first results for rotating gravastars are, the interior solution (\ref{intfinal}) contains a singularity at $r=0$,
which leads to divergences in the Weyl tensor at the origin proportional to $J^2/r^5$ in both the monopole and quadrupole terms.
For that reason (\ref{intfinal}) cannot be considered fully satisfactory. Since all the perturbations to order $J^2$ are fully determined
by the boundary conditions of asymptotic flatness, matching of the induced metric (\ref{surfmet}) on the null horizon surface,
and absence of $\del$-function singularities in the Komar $M$ or $J$ at the origin, the singular result at the origin cannot be
avoided in the Hartle-Thorne method, which requires that the eq.~of state of the rotating matter, in this
case $p =-\r$, be unchanged from the non-rotating solution. This assumption may be questioned, and indeed considerations
of rotating superfluid condensates suggests that a non-vanishing vortex density should arise in a rotating gravitational condensate
as well. This would require a different interior solution for a rotating gravastar, in which the $p=-\r$ condition 
of~\cite{Hartle:1967,HartleThorne:1968,ChandraMiller:1974} is relaxed. 

\begin{svgraybox}
To summarize the results of this section, the conditions of:
\begin{enumerate}[label=(\roman*)]
\item Asymptotic flatness,
\item Matching conditions on the slowly rotating null horizon,
\item Unchanged $p=-\rho$ eq.~of state from the non-rotating gravastar, and
\item Absence of $\del$-function singularities at the origin in the Komar mass and angular momentum,
\end{enumerate}
determine the slowly rotating gravastar solution to order $J^2$, up to one undetermined constant $C_{_I}$.
The exterior solution is identical to that of a Kerr BH to this order, including the moment
of inertia (\ref{MomIner}) which is consistent with all of the angular momentum being carried by the rotating null hypersurface 
at this lowest order in $J^2$. However the interior solution (\ref{intfinal}) under these four requirements is singular at the origin,  
and therefore not fully satisfactory. It appears that relaxing at least condition (iii) is necessary to remove the $1/r^5$ power
law singularity at $r=0$. It is possible that then allowing a distribution of angular momentum within the interior would also
modify the exterior solution for a rapidly rotating gravastar from that of a Kerr BH.
\end{svgraybox}

\section{Macroscopic Effects of the Quantum Conformal Anomaly}
\label{Sec:Anom}

The considerations of the previous sections are entirely classical, and show that an endpoint of complete gravitational collapse 
different from a BH is possible even in Einstein's classical GR, with no violation of the Equivalence Principle. However, it does not 
provide a mechanism by which a phase transition from one value of $\La_{\rm eff}$ outside a gravastar to another value inside 
can occur, elucidate the special role of the event horizon in triggering this transition, nor specify the physics determining 
the value of $\ve >0$. Addressing these questions requires consideration of the effects of the quantum theory of the 
matter/radiation stress tensor on the near-horizon geometry, and in particular, the macroscopic effects of the conformal anomaly 
of the quantum stress tensor source for Einstein's eqs. in the near-horizon region.

At the smallest microscopic scales probed, the principles of quantum field theory (QFT) hold, and matter under extreme 
pressures and densities in the standard model (SM) of particle physics is clearly quantum in nature. Yet Einstein's classical 
GR remains unreconciled to quantum theory, and the tension between quantum matter and classical gravity comes to the fore 
both in the puzzles and paradoxes of BHs reviewed in Sec.~\ref{Sec:QM}, and in the nature and value of cosmological dark energy 
driving the accelerated expansion of the universe~\cite{RiessSN:1998,PerlmutterSN:1999,DES:2019}. 

The problems of reconciling classical GR with QFT first appear with the stress-energy tensor $T^{\m\n}$, which is
treated as a completely classical source in Einstein's eqs., whereas $\hat T^{\m\n}$ is a UV divergent operator in QFT. 
Thus the minimal accomodation necessary to couple quantum matter to Einstein's theory is to replace the divergent
$\hat T^{\m\n}$ operator by its renormalized expectation value $\lag \hat T^{\m\n}\rag$, in a semi-classical approximation. 
Defining this expectation value by regularizing and renormalizing the contributions of matter loops leads inevitably to 
consideration of the effects of quantum matter on gravity itself, or in other words, to the effective field theory (EFT) treatment 
of gravity, keeping track of quantum corrections to GR. 

The quantum corrections are of two quite distinct kinds. The first are the strictly short distance/high energy corrections contained
in higher order local curvature invariants which are needed in any case to renormalize $\lag \hat T^{\m\n}\rag$~\cite{DeWitt:1975,BirDav}. 
The second are the effects of higher point stress tensor correlators, which can contain light cone singularities, that extend over macroscopic 
distance scales and provide quantum corrections to GR even at low energy scales, and in particular on null horizons.

Usually only quantum corrections of the first kind are considered~\cite{Don:1994,Bur:2004,Don:2012}, and the additions to the classical 
action of GR are assumed to involve only an expansion in the local curvature invariants, such as 
$R_{\a\b\m\n}R^{\a\b\m\n}, R_{\a\b}R^{\a\b}, R^2$.
Since these are fourth order in derivatives of the metric, they become significant only at the extreme UV Planck scale, but
are negligibly small at macroscopic distance scales or weak curvatures. The expansion in local invariants 
is based on the assumption of decoupling of UV degrees of freedom from the low energy EFT and strict separation of scales, 
familiar in other EFT approaches~\cite{AppCar:1975,Leut:1994,Alv:2012}.

On the other hand it has also been known for some time, even in flat space QFT, that anomalies are not captured by such an expansion 
in higher order local invariants, nor are they suppressed by any UV scale. Anomalies are associated instead with the fluctuations of 
massless fields which do not decouple, and which lead to $1/k^2$ poles in momentum space correlation functions, that grow large on the 
light cone $k^2 \to 0$ rather than the extreme UV regime $k^2 \sim  M_{\mathrm{Pl}}^2$. Such massless light cone poles are found in
explicit calculations of the triangle anomaly diagrams of $\lag  \hat J_5^\la \hat J^\a \hat J^\b\rag, \lag \hat T^{\m\n} \hat J^\a \hat J^\b\rag$
in massless QED$_4$~\cite{GiaEM:2009,ArmCorRose:2009}, and in the stress tensor three-point correlator 
$\lag  \hat T^{\a\b} \hat T^{\g\la}  \hat T^{\m\n}\rag$ of a general conformal field theory (CFT),  by solution of the conformal 
Ward Identities in momentum space~\cite{BzoMcFSken:2018,TTTCFT:2019}. 

Quite contrary to the decoupling hypothesis, quantum anomalies lead instead to the principle of {\it anomaly matching} 
from UV to low energy EFT~\cite{tHooft:1979}. In the strong interactions, the chiral anomaly of the UV theory, QCD, survives 
to low energies, requiring a specific Wess-Zumino (WZ) addition to the low energy 
meson EFT~\cite{WZ:1971,Leut:1994}, which is not suppressed by any high energy scale, and without which the low energy
$\p^0 \to 2\g$ decay rate, which helped establish QCD as the UV theory of the strong interactions, does not come out 
correctly~\cite{FritGellLeut:1973,TreiJackGross:2015,Bertlbook}.

In gravitation theory the relevant anomaly requiring attention is the conformal anomaly of $T_\m^{\ \n}$.
The classical stress-energy tensor of conformal matter or radiation is traceless $T_\m^{cl\,\m} = 0$. The conformal (or trace)
anomaly arises because it is impossible to maintain this traceless condition at the quantum level, in an arbitrary metric background, 
if the covariant conservation eq.~$\na_\n T_\m^{\ \n} = 0$ is also to be maintained. Since covariant conservation of $T_\m^{\ \n}$ 
is a necessary requirement for the consistent coupling of matter to Einstein's eqs., it is the trace condition on $T_\m^{\ \m}$ that is 
given up and which is `anomalous.' This clash of symmetries, forcing one to choose between them at the quantum level, is what is 
meant by an  `anomaly.' Although perhaps a surprise when they were first discovered, it eventually became 
clear that `anomalies' are a natural and necessary feature of QFT, and lead to interesting and essential connections between
the low energy (infrared:IR) and high energy (ultraviolet:UV) limits of the theory, both
in the SM and for GR and the EFT of low energy gravity.
 
Intuitively, a conformal symmetry of the 
classical theory cannot be maintained at the quantum level because QFT in all interesting cases involves regularization and 
renormalization of UV divergent Feynman diagrams, which introduce a scale, if only through logarithms, that necessarily breaks 
the conformal and scale invariance of the classical theory. This is in fact the reason that coupling `constants' run with 
energy scale~\cite{ChanEllis:1973,GiaEM:2009}, a well-verfied experimental fact~\cite{L3Running:2005}. 

\begin{svgraybox}
Since event horizons are null hypersurfaces with geometric significance as the locus of points at which the timelike Killing
field $K_{(t)}$ (or $K =K_{(t)} + \w K_{(\f)}$ in the rotating case \cite{BelGonEM1:2022}) become null, signaling
the critical condition for which a trapped surface can form if $K^2$ changes sign, the light cone singularities
of anomalous quantum correlators such $\lag  \hat T^{\a\b} \hat T^{\g\la}  \hat T^{\m\n}\rag$ can have
large macroscopic effects there. Such light cone singularities imply the existence of at least one additional light 
scalar ({\it a priori massless}) degree of freedom in the {\it low-energy} EFT of macroscopic gravity, that is clearly 
not accounted for in the classical action of GR, nor by the addition of higher order local curvature invariants to the effective action.

It is $f(r) = -K_{(t)}\cdot K_{(t)} \to 0$ (or $e^{2\n}= -K^2 \to 0$ in the rotating case of Sec.~(\ref{Sec:Rot})
which leads to the infinite blueshifting of frequencies (\ref{redshift}) or energies (\ref{Tol}), as soon as $\hslash \neq 0$
is admitted. The result is that the local quantum `vacuum' near a BH horizon is infinitely shifted in energy scales 
with respect to a state asymptotically far from the BH, violating a naive expectation of decoupling of UV from IR scales. 
The specification of the vacuum state involves {\it non-local} boundary conditions of the entire Cauchy constant time 
slice of (\ref{sphsta}), which are not determined by any simple expansion in the local curvature. On the contrary 
these boundary conditions span a large range of scales between IR and UV, and are sensitive to the conformal transformation 
properties of the effective action of quantum matter coupled to gravity. The enormous entropy~(\ref{SBH}), the Hawking 
effect upon which it is predicated, and the various BH paradoxes and conundra rely crucially upon the specification of the 
quantum vacuum state for arbitrarily high local frequencies and energies in the near-horizon region, precisely where masses 
become irrelevant compared to local energies typified by (\ref{Tol}), making all finite mass scales negligible and the 
conformal anomaly relevant there~\cite{EMZak:2010,AntMazEM:2012}.

These observations lead to the important step in an EFT of gravity by the taking into account the macroscopic effects of the
conformal anomaly on horizons in \cite{EMVau:2006}, where it was shown that the energy-momentum tensor derived 
from the effective action of the conformal trace anomaly of massless fields in curved space becomes large (indeed formally infinite) 
for {\it generic} quantum states at both the Schwarzschild BH and dS static horizons. 
\end{svgraybox}

\subsection{The Two Dimensional Conformal Anomaly and Stress Tensor}

In order to see the macroscopic effects of the conformal anomaly on horizons in a simple case, it is useful to consider
first $D=2$ spacetime dimensions. In 2D it is straightforward to show by any covariant method
preserving coordinate invariance~\cite{BirDav}, that the renormalized expectation value 
\be
\lag T_\m^{\ \m}\rag\Big\vert_{\rm 2D} =  \frac{N\hslash}{24\p} \ R
\label{Anom2D}
\ee
is non-vanishing in a general curved 2D spacetime, for $N = N_s + N_f$ the number of free conformal scalar and Dirac fermion fields,
{\it i.e.} in a conformal field theory (CFT). By taking a variation of (\ref{Anom2D}) with respect to the arbitrary background metric, 
and then evaluating the result in flat space, one can relate the anomaly to the existence of a massless $1/k^2$ pole in the one-loop 
vacuum polarization diagram of $\Pi_{\m\n\a\b} =i \lag T_{\m\n}T_{\a\b}\rag $, {\it cf.} Fig.~\ref{Fig:TT}. 

\begin{figure}[ht] 
\vspace{-5mm}
\hspace{9mm}
\includegraphics[width=9.5cm, trim=4.5cm 4mm 5cm 2mm, clip=false]{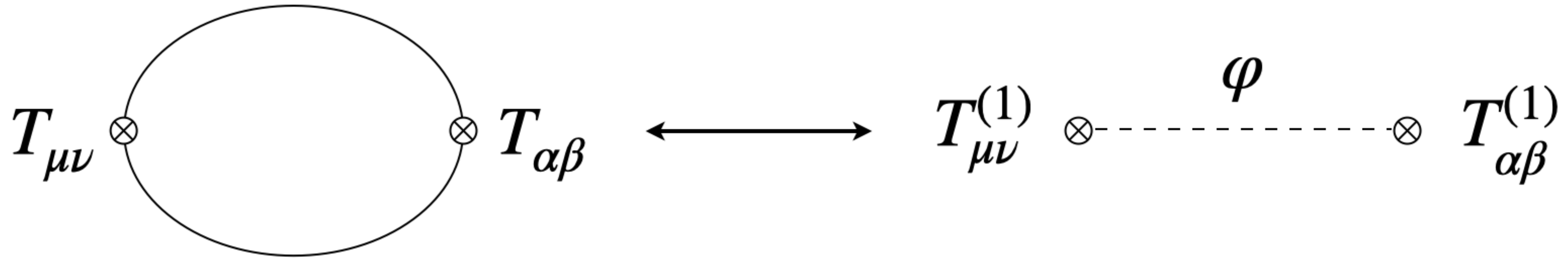}
\caption{The one-loop stress tensor vacuum polarization of a 2D CFT, which exhibits the massless $1/k^2$
pole of (\ref{Pi2D}). In the equivalent tree graph at right the conformalon scalar vertex $T^{(1)}_{\m\n}$ is given
by the term linear in $\vf$ in (\ref{anomT2}), and the $\vf$ propagator is obtained from (\ref{2Danom}), 
{\it cf.}~\cite{QuanGrav2D:2023}.}
\vspace{-5mm}
\label{Fig:TT} 
\end{figure}

This tensor has the form in momentum space
\bea
&\displaystyle{\Pi_{\m\n\a\b}(k)\Big\vert_{\rm 2D}  = (\h_{\m\n}k^2 - k_\m k_\n)(\h_{\a\b}k^2 - k_\a k_\b)\, \frac{N\hslash}{12\p k^2}} \label{Pi2D}\\
&\displaystyle{\Pi^{\ \m}_{\m\ \a\b}(k)\Big\vert_{\rm 2D}  = \frac{N\hslash}{12\p}\,  (\h_{\a\b}k^2 - k_\a k_\b)}\label{PiTr}
\label{TT2D}
\eea
showing that the non-zero trace and coefficient on the right side of (\ref{Anom2D}) is directly related to the existence
and residue of the $1/k^2$ pole in $\Pi_{\m\n\a\b}$. In fact, once the tensor index structure indicated in (\ref{Pi2D}) is fixed,
as required by symmetries and the covariant conservation law Ward identities $k^\m\Pi_{\m\n\a\b}(k)=0$ on any index,
the one-loop diagram of Fig.~\ref{Fig:TT} is {\it UV finite} and completely determined, with (\ref{TT2D}) the result~\cite{BertlKohl:2001}.
This shows that the conformal anomaly and pole is independent of the regularization scheme and detailed UV behavior
of the quantum theory, provided that the identities following from the covariant conservation law are maintained.

The essential point now is that the massless pole in (\ref{Pi2D}) is a lightlike singularity, indicating significant effects
on the light cone, which extends to macroscopic distance scales, and is particularly relevant on null horizons.
The $1/k^2$ pole can be expressed as the propagator of an effective scalar degree of freedom arising from the
fluctuations and correlations of massless (or sufficiently light) quantum fields in the vicinity of BH horizons.
Note that the classical theory of 2D gravity has no propagating degrees of freedom at all, so this $1/k^2$
propagator arises entirely from the quantum effect of the anomaly~\cite{EMZak:2010}.

This effective scalar degree of freedom and its consequences can be derived from the effective action corresponding to (\ref{Anom2D}),
{\it viz.}~\cite{Poly:1981,Poly:1987}
\be
S_{\!\rm anom}^{\rm 2D}[g]= - \frac{N\hslash}{96\p}\int\!d^2\!x\!\sqrt{-g(x)} \!\int\! d^2\!y\!\sqrt{-g(y)} \ R(x)\, \big(\!\sq^{-1}\big)_{xy} \, R(y)
\label{Poly2D}
\ee
where $\big(\!\sq^{-1}\big)_{xy}$ denotes the Green's function inverse of the scalar wave operator, that becomes the $1/k^2$
pole of (\ref{Pi2D}) in momentum space. This scalar degree of freedom can be made explicit by expressing the non-local anomaly 
effective action (\ref{Poly2D}) in the local form
\be
S_{\!\cA}^{\rm 2D}[g;\vf]= - \frac{N\hslash}{96\p}\int\!d^2\!x\!\sqrt{-g}\, \left\{g^{\m\n} (\na_{\!\m} \vf )(\na_{\!\n} \vf)- 2R \vf\right\}
\label{2Danom}
\ee
by the introduction of the scalar field $\vf$ describing the collective spin-0 degree of freedom, which is linearly coupled to $R$, and
whose massless propagator gives rise to the light cone singularities of the underlying massless CFT~\cite{BlaCabEM:2014}. 
Variation of (\ref{2Danom}) with respect to $\vf$ gives its eq.~of motion 
\be
-\sq \vf \!=\! R
\label{eom2D}
\ee
which when solved for $\vf\!=\! - \sq^{-1} R$ and substituted back into (\ref{2Danom}) returns the non-local form of the effective 
action (\ref{Poly2D}) up to a surface term. On the other hand variation of (\ref{2Danom}) with respect to the metric $g_{\m\n}$ yields 
the stress tensor 
\be
T^{\m\n}_{\!\cA}\Big\vert_{\rm 2D} = \frac{N\hslash}{48\pi}\left(2\na^\m\na^\n \vf - 2g^{\m\n}\, \sq\vf + \na^\m\vf\na^\n\vf
- \sdfrac{1}{2} g^{\m\n}\, \na_{\! \a}\vf\na^\a\vf\right)
\label{anomT2}
\ee
which is covariantly conserved in 2D, by use of (\ref{eom2D}) and by virtue of the vanishing of the Einstein tensor in two dimensions. 
The trace of this energy-momentum tensor gives the 2D trace anomaly (\ref{Anom2D}) by use of (\ref{eom2D}). 

To see the effect of the anomaly and $\vf$ on horizons, consider the 2D line element of the form
\be
ds^2 = -f(r) dt^2 + \frac{dr^2}{f(r)} = f(r) \big(-dt^2 + dx^2\big)\,,\qquad dx = \frac{dr}{f(r)}
\ee
where $x=r^*$ is the Regge-Wheeler tortoise coordinate. This metric has the static time Killing vector $K_{(t)}$ with
invariant norm $-K_{(t)}\cdot K_{(t)} = f(r)$, analogous to the Schwarzschild BH and dS static patch in coordinates (\ref{sphsta}). 
The  2D Ricci scalar is
\be
\displaystyle{R= - f'' = - \sdfrac{d^2\! f}{dr^2}}
\label{R2D}
\ee
and (\ref{eom2D}) in this case is
\be
\sq \vf = - \frac{1}{f} \frac{\pa^2\vf }{\pa t^2} +  \frac{\pa}{\pa r} \left(f  \frac{\pa \vf}{\pa r}\right)
= \frac{1}{f} \left(- \frac{\pa^2 }{\pa t^2} +  \frac{\pa^2}{\pa x^2}\right)\vf = f''\,.
\label{eomtx}
\ee
A particular solution to this linear inhomogeneous eq.~is $\vf =\ln f(r)$. The associated homogeneous wave eq.~has 
general wave solutions $e^{ik (x \pm t)}$. For stationary states $k=0$, which implies that only linear functions of $t$ 
and $x$ are allowed for such states. Thus the general stationary state solution of (\ref{eomtx}) is
\be
\vf = Pt + Q x  + \ln f
\label{phisoln}
\ee
where an irrelevant constant is set to zero because (\ref{2Danom}) and (\ref{anomT2}) depend only upon the
derivatives of $\vf$. The values of the integration constants $P,Q$ are dependent upon the boundary conditions
and quantum state of the underlying QFT that give rise to the anomaly~(\ref{Anom2D}).

Substituting the solution (\ref{phisoln}) into the stress tensor (\ref{anomT2}) gives
\bes
\begin{align}
T^{\ t}_t &= \frac{N \hslash}{24\p}\left\{ - \frac{1}{4f} \left(P^2 + Q^2 - f^{\prime\,2}\right) - f''\right\}\\
T^{\ t}_x& = -\frac{N \hslash}{48\p} \frac{PQ}{f}\\
T^{\ x}_x& =  \frac{N \hslash}{96\p} \frac{1}{f} \left(P^2 + Q^2 -  f^{\prime\,2}\right) 
\end{align}
\label{T2Dstat}\ees
in the $(t,x)$ coordinates. Hence if one takes as in (\ref{sphsta})
\be
f(r) = 1 - \frac{\rM}{\!r} \,,\qquad f' = \frac{\rM}{r^2}\,,\qquad f''= - \frac{2\rM\!}{r^3}
\ee
to model the 2D analog of the Schwarzschild geometry, (\ref{T2Dstat}) becomes
\bes
\begin{align}
T^{\ t}_t &= \frac{N \hslash}{24\p}
\left\{ - \frac{1}{4f} \left(\frac{p^2 + q^2}{\rM^2} - \frac{\rM^2}{r^4}\right) +  \frac{2\rM}{r^3} \right\}\\
T^{\ t}_x& = -\frac{N \hslash}{48\p \rM^2} \frac{pq}{f}\\
T^{\ x}_x& =  \frac{N \hslash}{96\pi} \frac{1}{f} \left(\frac{p^2 + q^2}{\rM^2} - \frac{\rM^2}{r^4}\right) 
\end{align}
\label{T2DSch}\ees
where the constants $P=p/\rM$ and $Q=q/\rM$, with $(p,q)$ dimensionless. This result for the 2D anomaly stress tensor shows 
divergences generically $\propto f^{-1}$ as $f\to 0$, and $r\to \rM$. The divergences can be arranged to cancel precisely on 
the future horizon by choosing $p=-q =\pm 1/2$, or on the past horizon by $p=q =\pm 1/2$, corresponding to the future or 
past Unruh states~\cite{Unruh:1976}, or on both horizons by $p=0, q= \pm 1$, corresponding to the Hartle-Hawking thermal 
state~\cite{HartHawk:1976,Israel:1976,GibPerry:1976} at the price of being non-vanishing as $r\to \infty$, and thermodynamically 
unstable due to negative heat capacity (\ref{dEdT}).

Any other values for $(p,q)$ result in divergences on the horizon. In particular, if one requires a time independent
truly static solution $p=0$, and asymptotic flatness as $r\to \infty$ so that $q=0$ also, then 
\be
\big(T^\m_{ \ \ \n}\big)_\cA \to -\frac{N\hslash}{96\p \rM^2 f}\ \left(\,\begin{array}{cl}\!\!\!-1\ & 0\\
\,0& 1 \end{array}\right) \to \infty \qquad {\rm as} \qquad r \to \rM
\label{Thorix2D}
\ee
which diverges on the two-dimensional horizon as $f \to 0$. There is no value of $q$ which yields a time independent regular 
solution for $\vf$ and (\ref{T2DSch}) on both the horizon and $r \to \infty$~\cite{QuanGrav2D:2023}.

\begin{svgraybox}
Thus the stress tensor of the 2D conformal anomaly has potentially large effects on a 2D model of a BH horizon 
in generic states which can be quickly and efficiently computed from the local form of the effective action (\ref{2Danom}) 
in terms of the independent conformalon scalar field $\vf$, by choosing different values of the integration constants $(p,q)$. 
Although attention has usually been focused solely on the special vacuum states for which the divergences in the stress tensor 
(\ref{T2DSch}) are arranged to cancel precisely as in \cite{DavFulUnr:1976,Unruh:1976}, inspection of (\ref{T2DSch}) shows 
that these special states are of measure zero and large backreaction effects on the horizon are to be expected in any other state, 
irrespective of the small value of the local curvature (\ref{R2D}) there. The $\vf$ field propagator is the $1/k^2$ light cone pole 
associated with the conformal anomaly, whose coherent classical solutions, which are bi-linear condensates of the underlying 
quantum theory~\cite{BlaCabEM:2014}, describe the macroscopic effects of the anomaly stress tensor on BH null horizons.
\end{svgraybox}

\subsection{The Conformal Anomaly Effective Action and Stress Tensor in Four Dimensions}

The general property of anomalies described in the 2D case carries over to four (and higher even) dimensions, although the algebraic details
becomes somewhat more complicated. The general form of the conformal anomaly in 4D in the presence of background curvature 
or gauge fields is~\cite{CapperDuff:1974,DeWitt:1975,Duff:1977,BirDav}:
\vspace{-1mm}
\be
\big\lag T_{\ \,\m}^{\m} \big\rag\  \equiv \ \frac{\cA}{\!\!\!\sqrt{-g}}\ =  \ b\, C^2 + b'\,\big( E - \tfrac{2}{3} \sq R\big) 
+ \sumi \ \b_i\, \cL_i
\label{tranom}
\vspace{-2.5mm}
\ee
again even when the classical theory is conformally invariant, and one might have expected $\big\lag T^\m_{\ \,\m} \big\rag$ to vanish. 
In (\ref{tranom})
\vspace{-1mm}
\be
\hspace{-2mm} E=R_{\a\b\g\la}R^{\a\b\g\la}\!\!-4R_{\a\b}R^{\a\b} + R^2\,,\hspace{6mm}
C^2\! = R_{\a\b\g\la}R^{\a\b\g\la}\!\!-2 R_{\a\b}R^{\a\b}  + \tfrac{1}{3}R^2
\label{ECdef}
\vspace{-.5mm}
\ee
are the Euler-Gauss-Bonnet (EGB) term and the square of the Weyl tensor respectively, and the $\cL_i$ are dimension-4 
invariants of gauge fields, with coefficients determined by the $\b$-function of the corresponding gauge coupling. The $b, b',\b_i$ 
coefficients are dimensionless numbers multiplied by $\hslash$ \cite{Duff:1977,BirDav}, so that the conformal anomaly is a quantum 
effect with no intrinsic length scale, applying at all scales at which any masses of the underlying quantum fields can be neglected, 
and determined by the light fields of the SM minimally coupled to gravity.

The non-local form of the effective action for the 4D conformal anomaly (\ref{tranom}) analogous to (\ref{Poly2D}) was found
first in~\cite{Rieg:1984}. The local form of the effective action of the 4D conformal anomaly analogous to (\ref{2Danom}) in 2D
requires the introduction of at least one scalar field, and is also given in a number of papers
\cite{Rieg:1984,AntEM:1992,BuchOdinShap,ShaJac:1994,StatesAntMazEM:1997,EMVau:2006}, the simplest form for which is
\cite{EMSGW:2017}, {\it i.e.}
\be
S\!\!_\cA [\vf]=  \frac{b'\!}{2\,}\!\!\int\!\!d^4\!x\sqrt{-g}\,\bigg\{\!\!-\big(\sq \vf\big)^2  
+ 2\,\Big(R^{\m\n}\! - \!\tfrac{1}{3} Rg^{\m\n}\Big) \,(\pa_\m \vf)\,(\pa_\n\vf)\bigg\}  
+ \frac{1}{2} \!\int\!\!d^4\!x\,\cA\,\vf   \,.
\label{Sanom}
\ee
in terms of a single scalar (conformalon) field $\vf$. This is quite analogous to the WZ action of the axial anomaly in QCD 
which must be added to the low energy EFT of mesons \cite{WZ:1971}. However its connection to $1/k^2$ poles
took some time to recognize, beginning with \cite{GiaEM:2009,ArmCorRose:2009,EMZak:2010} by analogy with the
axial anomaly in massless QED, which first appears in $D=4$ in $3$-point correlation function triangle diagrams.
This is because in $D=4$ the anomaly involves quadratic curvature invariants which require {\it two} variations with
respect to the metric to yield a non-zero result in flat space. The $1/k^2$ light cone singularities of the stress tensor 
three-point correlator  $\lag  \hat T^{\a\b} \hat T^{\g\la}  \hat T^{\m\n}\rag$ of a general CFT in 4D  has been 
verified explicitly only relatively recently by solution of the anomalous conformal Ward Identities in momentum 
space~\cite{BzoMcFSken:2018,TTTCFT:2019}, which are precisely generated by the anomaly effective action (\ref{Sanom}). 

It is important that $S\!\!_\cA$ as in $D=2$ is not purely a local functional of higher order curvature invariants (although $\cA$ itself is), and
either has a non-local form in terms of the original metric and curvature variables, or requires a new $\vf$ field to be
recast in the local form (\ref{Sanom}). As a result, the effects described by $\vf$ do {\it not} decouple and can have 
macroscopic effects on null horizons, depending on non-local boundary conditions, notwithstanding  the smallness 
of local curvatures there~\cite{EMVau:2006,EMEFT:2022}. 

The Euler-Lagrange eq.~for $\vf$ following from the variation of (\ref{Sanom}) is 
\be
\D_4 \vf = \sdfrac{1}{2}\left(E - \tfrac{2}{3}\sq R\right) + \sdfrac{1}{2b'}\, \left( b C^2 +  \sumi \,\b_i\cL_i\right) 
\label{phieom1}
\ee 
in which the fourth order Panietz-Riegert differential operator is~\cite{Panietz,Rieg:1984}
\be
\D_4 \equiv  \na_{\!\m}(\na^\m\na^\n +2R^{\m\n} - \tfrac{2}{3} R g^{\m\n}) \na_{\!\n}
=\sq^2 + 2 R^{\m\n}\na_{\!\m}\na_{\!\n} - \tfrac{2}{3} R \sq + \tfrac{1}{3} (\na^\m R)\na_{\!\n}
\label{Del4}
\ee
the unique fourth order scalar differential operator that is conformally covariant and the analog of the scalar
wave operator $\sq$ in two dimensions.

The full form of the 4D anomaly stress tensor following by variation of $S\!\!_\cA$ with respect to the
spacetime metric has been given in~\cite{EMVau:2006,EMSGW:2017}:
\be
T_{\!\cA}^{\,\m\n}\,[g;\vf] \equiv \frac{2\!}{\!\!\!\sqrt{-g}} \, \frac{\del}{\del g_{\m\n}} \ S\!\!_\cA[g;\vf] 
= b' E^{\m\n} + b\,C^{\m\n} + \sumi\ \b_i T^{(i)\,\m\n}
\label{Tphi1}
\ee
The first term here is 
\bea
&&E_{\m\n}=- 2\,(\na_{(\m}\vf) (\na_{\n)} \sq \vf)  + 2\na^\a \big[(\na_\a \vf)(\na_\m\na_\n\vf)\big]
- \tfrac{2}{3}\, \na_\m\na_\n\big[(\na_\a \vf)(\na^\a\vf)\big]\nn
&&\hspace{5mm}+\tfrac{2}{3}\,R_{\m\n}\, (\na_\a \vf)(\na^\a \vf) - 4\, R^\a_{\ (\m}\left[(\na_{\n)} \vf) (\na_\a \vf)\right]
 + \tfrac{2}{3}\,R \,(\na_{(\m} \vf) (\na_{\n)} \vf) \nn
&&+ \tfrac{1}{6}\, g_{\m\n}\, \left\{-3\, (\sq\vf)^2 + \sq \big[(\na_\a\vf)(\na^\a\vf)\big]
+ 2\, \big( 3R^{\a\b} - R g^{\a\b} \big) (\na_\a \vf)(\na_\b \vf)\right\}\nn
&& - \tfrac{2}{3}\, \na_\m\na_\n \sq \vf  - 4\, C_{\m\ \n}^{\ \,\a\ \b}\, \na_\a \na_\b \vf
- 4\, R^\a\!_{ (\m} \na_{\n)} \na_\a\vf  + \tfrac{8}{3}\, R_{\m\n}\, \sq \vf   +\tfrac {4}{3}\, R\, \na_\m\na_\n\vf\nn
&& - \tfrac{2}{3} (\na\!_{\!(\m}R) \!\na_{\!\n)}\vf + \tfrac{2}{3}g_{\m\n}\!\left[ \sq^2 \vf + 3R^{\a\b} \,\na_{\!\a}\na_{\!\b}\vf
- 2 R \sq \vf  +\tfrac{1}{2} (\na^\a R)\na_\a\vf\right]
\label{Eab}
\eea
which is the metric variation of all the $b'$ terms in (\ref{Sanom}), both quadratic and linear in 
$\vf$~\cite{EMVau:2006,EMZak:2010,EMSGW:2017}, while
\bes
\bea
&&C_{\m\n} \equiv  -\frac{1}{\sqrt{-g}\ } \frac {\del }{\del g^{\m\n}} \left\{\int d^4x\sqrt{-g}\,C^2\,\vf\right\}\nn
&&\hspace{1cm}=-4\,\na_\a\na_\b\,\Big( C_{(\m\ \n)}^{\ \ \a\ \ \b} \,\vf \Big)  -2\, C_{\mu\ \,\nu}^{\ \,\a\ \,\b}\, R_{\a\b}\, \vf\\
&& T^{(i)}_{\m\n}\equiv  -\frac{1}{\sqrt{-g}\ }\, \frac{\del}{\del g^{\m\n}}\left\{ \int d^4x\sqrt{-g}\,  \cL_i\, \vf\right\}
\eea
\label{CFterms}\ees
are the metric variations of the last two $b$ and $\b_i$ terms in (\ref{Tphi1}), both of which are linear in $\vf$.
With these explicit forms it is straightforward to compute the stress tensor in 4D geometries with horizons.

\subsection{The Stress Tensor of the Conformal Anomaly on the Schwarzschild and de Sitter Horizons}

The general solution to (\ref{phieom1}) for $\vf\!=\!\vf(r)$ and $\cL_i\!=\!0$ that is finite as $r\!\to \!\infty$ is readily
found by quadratures for the Schwarzschild metric (\ref{sphsta})-(\ref{Sch}) to be~\cite{EMVau:2006,EMZak:2010}
\bes
\bea
&&\hspace{-6mm}\displaystyle{\frac{d \vf_{\!_S}}{\!dr} =\frac{c_{\!_S} \rM}{r(r-\rM)}
 - \frac{2}{3\,\rM} \left(\frac{r}{\rM} + 1 + \frac{\rM}{r} \right) 
\ln   \left(1-\frac{\rM}{r}\right)   - \frac{2}{3\,\rM} - \frac{1}{r}}\\[1ex]
&&\hspace{-1.5cm}\displaystyle{\vf_{\!_S}(r)= c_{\!_S}\ln \left(1-\frac{\rM}{r}\right) 
+ \int_{r/\rM}^\infty \! dx \, \Bigg\{\frac{2}{3x}\left(x^2+ x+ 1\right)
 \ln \left(1-\frac{1}{x}\right) + \frac{2}{3} + \frac{1}{x}\Bigg\}}
\label{phiSint}
\eea
\label{phiS}\ees
in terms of the dimensionless integration constant $c_{\!_S}$. This solution has the limits
\be
\vf_{\!_S}(r) \to \left\{ \begin{array} {lr} c_{\!_S}\ln   \left(1-\sdfrac{\rM}{r}\right) + c_i 
- 2  \left(1-\sdfrac{\rM}{r}\right) \left[\ln  \left(1-\sdfrac{\rM}{r}\right) + \sdfrac{2}{3}\right] + \dots \ ,\quad & r \to \rM\\[2ex]
- \left(c_{\!_S} + \sdfrac{11}{9} \right) \sdfrac{\rM}{r} - \left(2 c_{_H} + \sdfrac{13}{9}\right) \sdfrac{\rM^2}{4r^4} + \dots\ , & r \to \infty
\end{array}\right.
\label{philim}
\ee
where the constant $c_i= -(7/6 +\p^2/9)=-2.26329...$ is the finite integral in (\ref{phiSint}) evaluated at the lower 
limit $r\!=\!\rM, x\!=\!1$. The addition of an arbitrary integration constant to (\ref{phiSint}) does not substantially change the results.
Substituting the solution (\ref{phiS}) into the anomaly stress tensor (\ref{Tphi1})-(\ref{Eab}), one finds 
\be
\big(T^\m_{ \ \,\n}\big)_\cA \to \frac{c_{_S}^2}{6\rM^4}\ \frac{b'}{f^2}\ 
 \left(\begin{array}{cccl}\!\! -3\ & 0\ & 0\ & 0\\
0 & 1 & 0& 0\\ 0 & 0 & 1& 0\\ 0 & 0 & 0 & 1 \end{array}\right) \to \infty \quad {\rm as} \quad r \to \rM
\label{ThorizM}
\ee
diverging quadartically ($\propto f^{-2}$) on the Schwarzshild BH horizon for any $c_{\!_S} \neq 0$. This is analogous to the
linear divergence ($\propto f^{-1}$) of (\ref{Thorix2D}) in the 2D case.

The divergent behavior of the local stress tensor (\ref{ThorizM}) shows that the anomaly stress tensor can become important near the
horizon of a BH and even dominate the classical terms in the Einstein eqs., the smallness of the curvature
tensor there notwithstanding. Even with $c_{\!_S}\!=\!0$ in (\ref{phiS}), which can be arranged by specific choice of the state 
of the underlying QFT, to remove the leading $f^{-2}$ divergence in (\ref{ThorizM}), there remain subleading divergences 
proportional to $f^{-1}, (\ln f)^2$ and $\ln f$. In fact, there is no solution of (\ref{phieom1}) in Schwarzschild spacetime with 
$\vf\!=\!\vf(r)$ only, corresponding to a fully Killing time $t$ invariant and spherically symmetric quantum state, with a finite stress
tensor at both singular points $r\!=\!\rM$ and $r=\infty$ of the differential eq.~(\ref{phieom1}). This result, following simply 
and directly from the conformal anomaly effective action, confirms results of previous studies of the stress tensor expectation value in
specific states in Schwarzschild spacetime~\cite{ChrFul:1977}. 

The divergences on either the future or past BH horizon (but not both) can be cancelled by allowing linear time dependent
solutions of (\ref{phieom1}), which give rise to a Hawking flux $\lag T^t_{\ \,r}\rag$~\cite{EMVau:2006}, such as in the 
Unruh states~\cite{Unruh:1976};  or by relaxing the regularity condition at infinity which gives rise to a non-zero stress tensor 
there, as in the Hartle-Hawking state~\cite{HartHawk:1976,Israel:1976,GibPerry:1976}. This thermal state is both incompatible 
with asymptotically flat boundary conditions and unstable due to its negative heat capacity (\ref{dEdT})~\cite{HawkSpecHeat:1976}. 
The only other possibility is that divergence on the fixed Schwarzschild BH horizon (\ref{ThorizM}) is uncancelled. 

\begin{svgraybox}
Usually the {\it assumption} of regularity of the semi-classical $\lag T_\m^{\ \,\n}\rag$ on the horizon is used to argue for the necessity 
of Hawking radiation flux $\lag T_r^{\ \,t}\rag\!>\!0$~\cite{FredHag:1990,RobinsWilz:2005}. However the converse is also true, namely if 
the final state of gravitational collapse is: 
\begin{enumerate}
\vspace{-2mm}
\item asymptotically flat,
\item truly static (implying with $\lag T^t_{\ \,r}\rag\!=\!0$), and
\item thermodynamically stable, 
\vspace{-2mm}
\end{enumerate}
then quantum effects at the horizon imply the breakdown of regularity there. 

The stringency of these conditions and the non-local topological obstruction to regular behavior on the horizon they imply 
is well illustrated by the impossibility of finding a static solution only of $r$ of the conformalon eq.~(\ref{phieom1}) that
satisfies the three conditions above and is also regular on the horizon in the Schwarzschild background.
It is natural that the effective action of the conformal anomaly contains information about the global quantum state 
properties in the solutions of (\ref{phieom1}) that relates the behavior of the state on the horizon with that at infinity.
The definition of the vacuum state in QFT relies on a clear and Lorentz invariant separation between positive and negative
energy states, whereas all finite energy scales on the horizon are redshifted to zero asymptotic energy by (\ref{redshift}). 
This means that the mass gap $2mc^2$ between the positive and negative energy excitations goes to zero 
and the local vacuum at the horizon is radically different from that in the asymptotic region, and susceptible to breakdown.
The coordinate invariant large quantum effects in the energy-momentum-stress tensor $T\!_\cA^{\ \m\n}$ due to the conformal 
anomaly in (\ref{ThorizM}) is the signal of this vacuum breakdown--and an incipient quantum phase transition. 
A large stress tensor would necessarily produce large backreaction effects on the near-horizon geometry, and could 
in principle prevent the classical BH horizon or any trapped surface from ever forming.
\end{svgraybox}

The breakdown of regularity on the horizon and potentially large quantum effects on the Schwarzschild BH horizon, were first 
found by D. Boulware~\cite{Boul:1975}, by defining the vacuum state by the positive frequency modes with respect to the 
Schwarzschild Killing time $t$. Rather than being special to the Boulware state, or `pathological,' the anomaly stress tensor 
(\ref{Tphi1}) makes it clear that vacuum stresses growing without bound as $r \to \rM$, as in (\ref{ThorizM}), are the {\it generic} 
case for the static geometry, while states for which these divergences are cancelled and are regular on the horizon
with a Hawking flux are the special ones and a set of measure zero. 

The $f^{-2}$ behavior of (\ref{ThorizM}), where $\ka_+=1/2\rM $ is the surface gravity, can be understood on dimensional
grounds from the $T_{\rm loc}^4$ conformal behavior of the local Tolman temperature (\ref{Tol}) of virtual massless fields 
near the horizon, where all but the heaviest fields may be treated as effectively massless. Indeed the near-horizon geometry
of the static metric (\ref{sphsta}) at constant $t$ is isomorphic to three dimensional Euclidean anti-dS (\L obachewsky) space 
EAdS$_3$~\cite{SacSol:2001,MazEMWeyl:2001,AntMazEM:2012}, where operators of conformal weight $w$ behave 
as $f^{-\frac{w}{2}}$. The stress energy tensor being of weight $w=D$ in $D$ dimensions has the power law behavior 
$f^{-1}$ in $D=2$ as in (\ref{T2DSch}), $f^{-2}$ in $D=4$ as in (\ref{ThorizM}). The effective action of the anomaly
with the $\ln f$ behavior of the dimensional conformalon field $\vf$ carries this conformal behavior to the stress tensor
in the horizon region. No other part of the quantum effective action has this conformal property \cite{MazEMWeyl:2001}. This 
is another way of seeing that the anomaly $S\!\!_\cA$ is relevant for BH physics~\cite{EMZak:2010}, at the macroscopic horizon 
scale, no matter how small the local curvature is there.

As the importance of the anomaly stress tensor on the exterior BH horizon is demonstrated by (\ref{ThorizM}),
a similar behavior is observed in dS spacetime, approaching the static dS horizon from the interior of a gravastar.
Indeed the $f^{-2}$ behavior in any spherically symmetric static spacetime (\ref{sphsta}) with a horizon at which 
$-K^\m K_\m \!=\! f(r)\! \to\! 0$~\cite{AndHisSam:1995}. With $\La$ positive, the static patch of dS space is of
this form with $f(r)\! \propto \!h(r) \!=\! 1\!-\!H^2r^2\!=\! 1\! -\! \La r^2/3$. The general spherically symmetric static 
solution of (\ref{phieom1}) for $\vf \!=\! \vf(r)$ which is regular at the origin in this case is~\cite{EMVau:2006,EMZak:2010}
\bea
&&\vf_{_{dS}}(r)= \ln\left(1-H^2r^2\right) + c_0 + \frac{q}{2} 
\ln\left(\frac{1-Hr}{1+Hr}\right) + \frac{2c_{_H} - 2 - q}{2Hr} \ln\left(\frac{1-Hr}{1+Hr}\right)\nn
&& \to \Big[c_{_H} + \Big(c_{_H} -1 - \sdfrac{q}{2}\Big)  \big(1-Hr\big) + \dots \ \Big ]\, \ln\big(1-Hr\big) + c' + \cO\, (1-Hr)
\label{phidS}
\eea
where the constant $c'=  c_0 +  (2-c_{_H})\ln 2$. Substituting this into the anomaly stress tensor (\ref{Tphi1}) gives 
\be
\big(T^\m_{ \ \,\n}\big)_{\!\cA} \to \frac{2}{3}\, c_{_H}^2\, H^4 \frac{b'}{(1-Hr)^2}\ 
 \left(\begin{array}{cccl}\!\! -3\ & 0\ & 0\ & 0\\
0 & 1 & 0& 0\\ 0 & 0 & 1& 0\\ 0 & 0 & 0 & 1 \end{array}\right) \to \infty \quad {\rm as}\quad  r \to \rH \equiv H^{-1}
\label{ThorizH}
\ee
which also diverges as $f^{-2}$ for any $c_{_H} \!\neq \!0$ as the dS static horizon is approached. 

As in the Schwarzschild case this divergence can be removed if a thermal state is considered, but then only if the temperature is 
precisely matched to the Hawking temperature $T_{_H}\!=\!H/2\p$ associated with the horizon, which in the dS case 
leads to the maximally $O(4,1)$ symmetric state~\cite{CherTag:1968,BunDav:1978}. However, this state is not a vacuum 
state of QFT and is unstable to particle pair creation, much as a uniform, constant electric field
is~\cite{EM:1985,Polyakov:2008,AndEM:2014a,AndEM:2014b,AndEM:2018}, and for essentially the same reason. Due to the 
non-existence of a global static time by which positive and negative frequency (particle and anti-particle) solutions can be invariantly 
distinguished, a time independent Hamiltonian bounded from below and stable vacuum state cannot be defined. This is similar
to the non-existence of a globally static time in the full Kruskal extension of the Schwarzschild solution in
Fig.~\ref{Fig:SchwKruskal}.

\begin{wrapfigure}{tr}{.5\textwidth}
\vspace{-2mm}
\includegraphics[width=5.6cm, viewport=25 92 555 590,clip]{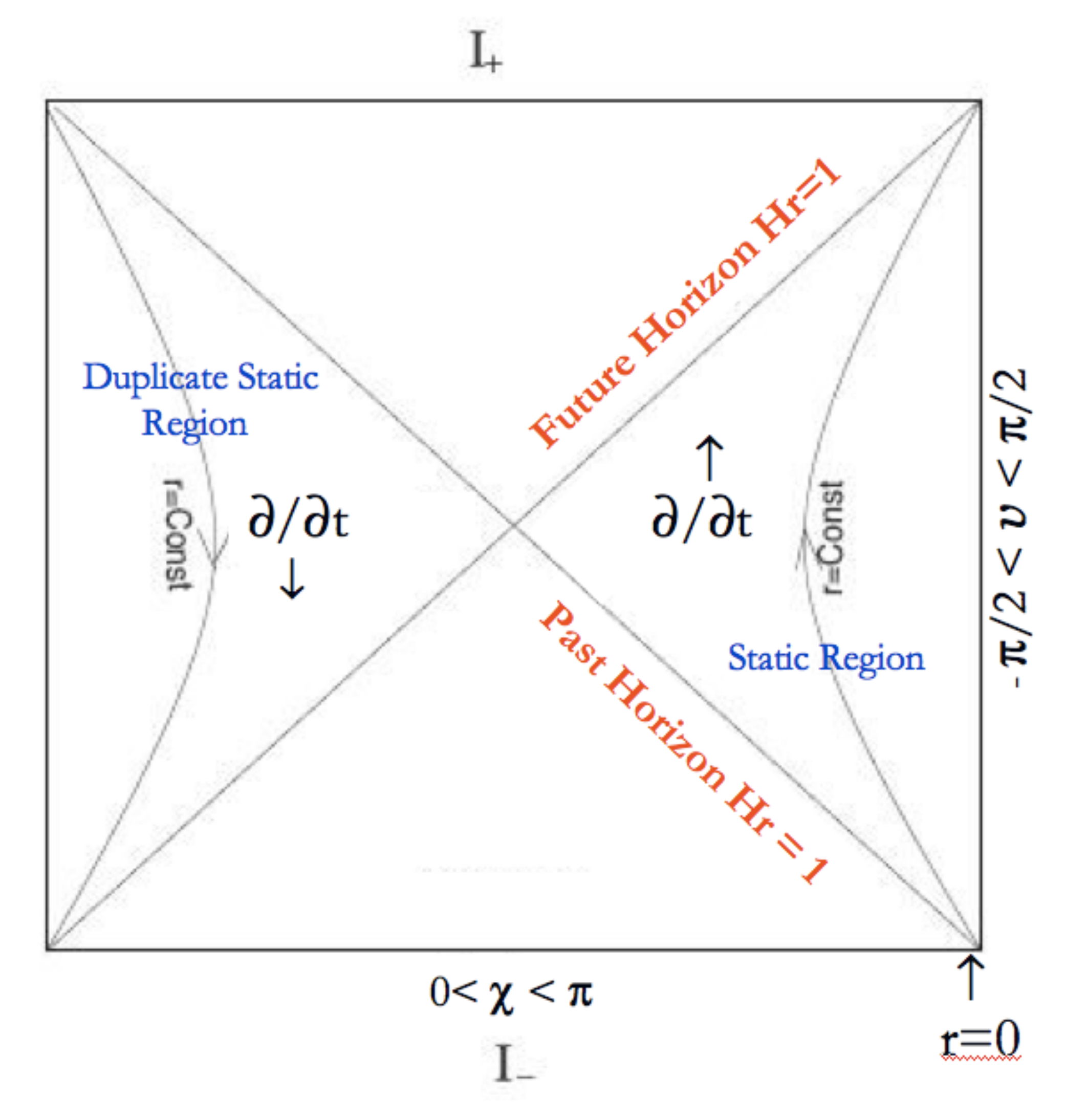}
\vspace{-2mm}
\caption{The Carter-Penrose conformal diagram of the maximal extension of de Sitter space, showing the two static 
patches where the static time Killing field $\pa_t$ runs in opposite directions. As in the maximal analytic extension of
the Schwarzschild geometry of Fig.~\ref{Fig:SchwKruskal}, this renders it impossible to define a conserved quantum Hamiltonian,
invariantly separating positive and negative frequencies that is bounded from below.}
\label{Fig:dSPenrose}
\vspace{-9mm}
\end{wrapfigure}

In both cases the horizon where the Killing vector of time translation $\pa_t$ becomes null and $f(r)=0$ in either fully analytically
extended Schwarzschild or dS space is the sign of this, so that $\La$ cannot be globally constant and positive everywhere in space in QFT. 
The conformal anomaly shows this through its sensitivity to lightlike correlations on the horizon and non-local boundary conditions 
on the quantum state. In each case one is restricted to one of the static patches only, and their joining together at their
mutual horizons as in the realization of the gravastar as a limiting case of the Schwarzschild constant density solution
in Sec.~\ref{Sec:SchwInt}. This resolves the conflict each of these globally extended solutions of Einstein's eqs.~has with QFT.

\subsection{Determination of $\ve$ and $\ell$  for a Gravastar}

It is important to recognize that in a full solution the stress tensor will not diverge. Instead the $1/f^2$ behavior of the
anomaly stress tensor will grow until (\ref{ThorizM}) becomes large enough to affect the classical Schwarzschild

geometry (\ref{sphsta})-({\ref{Sch}). At that point their backreaction effects on the geometry must be taken into
account in a self-consistent solution of the Einstein eqs.~together with the conformalon field $\vf(r)$. As long as $f(r)$
is finite, the states with large stress tensors on the horizon are continuously connected to the Hilbert space of the usual
Minkowski vacuum. 

Since the quantum anomaly terms are parametrically suppressed by a factor of $L_{\rm Pl}^2/\rM^2$ compared to classical terms, 
only when 
\be
\displaystyle{f \sim \ve\equiv \frac{M_{\rm {Pl}}}{M} \simeq 1.1 \times 10^{-38}\ \frac{M_\odot\!}{\!M} \ll 1}\hspace{-2mm}
\label{eps}
\ee 
or at $\D r = r-\rM \sim L_{\rm Pl}$,  does the anomaly stress tensor become comparable to the classical terms, for $c_{\!_S}$ in (\ref{ThorizM}) 
of ${\cal O}(1)$. This gives a {\it physical} thickness of the critical boundary layer of 
\be
\hspace{-5mm}\ell\! \sim\! \sdfrac{\D r\!\!}{\!\!\sqrt{h(r)}}\bigg\vert_{r=\rM + \D r}\hspace{-5mm} \simeq \!\sqrt{\ve}\, \rM 
\!\simeq\! 2.2\,\times\,10^{-14}\!\sqrt{\sdfrac{\! M}{M_\odot}\!} \ \, {\rm cm}
\label{ellest}
\ee
a factor of $10^{19}$ greater than the Planck scale $L_{\rm Pl}$, making a mean field EFT treatment possible, but well justifying the
thin shell approximation of the original gravastar model reviewed in Sec.~\ref{Sec:gstar1}, and determining $\ell$ in (\ref{thickness}).

\begin{svgraybox}
The result of these considerations is that the effective action $S\!\!_\cA$ of (\ref{Sanom}) amounts to a specific addition to Einstein's GR,  
consistent with, and in fact required by first principles of QFT, general covariance, and the general form of the conformal anomaly (\ref{tranom}). 
It is a relevant addition in both the mathematical and physical sense~\cite{MazEMWeyl:2001}, capturing the macroscopic light cone 
singularities of anomalous correlation functions, and a necessary part of the low energy EFT of gravity, particularly relevant
for spacetimes with BH or cosmological horizons. It therefore should be added to the Einstein-Hilbert action of classical GR, much as the 
WZ term must be added to the low energy meson theory to account for the chiral anomaly of QCD
\cite{WZ:1971,Bertlbook,MazEMWeyl:2001, EMVau:2006, GiaEM:2009, EMZak:2010, EMSGW:2017,EMEFT:2022}. 
The effective action (\ref{Sanom}) and stress tensor (\ref{Tphi1}) generated by it generically produce the large quantum 
(but still semi-classical) backreaction effects on BH and dS horizons, necessary to trigger the quantum phase transition 
hypothesized in the original gravastar paper. 
\end{svgraybox}

\section{The EFT of Gravity, and Dynamical $\La$ Vacuum Energy}
\label{Sec:EFT}

In addition to the conformal anomaly, the second essential element needed for the formation of a gravastar is the ability of vacuum energy, or
effective cosmological term $\La_{\rm eff}$, to change rapidly at the horizon. The EFT that describes this quantum phase transition
boundary layer at the horizon also provides a resolution of the `naturalness' problem of $\La$ dark energy in 
cosmology~\cite{CosmicProbesSnowmass,TF1Snowmass,EMEFT:2022}. That the physics of BHs is closely related to the `fine tuning' 
issue of cosmological vacuum energy is inherent in the gravastar proposal,  which requires a dynamical vacuum energy 
$\rho_{_V} = 3H^2/8\p G$ in the dS interior to adjust to the total mass $M$ of the gravastar to be joined at their mutual horizons $H^{-1} = 2GM$.

The basis of the solution to both problems begins with the observation that the constant $\La$ term can be reformulated in terms 
of an abelian gauge theory of a $4$-form field strength~\cite{Aurilia:1978,DuffvanN:1980,AurNicTown:1980,HenTeit:1986}
\be
F = \sdfrac{1}{\,4!}\, F\!_{\a\b\g\la}\, dx^\a \wedge dx^\b \wedge dx^\g \wedge dx^\la
\label{F4}
\ee
which is the curl of a totally anti-symmetric $3$-form gauge potential
\be
F\!_{\a\b\g\la} =4\, \pa_{[\a} A_{\b\g\la]} 
= 4\, \na_{[\a} A_{\b\g\la]} = \na_{\a} A_{\b\g\la} - \na_{\b} A_{\a\g\la}+  \na_{\g}A_{\a\b\la}- \na_{\la} A_{\a\b\g}
\label{divF}
\ee
that is, $F$ is an exact $4$-form
\be
F= dA\,,\qquad A= \sdfrac{1}{\,3!\!} \, A_{\a\b\g} \ dx^\a \wedge dx^\b \wedge dx^\g\,.
\label{FdA}
\ee
As a natural generalization of ordinary electromagnetism where $F\!=\!dA$ is an exact $2$-form exterior 
derivative of the $1$-form vector gauge potential $A\!=\!A_\m \,dx^\m$, let $F$ be provided with the `Maxwell' action
\be
S\!_F =  -\sdfrac{1}{2\vk^4}\int F \wedge \st F=  -\sdfrac{1}{\,48\, \vk^4}\! \int d^4x \sqrt{-g} \  F\!_{\a\b\g\la}F^{\a\b\g\la}
= \sdfrac{1}{2\vk^4}\!\int d^4x \sqrt{-g}\, \tF^{\,2}
\label{Maxw}
\ee         
where
\be
\tF \equiv \st F = \frac{1}{\,4!}\, \ve_{\a\b\g\la}\, F^{\a\b\g\la}\,, \qquad F\!_{\a\b\g\la} = -\ve_{\a\b\g\la}\,\tF
\label{Fdual}
\ee
is the scalar Hodge star $\st$ dual to $F$,  and $\vk$ is a free parameter whose significance  as a topological susceptibility 
of the gravitational vacuum is discussed in \cite{EMEFT:2022}.  

When the rank of the $D$-form is matched to the number of $D\!=\!4$ spacetime dimensions in this way, the free
`Maxwell' theory (\ref{Maxw}) has two very special properties: 
\begin{enumerate}[label= (\roman*)]
\vspace{-1mm}
\item  $\tF$ is constrained to be a constant, with no propagating degrees of freedom, and
\item  Its stress tensor $T^{\m\n}_{\! F}$ is proportional to the metric $g^{\m\n}$, hence equivalent to a cosmological $\La$
term.
\end{enumerate}

That $F$ and its dual $\tF$ are constant follows from the {\it sourcefree} `Maxwell' eq.~obtained by variation of~(\ref{Maxw}) 
with respect to $A_{\a\b\g}$, {\it viz.}
\begin{equation}
\na\!_\la F_{ \a\b\g}\!^{\,\la}  =  0 \,,\qquad {\rm for} \qquad J^{\a\b\g}=0
\label{FnoJ}
\end{equation}
and $\pa_\la \tF\!=\!0$, so that $\tF \!= \!\tF_0$ is a spacetime constant -- in the complete absence of any sources $J = 0$.
At the same time the stress tensor corresponding to~(\ref{Maxw})
\begin{equation}
T^{\m\n}_{  F} = \frac{2\!}{\!\!\sqrt{-g}} \frac{\del S\!_F}{\del g_{\m\n}} =
-\frac{1}{4!\, \vk^4}\left(\sdfrac{1}{2} g^{\m\n} F^{\a\b\g\la}F\!_{\a\b\g\la} - 4 F^{\m\a\b\g}F^{\n}\!_{\a\b\g}\right)
= -\frac{1} {2\vk^4}\, g^{\m\n} \,\tF^{\,2}
\label{TF}
\end{equation}
is proportional to the metric tensor, if the convention that $F\!_{\a\b\g\la}$ with all lower indices is independent of the metric is
adopted.

Hence~(\ref{F4})--(\ref{TF}) are completely equivalent to a cosmological term in Einstein's eqs.~in $D = 4$
dimensions, with the identification
\begin{equation}
\La_{\rm eff} = \frac{4\p G}{\vk^4}\, \tF^{\,2} \ge 0
\label{Lameff}
\end{equation}
the effective (necessarily non-negative) cosmological constant term for $\vk$ and $\tF$ real constants.
In this way one can freely trade a positive cosmological constant $\La$ of classical GR for a new fundamental constant $\vk$ 
of the low energy EFT, together with an integration constant $\tF_0$ of the constraint $\pa_\la \tF= 0$.

The integration constant $\tF_0$ can be fixed by a classical global boundary condition in flat space, without any reference 
to quantum zero-point energy, UV divergences, or cutoffs. A vanishing $\La_{\rm eff}$ corresponds instead to the vanishing of the 
sourcefree `electric' field strength $\tF = F^{0123}$ in infinite $3$-dimensional empty flat space, analogous to the vanishing  
of the electric field in the vacuum of ordinary electromagnetism in the absence of sources. In either case this is simply the classical 
state of lowest energy, as well as the unique state that is even under the discrete symmetry of space parity inversion.

\begin{svgraybox}
Moreover the setting of the value of the free constant $\tF = \tF_0$, which is {\it a priori} independent of geometry, to zero in empty
flat space, is {\it required} by the sourcefree Einstein's~eqs.
\begin{equation}
\left[R_{\m\n} - \sdfrac{R}{2} g_{\m\n}\right]_{\rm flat} = 0 =-\La_{\rm eff}\Big\vert_{\rm flat}\,\h_{\m\n}
\label{Einflat}
\end{equation}
viewed as a low energy EFT. This shows that flat space QFT estimates of vacuum energy in any way dependent upon UV cutoffs 
or heavy mass scales are {\it inconsistent} with Einstein's eqs. It is well-known that QFT in flat space is sensitive only to 
{\it differences} in energy. Hence the absolute value of quantum zero point energy in flat space, and its dependence upon cutoffs 
or UV regularization schemes is arbitrary and of no physical significance~\cite{Sola:2022}. The value of $\La_{\rm eff}$ is significant only through 
its gravitational effects, and hence cannot be evaluated in isolation, but only within the context of  a gravitational EFT, `on shell' 
as in~(\ref{Einflat}), and only if each side of (\ref{Einflat}) can be evaluated {\it independently}. This becomes possible only 
if $\La_{\mathrm{eff}}$ is a {\it free} constant of integration, as it is in (\ref{Lameff}), and not a fixed parameter of the Lagrangian.

For the longstanding problem of the cosmological term when QFT is coupled to gravity,  the classical state of minimum energy is 
that of vanishing $4$-form classical coherent field or condensate: $\tF\! =\! \tF_0\! =\! 0$. By the identification of $\La_{\rm eff}$
in~(\ref{Lameff}), the condition (\ref{Einflat}) automatically sets the value of the cosmological term to zero in infinite flat Minkowski space.
By simply allowing a consistent flat Minkowski solution, this removes one oft-stated obstacle and `no-go theorem' to solution of the `cosmological
constant problem'~\cite{WeinbergRMP:1989,KlinkVol:2010}. 

In isolation this reformulation of the cosmological constant $\La_{\rm eff}$ in terms of $\tF =\tF_0$ and $\vk$ shifts the
consideration of cosmological vacuum energy away from the UV divergences of QFT to a macroscopic (IR) boundary condition
solving the classical constraint eq.~(\ref{FnoJ}) and minimization of energy in flat space. In addition to removing the fine tuning 
or naturalness problem of $\La$, introducing an independent $4$-form field $F$ in place of constant $\La$ also allows for the 
introduction of sources in~(\ref{FnoJ}) that will enable $F$ (and hence $\La_{\rm eff}$) to change, departing from its zero value 
in infinite sourcefree flat space in finite calculable ways, and eventually to become a full-fledged dynamical variable of the low energy 
EFT of gravity in its own right.
\end{svgraybox}

This is exactly what occurs if the $3$-form potential $A$ is identified with the torsional part of the $3$-form Chern-Simons
potential naturally defined by the Euler-Gauss-Bonnet (EGB) term $E$~(\ref{ECdef}) in the trace anomaly. This term 
is distinguished by its topological character. Its integral is a topological invariant insensitive to local variations, and
is also associated with a $4$-form gauge field 
\begin{equation}
\mathsf{F}\equiv \ve_{abcd}\, \mathsf{R}^{ab}\wedge  \mathsf{R}^{cd} = \sdfrac{1}{4} \e_{abcd}\, R^{ab}_{\ \ \,\a\b}\, R^{cd}_{\ \ \,\g\la}
\, dx^\a \wedge dx^\b \wedge dx^\g \wedge dx^\la
\label{Fdef}
\ee
which is also exact, {\it i.e.}~$\mathsf{F}=d \mathsf{A}$,  where $\mathsf{A}$ is the $S\!O(3,\!1)$ Lorentz frame dependent Chern-Simons 
$3$-form~\cite{PadYal:2011}
\begin{equation}
\mathsf{A}= \e_{abcd}  \left(\w^{ab} \wedge d \w^{cd} + \sdfrac{2}{3}\, \w^{ab} \wedge \w^{ce}\wedge \w^{fd}\, \h_{ef}\right)
\label{eA}
\end{equation}
defined in terms of the $S\!O(3,\!1)$ spin connection $\w^{ab}$. The spin connection is {\it a priori} independent of the spacetime metric,
becoming locked to it only by the condition of zero torsion, in which case $\w^{ab}$ becomes the usual Christoffel connection in
holonomic coordinates~\cite{EguGilHan:1980,Gockeler}. In Einstein-Cartan geometries non-zero torsion as well as non-zero curvature 
are allowed, as general considerations of minimal coupling of fermions to gravity 
require~\cite{Cartan:1922,Utiyama:1956,Sciama:1964,Hehl:1976}. Then $\w^{ab}$ and hence $\mathsf{A}=\mathsf{A_R}+ \mathsf{A_T}$
acquire a non-Riemannian torsion dependent contribution $\mathsf{A_T}$ independent of the metric, in addition to the Riemanian part 
$\mathsf{A_R}$. The identification of $\mathsf{A_T}=A$ with the 3-form potential of (\ref{divF})-(\ref{FdA}), which must be varied
independently of the metric, then couples the anomaly scalar $\vf$ to $\La_{\rm eff}$, making it spacetime dependent.

This is established as follows. Since $\mathsf{F}$ is related to $E$ by the Hodge star $\st$ dual
\begin{align}
\st \mathsf{F}& =  \ve^{\a\b\g\la} \left(\sdfrac{1}{4} \e_{abcd}\, R^{ab}_{\ \ \,\a\b}\, R^{cd}_{\ \ \,\g\la}\right)
=  \sdfrac{1}{4} \e_{abcd}\,\e^{mnrs}\, R^{ab}_{\ \ \ mn}R^{cd}_{\ \ \ rs}   \nn
& =-\left(R_{abcd}R^{abcd} - 4 R_{ab}R^{ab}+ R^2\right)  =  -E
\label{starF}
\end{align}
which in component form means that
\begin{equation}
E =  -  \big(\st d\mathsf{A}\big) = \sdfrac{1}{\,3! } \, \ve^{\a\b\g\m} \pa_\m \mathsf{A}_{\a\b\g}
= \sdfrac{1}{\,3! } \,\na\!_\m \left( \ve^{\a\b\g\m}\mathsf{A}_{\a\b\g}\right)
\label{EdA}
\end{equation}
is a total derivative, the torsional part of the $E\vf$ term in the anomaly effective action is first separated off, by
replacing $E$ in (\ref{tranom}) and the anomaly effective action (\ref{Sanom}) with $E -(\st d\mathsf{A_T})= E -\tF$, 
where $E$ is now defined to be the strictly Riemannian non-torsional part. The torsion dependent term in (\ref{Sanom})
can then be integrated by parts
\begin{equation}
-\sdfrac{b'}{2\,} \int d^4 x \sqrt{-g}\, \tF\,\vf =
-\sdfrac{b' }{2} \sdfrac{1}{\,3! } \,\int   d^4x \sqrt {-g}\,  A_{\a\b\g}\, \ve^{\a\b\g \m}\, \pa_\m \vf
\label{Ephi}
\end{equation}
up to a surface term which does not affect local variations. Defining the $3$-current
\begin{equation}
J^{\a\b\g} \equiv  -\sdfrac{\,b' }{2}\, \ve^{\a\b\g \m}\, \pa_\m \vf
\label{Jphi}
\end{equation}
the last term in~(\ref{Ephi}) can be expressed in the form
\begin{equation}
S_{\rm int}[\vf, A] = \sdfrac{1}{\,3! }\int  d^4x \sqrt {-g}\ J^{\a\b\g} A_{\a\b\g}
\label{Sint}
\end{equation}
analogous to a $J\cdot A$ interaction of ordinary electromagnetism. Since the `Maxwell' action~(\ref{Maxw}) and interaction 
term~(\ref{Sint}) must be stationary with respect to independent variation of $A_{\a\b\g}$,  with $g_{\m\n}$ and $\vf$ fixed, 
(\ref{Sint}) provides a source term for the `Maxwell' eq.~of the $4$-form gauge field
\begin{equation}
\na_{\!\la} F^{\a\b\g\la} = \vk^4  J^{\a\b\g} = -\sdfrac{\vk^4 b'}{2}\,\ve^{\a\b\g\la}\,\pa_\la\vf
\label{Maxeq}
\end{equation}
with the source current~(\ref{Jphi}).  Upon taking its dual, with~(\ref{Fdual}), this becomes
\begin{equation}
\pa_\la\left(\tF - \sdfrac{\vk^4 b'}{2}\, \vf \right) = 0
\label{paF}
\end{equation}
which is an eq.~of constraint  that is immediately solved by
\begin{equation}
\tF = \sdfrac{ \vk^4 b' }{2} \,\vf + \tF_0
\label{Fsoln}
\end{equation}
in which $\tF_0$ is a spacetime constant. The result is that $\tF$ given by (\ref{Fsoln}) is no longer a constant, and will change, 
as will the effective cosmological term $\La_{\rm eff}$ in (\ref{Lameff}), when $\vf$ changes according to (\ref{phieom1}). From 
(\ref{phiS})-(\ref{ThorizM}) and (\ref{phidS})-(\ref{ThorizH}), this most rapid change occurs at the BH and dS horizons.
The blueshifting of local frequencies $\sim  f^{-\frac{1}{2}}$ as in~(\ref{redshift}) leads to the $\vf$ field having an increasingly 
large radial derivative in the vicinity of $\rM\simeq \rH$, so that (\ref{Jphi}) and (\ref{Tphi1}) and the torsional effects become significant 
there. Thus the addition of the $4$-form gauge field `Maxwell' action (\ref{Maxw}) to the effective action of gravity, which exactly reproduces
a positive cosmological term (\ref{Lameff}) in the purely classical theory, allows (and requires) the vacuum energy to change
when its torsional part is coupled the conformalon scalar and effective action of the quantum conformal anomaly by (\ref{Sint}).

The scalar $\tF$, dual to the $4$-form field strength $F$ is a classical coherent field that provides an explicit realization 
of a gravitational condensate $\La_{\rm eff}$ interior. When coupled to the conformalon scalar through the $3$-form abelian 
current $J^{\a\b\g}$, concentrated on a three-dimensional extended world tube of topology ${\mathbb R} \times {\mathbb S^2}$ 
where $\pa_\m\vf$ grows large, $\vf, \tF$ and the condensate $\La_{\rm eff}$ all change rapidly in the radial direction. This is precisely 
the appropriate description of a thin shell phase boundary layer of a gravastar with $\mathbb S^2$ spatial topology sweeping out 
a world tube in spacetime. 

The resulting low energy EFT of gravity contains two additional degrees of freedom in addition to the 
metric, which lead to a coupled set of eqs.~relevant for the near-horizon geometry of a gravastar where $\vf$ and $\La_{\rm eff}$ change 
rapidly~\cite{EMEFT:2022}. This realizes the quantum phase transition at the horizon hypothesized in the original gravastar
proposal~\cite{gravastar:2001,Grav_Univ:2023, MazEMPNAS:2004}.

That the extension of GR to Einstein-Cartan spacetime with torsion should occur where the vierbein `soldering form'
$e^0_{\ t}$ vanishes, and the locking together of the $S\!O(3,\!1)$ tangent space gauge group and $GL(4,\mathbb{R})$ group 
of coordinates transformations is broken, is related to some earlier studies~\cite{DAuriaRegge:1982}. However the specific
relation to the conformal anomaly and its macroscopic effects on horizons as the locus of where these torsional effects
should first appear is introduced only recently in \cite{EMEFT:2022}.

In the resulting EFT of gravity one can now search for static, rotationally invariant solutions of the EFT eqs.~which are 
externally Schwarzshild with $\La_{\rm eff}$ vanishing in the exterior region, changing rapidly but continuously
near the Schwarzschild $\rM$ or dS $\rH$ classical horizons by (\ref{Fsoln}) and (\ref{phieom1}) 
in the phase boundary region, and then remaining nearly constant $3H^2$ in the interior region. 
The stability and normal modes of oscillations about this solution can be studied in a systematic way
within a Lagrangian framework, for the first time, without reliance upon fluid ans\"atze. 
The interactions of SM fields with the quantum phase boundary surface layer of thickness given by (\ref{thickness})
can be studied in this EFT as well, and the effects of this surface layer and regular dS interior on accretion,
binary BH mergers, gravitational waves, ringdown and `echoes' investigated.
The effective Lagrangian framework for gravity also will permit studies of gravastar formation by
the triggering of a quantum phase transition in the EFT before any trapped surface can form.

Clearly there is considerable work remaining to be done, but the path seems open to the goal 
of establishing gravitational condensate stars as the stable endpoint of gravitational collapse
consistent with, in fact relying upon quantum field theory.

\section*{Appendix A: Thin Shell vs. Thick Shell}

\label{Sec:Shell}

The distinguishing feature of the original gravitational condensate star proposal of \cite{gravastar:2001,Grav_Univ:2023} 
of the main text is the abrupt change in ground state vacuum energy at the horizon, characteristic of a quantum phase transition there. 
This should be clear from the essential role of the horizon as a infinite red shift surface in both \cite{ChaplineHLS:2001} and 
\cite{gravastar:2001,Grav_Univ:2023}, the assumption of $\ve \ll 1$ and the estimate of $\ve$ and $\ell$ in \ref{Sec:Anom}
from the conformal anomaly. The proper length $\ell$ determined by the stress tensor of the conformal anomaly takes the place of the 
`healing length' introduced, but left undetermined in the analogy of the horizon in GR to the non-relativistic quantum critical surface 
of a sound horizon in \cite{ChaplineHLS:2001}. Thus the term `gravastar' should apply only to the gravitational condensate star model of 
\cite{gravastar:2001,Grav_Univ:2023} in the text, described also in \cite{MazEMPNAS:2004}, and further refined in \cite{MazEM:2015}, 
where the lightlike null horizon clearly plays a privileged role as the locus of joining of interior and exterior classical geometries, with equal 
and opposite surface gravities, and where the conformal anomaly stress tensor also grows large, and a quantum phase transition can occur.

Despite this physically privileged role of the horizon in the original gravastar proposal~\cite{gravastar:2001,Grav_Univ:2023}, a number of papers 
appeared subsequently that discussed what may be called `generalized gravastars,' or regular solutions with macroscopically 
large or `thick' shells, comparable to the gravitational radius $\rM$ itself, with compactness $GM/r$ differing from the maximal value of $1/2$, 
by order unity, some time varying, or with timelike surfaces displaced from the Schwarzschild or dS horizons by finite amounts 
\cite{VisserWilt:2003,DymnGalak:2005,Lobo:2006,DeBenHorvat:2006,ChirRez:2007,HorvIlij:2007,BertiCardPani:2009,BertiCardPani:2010,
Mar-MorGarLoboVisser:2012,SakaiGravShadow:2014,Pani:2015,Uchikata:2015,ChirRez:2016,UchiPani:2016,VolkKok:2017,RaySenNim:2020,SenGhoRayMisTri:2020}.
Several authors proceeded to discuss both ergoregion instabilities and observational bounds on such hypothetical objects, with various 
assumptions about boundary conditions and surface matchings \cite{ChirRez:2008,CardPaniCadCavPRD:2008, CardPaniCadCavCQG:2008}.
It should be clear that these instabilities or observational bounds do not apply to gravastars, which {\it by definition} are static 
configurations with a {\it infinitesimally thin shell located at the horizon}, for $\ve =0$, or straddling and replacing the would-be 
classical Schwarzschild and dS horizons for very small finite $\ve$, with metric functions $f \sim h = \cO(\ve)$ there. Any other regular QBH
is not a gravastar.

\section*{Appendix B: Observational Bounds, and  Claims of `Proofs' of Horizons}

Because the exterior geometry of a gravastar is identical to that of a classical BH down to the scales of its very thin shell surface
boundary layer at $\ell$ given by (\ref{thickness}) above the would-be classical horizon, it should be clear that a gravastar will be cold, 
dark and indistinguishable from a BH by almost all traditional astronomical observations. The first images of the `BHs'
at the center of M87 and Sgr A$^*$ at the center of own Milky Way galaxy by the Event Horizon Telescope, impressive as they are, 
lack the angular resolution to resolve the very near $r=\rM$ horizon region, and are principally viewing the electromagnetic radiation 
from the light ring or photon sphere ($r=3GM/c^2$ for a non-rotating BH), significantly farther away from the horizon
than the very small $\ell$ of (\ref{thickness})~\cite{GraHolzWald:2019}. If light at these mm wavelengths from behind the `BH' 
is either too faint, absorbed by the surface or accreting material, or too defocused to be observed, the EHT images will not be able 
to distinguish a gravastar from a BH.

\begin{svgraybox}
If there is a surface which is as deeply redshifted as the semi-classical estimates of $\ve$ and $\ell$ would imply,
any radiation emitted from the surface can escape to infinity only if emitted from a tiny `pinhole' solid angle less than 
of order $\ve$ from the perpendicular, or it will fall back onto the surface. Attempts to `prove' the existence of a BH horizon 
or absence of a surface from the absence of thermal radiation and/or absence of X-ray bursts which would be expected {\it if} 
the surface is composed of conventional matter, and {\it if} any advected matter deposited onto the surface is re-radiated rather 
than absorbed, are therefore bound to fail. This point was succintly made in~\cite{AbramKluzLas:2002}, soon after the gravastar 
proposal of \cite{gravastar:2001,Grav_Univ:2023}. The authors of~\cite{AbramKluzLas:2002} also recognized that any 
surface of an ultracompact QBH was bound {\it not} to be composed of conventional matter, such as a neutron star crust, needed 
for the thermonuclear reactions that give rise to X-ray bursts. Moreover, in order for the gravastar proposal to be a viable alternative 
for a BH of {\it any} mass, a gravastar must be able to absorb accreting baryonic matter and convert it to the interior condensate, 
thereby growing its mass to any larger value. Any substantial efficiency of absorption and conversion of energy to interior condensate 
would reduce the energy re-radiated and make the object dark in most if not all the observable electromagnetic spectrum. 
\end{svgraybox}

The authors of \cite{BrodNary:2007} argued for quite stringent limits on what they called `gravastar' models, in the context of a certain 
specialized assumption of internal energy of the `matter' composing the QBH, assuming a conventional thermalization of accreting matter 
in a steady state emission. Aside from not accounting for the relativistic `pinhole' effect suppressing all emission from a deeply redshifted 
surface, and ignoring the possibility of near total absorption of accreting matter without any heating of the QBH, which would all but eliminate 
any thermal re-emission whatsoever, the observational bounds of \cite{BrodNary:2007} are attempts to constrain the condensed matter 
analog model of Chapline {\it et al.} \cite{Chap:2003,Chap:2005}, which in any case is not the gravastar described 
in \cite{Grav_Univ:2023,MazEMPNAS:2004}, this article, or \cite{MazEM:2015}.  

\begin{svgraybox}
Similar arguments based on thermalization and steady state re-emission of radiation, again ignoring the possibility of absorption by the 
QBH surface, with claims of strong observational bounds were made in \cite{NarayMc:2008,BrodNarayDoel_ApJ:2015}. These and
similar unjustifiably strong claims of `proof' of BH horizons and the assumptions upon which they are based have been critically examined 
by several authors~\cite{CardPani:2017,CarbDiFiLibViss:2018,CardPani:2019,CarbRubDiFilLibVis:2022}, and shown to be flawed. First, the
assumption that thermodynamic and dynamic equilibrium can be established between an accretion disk and the QBH on a 
reasonably short timescale is incorrect for a deeply redshifted surface for $\ve \to 0$, due to the gravitational lensing `pinhole' effect,
already pointed out in~\cite{AbramKluzLas:2002}. The best limits one can obtain from the observations of M87 or Sgr A* when
this classical GR lensing is taken into account is in the range of $\ve < 10^{-15}\ {\rm to}\ 10^{-17}$, impressive, but still many
orders of magnitude short of $10^{-38}$ expected for a gravastar. Second, the energy emitted was assumed to be electromagnetic
in order to be observable, whereas a sizable fraction of any re-emitted energy could be in the form of neutrinos or other unobserved
radiation~\cite{CardPani:2017,CardPani:2019}. Third, and most importantly, as already mentioned, a sizable fraction even approaching 
unity of the accreting matter may be absorbed by the gravastar, with virtually no re-emission whatsoever. As a result, there are no 
useful bounds from the non-observation of electromagnetic emission from any astrophysical QBH, and the possibility that they may all be gravastars with $\ve \ll 10^{-17}$ remains open.
\end{svgraybox}

The converse claim of a {\it lower} bound of $\ve \gtrsim 10^{-24}$ in \cite{CarbKumarLu:2018} is based on a strong assumption 
of the restrictive form of the Vaidya metric and stress tensor in the vicinity of the QBH surface, setting to zero all of its components 
except $T_{vv}$ in advanced null coordinates. This bound also disappears if the assumption upon which it is based is relaxed,
as it almost certainly should be.

\section*{Appendix C: Gravitational Waves and Echoes}

The observation of gravitational waves (GWs) by LIGO/LSC \cite{LIGO1:2016} has opened up a new window on the universe that among 
many other interesting possibilities provides perhaps the best opportunity for observational tests of the gravastar proposal. The GW data is
not yet accurate enough to test the prediction of a {\it discrete} spectrum of ringdown modes from a non-singular gravastar with a surface 
made in \cite{MazEM:2015}. Indeed it was quickly realized that sensitivity to the nature of a very compact QBH with $\ve \ll 1$ is obtained 
only some delay time after the initial GW merger signal, in the ringdown phase \cite{CardFranPani:2016}, where the signal/noise ratio is 
very much lower. Nevertheless a regular QBH such as a gravastar could produce a GW `echo' at multiples of the characteristic time 
\be
\D t \sim 2GM\, \ln (1/\ve)
\label{Delt}
\ee
after the compact object merger event~\cite{CardHopMacPalPani:2016}. These may be observable with the improved sensitivities of
Advanced LIGO and future detectors. 

The basis for such echoes is the expectation that GWs produced in the merger could reflect from the internal centrifugal barrier 
of a gravastar and re-emerge with a logarithmically long time delay for $\ve \ll 1$, thus in principle opening up the possibility of 
testing GR and the nature of QBHs on scales very close to the would-be horizon, and their interior. A somewhat different scenario
was considered in~\cite{AbediDykAfsh:2017}, with a claim of tentative evidence for an echo signal in the LIGO data.
However, an analysis of the same data by members of the LIGO/LSC collaboration concluded that the echo signal was just $1.5 \s$ above the 
noise level~\cite{WesterNiel:2018}. The subject of GW echoes from QBHs such as gravastars continues as an area of active research
\cite{BarCarGar:2017,VolkKok:2017,MasVolkKok:2017,WangOshAfs:2020}, requiring substantially more data from Advanced LIGO and 
successor detectors to resolve this question or possibly provide the first evidence of deviation in the universe from the mathematical
BHs of present textbooks~\cite{CardPani:2019}. 


\begin{acknowledgement}
It is a pleasure to acknowledge the seminal contributions of Pawel O. Mazur  to the inception of the proposal of gravitational
condensate stars in \cite{gravastar:2001,Grav_Univ:2023,MazEMPNAS:2004}, as well as numerous insights on the 
conformal anomaly, revisiting the Schwarzschild interior solution and the bringing ref.~\cite{DAuriaRegge:1982} to the author's 
attention. The author also gratefully acknowledges the other colleagues with whom he has collaborated on various aspects of
gravitational condensate stars, de Sitter space and the effective action of the conformal anomaly, including P. R.~Anderson,
I. Antoniadis, P. Beltracchi, D. Blaschke, R. Carballo-Rubio, G. Chapline, C. Corian{\`o}, M. Giannotti, P. Gondolo, M. M. Maglio, 
C. Molina-Par{\` i}s, I. L. Shapiro and R. Vaulin. The recent hospitality of the gravity group at SISSA of S. Liberati, and gravity wave 
research group at SISSA of E. Barausse in the organization of the workshop ``Quantum Effective Field Theory and Black Hole 
Tests of Einstein Gravity" [https://grams-815673.wixsite.com/september12-16] in September, 2022 is also gratefully acknowledged.
\end{acknowledgement}

\bibliographystyle{myspringer}
\raggedright
\bibliography{gravity21Aug}


\end{document}